\documentclass{aa}
\usepackage[varg]{txfonts}
\usepackage[T1]{fontenc}
\usepackage[final]{microtype}
\usepackage{graphicx}	
\usepackage{xcolor}	
\usepackage{amsmath}	
\usepackage{amssymb}	
\usepackage{caption}    
\usepackage{subcaption} 
\usepackage{multicol}  
\usepackage{booktabs}  
\usepackage{longtable}
\usepackage{footmisc}
\usepackage[breaklinks=true,colorlinks=true,linkcolor=blue,citecolor=blue,urlcolor=blue]{hyperref}
\usepackage{upgreek}
\usepackage{orcidlink}
\usepackage{placeins}
\bibpunct{(}{)}{;}{a}{}{,} 
\usepackage{soul} 
\usepackage{ulem} 
\usepackage{dblfloatfix}
\usepackage{float}
\usepackage{wrapfig}
\usepackage{textcomp}

\newcommand{\Msun}{{\rm M_{\sun}}}

\newcommand{\kpc}{{\rm kpc}}

\newcommand{\stream}{\rm s}
\newcommand{\cgm}{\rm cgm}

\allowdisplaybreaks[2]

\begin{document}
\title{Cosmic ray heating of cold streams}
\subtitle{Implications for the gas supply and growth of massive galaxies}

\author{Ellis R. Owen\orcidlink{0000-0003-1052-6439}\inst{1, 2}
\and 
Nicolas Ledos \orcidlink{0000-0001-9699-8941}\inst{3} 
\and 
Evangelia Ntormousi\orcidlink{0000-0002-4324-0034}\inst{4}
\and \\
Shinsuke Takasao\orcidlink{0000-0003-3882-3945}\inst{5}
\and 
Kentaro Nagamine\orcidlink{0000-0001-7457-8487}\inst{2,6,7,8,9}
\and 
Sebastiano Cantalupo\orcidlink{0000-0001-5804-1428}\inst{3}
}

\offprints{E. R. Owen, \email{ellis.owen@riken.jp}; N. Ledos, \email{nicolas.ledos@unimib.it}}

\institute{
Astrophysical Big Bang Laboratory (ABBL), RIKEN Pioneering Research Institute (PRI), Wak\={o}-shi, Saitama 351-0198, Japan 
\and 
Theoretical Astrophysics, Department of Earth and Space Science, The University of Osaka, 1-1 Machikaneyama, Toyonaka, Osaka 560-0043, Japan
\and
Dipartimento di Fisica "G. Occhialini”, Universit\`{a} degli Studi di Milano-Bicocca, Piazza della Scienza 3, 20126 Milano, Italy
\and
Scuola Normale Superiore di Pisa, Piazza dei Cavalieri 7, 56126 Pisa, Italy
\and 
Humanities and Sciences/Museum Careers, Musashino Art University, Tokyo 187-8505, Japan 
\and 
Theoretical Joint Research, Forefront Research Center, Graduate School of Science, The University of Osaka, 1-1 Machikaneyama, Toyonaka, Osaka 560-0043, Japan 
\and 
Kavli IPMU (WPI), UTIAS, The University of Tokyo, 5-1-5 Kashiwanoha, Kashiwa, Chiba 277-8583, Japan
\and 
Department of Physics and Astronomy, University of Nevada, Las Vegas, 4505 S. Maryland Pkwy, Las Vegas, NV 89154-4002, USA
\and
Nevada Center for Astrophysics, University of Nevada, Las Vegas, 4505 S. Maryland Pkwy, Las Vegas, NV 89154-4002, USA
}

\date{Received -- / Accepted --}

\abstract{Recent observations have demonstrated the presence of cosmic rays (CRs) in cosmic-web filaments. 
Cold streams supply gas inflows from these filaments into massive galaxies during the cosmic noon. As these streams are expected to be magnetised, external cosmic-web CRs may become entrained with this inflowing gas.} 
{We aim to determine whether this externally-supplied CR population can deposit energy to alter or disrupt the supply of cold gas to galaxies. } 
{We couple a 
spectrally-resolved 
CR transport calculation to a redshift-dependent analytical model of magnetised cold streams in galaxy haloes and investigate whether externally-supplied CRs can modify gas supply through this channel.} 
{We find that CR energy deposition can alter the thermal state of cold streams. Dense stream cores remain largely resilient and only experience weak heating. Their temperature is raised by less than a factor of 10, which is insufficient to overcome radiative cooling at the stream-CGM interface. In more diffuse streams, and in partially mixed interface gas of the most massive haloes near the virial radius, CR heating becomes strong enough that radiative cooling can no longer balance it, and the gas is heated toward or above the mixing-layer temperature. This weakens the stability of the stream, making it more susceptible to disruption. Complete evaporation is possible only in extreme cases. In a cosmological context, cold streams are therefore more vulnerable to CR heating at larger galactocentric radii, higher halo masses, and in more diffuse or partially-mixed stream material.} 
{By preferentially heating diffuse or partially mixed gas at stream-CGM interfaces, externally supplied CRs may introduce additional selectivity into cold-gas accretion that modifies the gas supply and growth of massive galaxies. These CRs weaken fragile streams and erode their cold envelope, and may cause surviving cold flow components to appear thinner and more sharply confined far into galaxy haloes.} 
 
\keywords{ Galaxies: evolution --  Galaxies: haloes -- (Galaxies:) intergalactic medium -- Magnetic fields -- (ISM:) cosmic rays -- Methods: analytical -- Methods: numerical}
\maketitle

\section{Introduction}
\label{sec:introduction}

Cosmic rays (CRs) are an important non-thermal component of astrophysical environments across a broad range of scales.
They are most readily observed in collapsed baryonic structures, where they are often associated with astrophysical feedback and can influence gas dynamics, thermal balance, and non-thermal emission. 
Their effects have been studied extensively in galaxy clusters~\citep[e.g.][]{Ackermann2012ApJ,Brunetti2014IJMPD,Guo2008MNRAS,Ruszkowski2017ApJ,vanWeeren2019SSRv,Rajpurohit2022ApJ,Lin2023MNRAS,Ruszkowski2023A&ARv,Globus2025ARA&A}, galaxies~\citep[for reviews, see][]{Blasi2013A&ARv,Gabici2019IJMPD,Owen2023Galax,Ruszkowski2023A&ARv}, and in circum-galactic media/galaxy haloes~\citep{Ji2020MNRAS,Quataert2025OJAp,Romano2025A&A,Roy2025arXiv,Ponnada2026ApJ}. 
More recently, observations have also provided evidence for CRs in larger-scale cosmic-web structures.
Stacking analyses of close cluster pairs at $z\lesssim0.7$ have revealed faint, polarised synchrotron emission extending between neighbouring clusters~\citep{Vernstrom2021MNRAS}.
This emission provides statistical evidence for a relativistic CR electron population in intergalactic filaments and their associated accretion shocks~\citep{Vernstrom2021MNRAS,Vernstrom2023SciA}. 

The faintness of the observed synchrotron signal means that evidence for CR electron populations in filaments is currently limited to low redshift.
However, the shock-acceleration processes responsible for these CRs should also operate during the earlier stages of structure formation.
External accretion shocks around filaments can reach high Mach numbers as they decelerate cold, previously unshocked gas~\citep{Ryu2003ApJ}, and may convert a larger fraction of their dissipated energy into CRs than the lower-Mach shocks that dominate at later times~\citep{Jubelgas2008A&A}.
Such structure-formation shocks are expected to be among the most volume-filling sources of CRs in the Universe, and may dominate the CR energy budget of large-scale structure compared to the more localised contribution from astrophysical feedback~\citep{Vazza2025A&A}.
The high-redshift cosmic web may therefore host a relic CR population generated during structure assembly~\citep{Miniati2000ApJ,Miniati2001ApJ,Loeb2000Natur}. This population is distinct from the CRs supplied by astrophysical feedback from galaxies. It is present in the intergalactic environment, and may be the dominant CR population present in gas supplied to galaxies. In contrast, feedback-produced CRs originate within galaxies and must propagate out of the denser star-forming gas before reaching the surrounding halo~\citep[e.g.][]{Hopkins2020MNRAS}. Their different origins and propagation histories may therefore lead to different spatial and spectral distributions, with corresponding differences in their astrophysical effects. 

The observed synchrotron emission from the cosmic web only probes CR electrons, but accretion and structure-formation shocks should also accelerate protons at least as efficiently~\citep{Park2015PhRvL,Gupta2024ApJ}.
Unlike rapidly cooling CR electrons, CR protons in large-scale structure conditions lose energy comparatively slowly.
High-energy protons can survive for Gyr timescales or longer against pion-producing inelastic interactions~\citep{Berezinsky1997ApJ,Wu2024Univ}, while lower-energy protons are similarly long-lived against Coulomb and ionisation losses~\citep{Brunetti2014IJMPD,Sazonov2015MNRAS,Leite2017MNRAS}. 
The synchrotron-emitting CR electrons therefore only trace the short-lived component of a substantially larger, longer-lived reservoir of CR protons in the cosmic web.  

If the cosmic web has hosted a long-lived CR reservoir since early epochs, it would have consequences for galaxy evolution.
During the cosmic noon, massive galaxies are fuelled in part by streams of dense, $\sim 10^4 \,{\rm K}$ gas that penetrate the $\gtrsim 10^6 \,{\rm K}$ hot circumgalactic medium (CGM) and deliver material toward central star-forming regions. 
Although this scenario is motivated primarily by numerical simulations~\citep[e.g.][]{Fardal2001,Keres2005,Dekel2006,Dekel2009,Waterval2025}, observations have also revealed dense ($n_{\rm H}>1 \,{\rm cm}^{-3}$), cold ($10^4 \,{\rm K}$) emitting halo gas at $z>2$ that could fuel galaxy and black-hole growth~\citep[for a review, see][]{Cantalupo2017ASSL}. 
Because cold streams are fed by gas from the wider intergalactic environment~\citep{Dekel2006,Dekel2009,Keres2009MNRAS,Emonts2023Sci}, an external cosmic-web CR reservoir could become associated with this inflowing material before it enters the halo. 

How these CRs then enter and move through the halo is then governed by the magnetic structure of the stream. 
Recent studies have shown that, despite the initially weak magnetic field of the intergalactic medium, cold streams can amplify magnetic fields to micro-Gauss ($\mu{\rm G}$) strengths through their interaction with the surrounding hot CGM. 
Amplification can arise from turbulent stretching in the cold-hot mixing layer surrounding the stream, where cold stream gas interacts with the ambient hot CGM~\citep{Ledos2024a,Ledos2024b,Kaul2025,Das2024MNRAS}, while magnetic draping around cold CGM structures may further regulate transport across cold-hot interfaces~\citep{Ramesh2024A&A}. 
This magnetisation therefore allows a CR population associated with the upstream gas to remain tied to the stream as it falls inward through the halo.  

In contrast to CRs produced by feedback, the astrophysical effects of externally supplied CRs from the cosmic web have received little attention. 
Here, we focus on a distinct scenario where cosmic-web CRs become entrained into magnetised cold streams and are transported inward through young galaxy haloes.
We investigate this process by coupling a spectrally resolved hadro-leptonic CR transport calculation to a redshift-dependent model of magnetised cold streams. 
This framework allows us to determine whether externally supplied CRs can alter the thermodynamic state of cold streams, weaken or disrupt them, and survive transport toward the central galaxy.
To our knowledge, this is the first study to consider how cosmic-web CRs affect cold-stream accretion and its role in galaxy growth.  

This paper is organised as follows. 
In Sec.~\ref{sec:models}, we describe the entrainment of externally supplied CRs into magnetised cold streams, introduce the cold-stream model and CR transport calculation, and discuss the relevant physical timescales. 
In Sec.~\ref{sec:results}, we present the impact of CRs on cold streams using a representative fiducial model, then extend the analysis across halo mass, redshift, and stream density. We then discuss the broader implications of our findings. Finally, we present our conclusions in Sec.~\ref{sec:summary_conclusions}. 

\section{Theoretical framework and model}
\label{sec:models} 

\subsection{Physical scenario}
\label{sec:picture}

\begin{figure*}
    \centering
    \includegraphics[width=1.\linewidth]{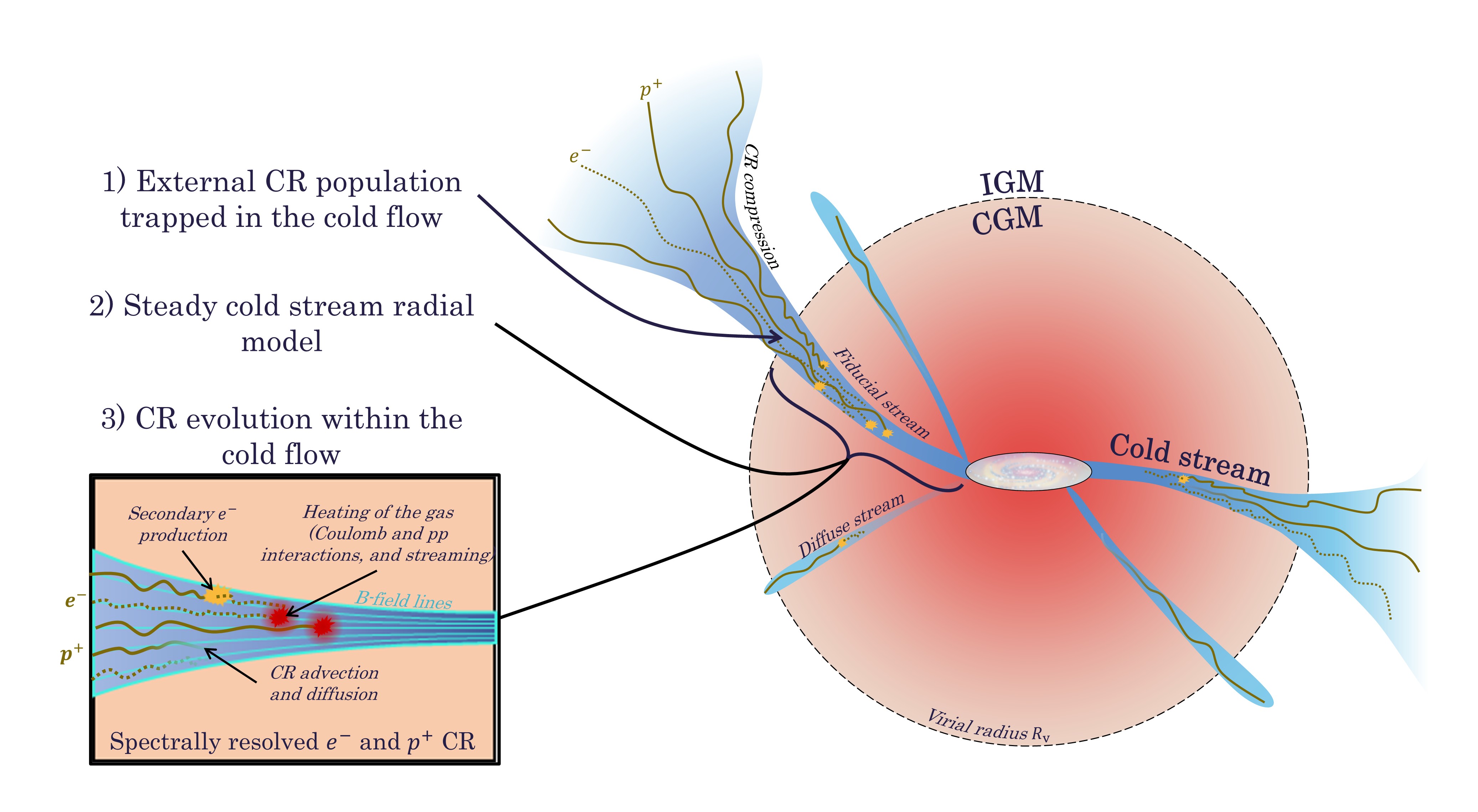}
    \caption{Schematic illustration of the physical picture considered in this work (not to scale). A massive galaxy is embedded in hot halo gas, and is fed by cold streams. We follow individual representative streams, and invoke an externally supplied CR population which becomes entrained into the flow as it approaches the virial radius (indicated by the dashed line separating the CGM from the IGM). These CRs are transported inward along the magnetised stream, where they deposit energy into the gas.}\label{fig:illustration}
\end{figure*}

The physical picture adopted in this work is illustrated in Fig.~\ref{fig:illustration}.  
We consider a cold stream that supplies gas to a young massive galaxy from the wider cosmic-web environment. As the stream propagates through the hot CGM, shear and thermal interactions with the surrounding halo gas produce a partially-mixed layer of stream and hot halo material (referred to as the mixing layer).  
Near the virial radius, the stream carries an entrained fraction of a pre-existing external CR reservoir associated with the surrounding filamentary or outer-CGM environment.  
The entrainment and transport of this CR population depend on its coupling to the magnetised inflow.
Within magnetised gas, CR transport is governed by diffusion along magnetic-field lines, advection with the bulk flow, and streaming relative to the gas.
It is therefore sensitive to the magnetic-field geometry, turbulence, scattering physics, and gas motions.
In this work, we treat the stream as a coherent magnetised channel. We then consider whether externally supplied CRs can remain coupled to the inflow, survive transport through the halo, or deposit sufficient energy to modify the thermal state of the gas before it reaches the central galaxy.  
CR heating is not expected to be the only thermal process acting on cold streams.
Photoheating, radiative cooling, conduction, turbulent mixing, and hydrodynamical dissipation may also operate.
However, in this study, our intention is to isolate the additional heating contribution provided by externally supplied CRs, rather than attempting a self-consistent calculation of all heating, cooling, and dissipation channels.  

\subsection{External cosmic ray reservoir}
\label{sec:boundary_condition}

We model the upstream CR population as a pre-existing component associated with a filament IGM 
environment surrounding the host galaxy halo. This population represents a reservoir supplied from outside the host-galaxy feedback cycle. It is not generated by feedback from the central galaxy, nor by acceleration inside the cold stream itself. We distinguish this external population from CRs produced within the central galaxy, which would propagate outward from the inner halo and would have to enter the stream laterally across the stream-CGM interface. Magnetic draping and field amplification around cold CGM structures would restrict such transverse transport~\citep{Ledos2024b, Ledos2024a,Ramesh2024A&A,Kaul2025}. We therefore prescribe an upstream energy density for a CR reservoir already associated with the inflowing gas at the virial boundary. We do not model the origin or time evolution of this population consistently, but focus on the consequences for galaxy growth if a fraction of the cosmic-web CR reservoir becomes entrained into a coherent, magnetised cold stream.  

As cold streams connect galactic haloes to gas arriving from the wider intergalactic environment~\citep{Dekel2006, Dekel2009, Keres2009MNRAS,Waterval2025}, 
we consider that they may advect CRs from this pre-existing reservoir into the halo.  
Even when the magnetic pressure remains dynamically subdominant, background CRs can be captured by a converging magnetised inflow if their gyroradii are smaller than the characteristic stream width, and if pitch-angle scattering is efficient enough to keep the CR population approximately coupled to the flow over a stream-compression timescale. The first condition is easily satisfied for CRs up to at least $\sim 100$ TeV under typical stream conditions. The second condition is more uncertain, as it requires the effective CR scattering mean free path to remain shorter than the characteristic stream-compression scale, and the perpendicular escape time to exceed the stream-compression time. In this limit, CRs remain coupled to the stream magnetic field, and are sufficiently scattered to be advected with the converging flow, rather than rapidly escaping across the stream boundary. We therefore treat entrainment as a controlled assumption for coherent, magnetised streams, while noting that it may break down if turbulence or hydrodynamical disruption make the stream-CGM boundary permeable to CRs. 

The characteristic normalisation of the filament CR population is uncertain. However, 
a useful guide is provided by cosmological calculations that include CR injection at structure-formation shocks, which can channel a dynamically important fraction of energy into CRs~\citep{Ryu2003ApJ}. 
These suggest that in shocked gas near the outskirts of collapsed haloes, groups, and large-scale-structure nodes, where these systems interface with filaments and the lower-density intergalactic medium, CR pressure support can reach $\sim 10$--$20\%$ of the thermal gas pressure~\citep{Vazza2014MNRAS}. 
To set the thermal-pressure scale of the upstream filamentary environment, we use the filament-core pressure profiles of~\citet{GalarragaEpinosa2021A&A}, obtained from cosmic-web filaments identified in the TNG300-1 simulation at $z=0$.
They found characteristic filament-core pressures of $\sim 4-12 \times 10^{-4}$ eV cm$^{-3}$. We therefore use $P_{\rm fil}\sim10^{-3} \; {\rm eV \; cm^{-3}}$ as a representative order-of-magnitude thermal-pressure scale for filament cores at low redshift.
Combined with the above CR-to-thermal pressure fraction, this motivates a conservative baseline CR energy density  
of $U_{\mathrm{CR,bg}} \approx 10^{-4}$\; eV \; cm$^{-3}$.\footnote{For comparison, this 
is well below the dynamical energy density of filament gas ($U_{\rm dyn}\sim 10^{-2}$ \,eV\,cm$^{-3}$ for densities and velocities typical of simulated filament gas; see e.g.~\citealt{Zhu2015ApJ, Vurm2023A&A}), 
but is comparable to the magnetic energy density of cosmic filaments if their magnetic-field strengths are of order tens of nG~\citep{Vacca2018Galax, Vernstrom2021MNRAS}.}  This normalisation should be regarded as uncertain at least at the factor-of-few level, since filament pressures depend on environment, redshift, halo contamination, and the adopted filament definition. 
In our model, changing this baseline value rescales the supplied CR energy density, but does not significantly change the qualitative dependence of CR heating on stream density, halo mass, redshift, and transport losses. 

For a relativistic CR population that remains approximately coupled to the flow during adiabatic compression,  
$U_{\mathrm{CR}} \propto n_{\mathrm{H}}^{4/3}$, where $n_{\mathrm{H}}$ is the ambient gas density. 
In the cosmic-web environment, this ambient gas density $n_{\mathrm{H,web}}$ evolves with redshift, as 
  $n_{\mathrm{H,web}}(z)
    = \delta_{\mathrm{fil}}\,\bar{n}_{\mathrm{H},0}\,(1+z)^{3}$, 
where $\bar{n}_{\mathrm{H},0}$ 
is the
present-day cosmological mean hydrogen number density  
and $\delta_{\mathrm{fil}}$ is the 
filament overdensity relative to the cosmic mean. 
Cosmological simulations and observations indicate $\delta_{\mathrm{fil}} \sim
10$--$100$ for the cores of filaments 
\citep[e.g.][]{Tanimura2020A&A, GalarragaEpinosa2021A&A, Vurm2023A&A, Migkas2025A&A}. We adopt $\delta_{\mathrm{fil}} =
20$ as a fiducial value, which gives 
$n_{\mathrm{H,web}} \approx 10^{-4}$\,cm$^{-3}$ at $z \sim 2$. 
By comparing this background density to the cold-stream density at the virial radius, we estimate the energy-density compression factor experienced by the entrained CR population as  
\begin{align}   
\label{eq:cr_compression_factor}
  C(z,M_{\mathrm{h}})
    \approx 4000
    \left(\frac{n_{\mathrm{H,s}}(R_{\mathrm{v}})}{5\times10^{-2}\,
      \mathrm{cm^{-3}}}\right)^{4/3}
    \left(\frac{n_{\mathrm{H,web}}(z)}{10^{-4}\,
      \mathrm{cm^{-3}}}\right)^{-4/3} ,
\end{align} 
where $n_{\mathrm{H,s}}(R_{\mathrm{v}})$ is the gas number density of the cold stream at the virial radius. The chosen normalisation corresponds to the typical compression expected for a representative stream entering a halo of $M_{\mathrm{h}} = 10^{12}\,\Msun$ at 
$z = 2$, with an 
upstream representative cosmic-web density set by $\delta_{\rm fil}=20$.  
This compression factor should be interpreted as an upstream boundary estimate rather than a self-consistent calculation of the full capture and entrainment process. CRs with large gyroradii, inefficient scattering, or efficient perpendicular escape in a particular flow configuration would experience a smaller enhancement than invoked here. The CR energy density supplied to the stream at its outer boundary, taken to be at virial radius, is then 
  $U_{\mathrm{CR,ext}}(z, M_{\mathrm{h}})
    = C(z, M_{\mathrm{h}})\; U_{\mathrm{CR,bg}}$. 
For the representative stream above (with $M_{\mathrm{h}} = 10^{12}\,\Msun$ at 
$z = 2$), this gives 
$U_{\mathrm{CR,ext}} \approx 0.4$\,eV\,cm$^{-3}$. Further boundary choices required in the CR model are the spectral form of the CR flux, the minimum and maximum CR energies considered, and the partition of the total boundary CR energy density between protons and electrons. These choices are detailed in Appendix~\ref{app:cr_profile} and summarized in Table~\ref{tab:fiducial_cr_stream}. Once CRs are entrained, the convergence of the flow produces adiabatic energy changes in the CR population. These are modelled using a parametrised adiabatic energy-change term applied to all CR species (as described in Appendix~\ref{sec:CR_processes} and Eq.~\ref{eq:adiabatic}).  

\subsection{Cosmic ray transport and heating along streams}
\label{sec:cr_transport}

In our approach, we assume that cold streams are approximately coherent, allowing us 
to model CR propagation using effective one-dimensional transport along the stream axis. 
We adopt a macroscopic diffusion-advection-loss prescription for the pitch-angle-averaged CR distribution.
This transport-level description is appropriate for our aim to follow the halo-scale evolution of the CR spectrum, species-dependent losses, secondary production, and heating, without resolving the gyro-scale plasma processes that determine the local CR scattering rate~\citep[for a review, see][]{Ruszkowski2023A&ARv}.
We therefore assume that CRs are sufficiently magnetised and scattered for their unresolved pitch-angle dynamics to be represented statistically through effective transport coefficients.
The net effects of kinetic instabilities, self-confinement, wave damping, and CR confinement are absorbed into these coefficients, while a first-principles calculation of the coefficients themselves is beyond the scope of this work.
This allows the CR population to be treated spectrally and by species, while estimating interaction-level heating channels through the adopted density and magnetic-field strength of the stream.  
Once coupled to the gas, CRs can heat their ambient medium through collisional losses, wave-mediated streaming losses, and hadronic interactions.
In our model, the gas heating channels are proton Coulomb losses, direct thermalisation in hadronic proton-proton (pp) interactions, proton streaming losses mediated by self-confinement and Alfv\'{e}n-wave excitation, and collisional heating by CR electrons. 

An effective macroscopic description of the transport and evolution of CRs entrained into a cold stream can then be written, for a CR species $j\in\{p,e\}$, as 
\begin{align}
\label{eq:transport}
\frac{\partial n_j}{\partial t}
-\nabla\cdot\left(D \nabla n_j\right)
&+\nabla\cdot\left(\mathbf{u}_{\rm eff}\,n_j\right) \\ \nonumber
&-\frac{\partial}{\partial E}\!\left[b_j(E,\mathbf{r})\,n_j\right] 
 =
S_j(E,\mathbf{r})
-{\lambda}_j(E,\mathbf{r})\,n_j \ , 
\end{align}
where \(n_j=n_j(E,\mathbf{r},t) \) is the differential number density of CR particles of species \(j\), per unit energy and volume, at position $\mathbf{r}$, and the boundary conditions are described in Appendix~\ref{app:cr_profile}. The particle kinetic energy is 
\(E=(\gamma_j-1) m_jc^2\), where $\gamma_j$ as the Lorentz factor, $m_j$ is the particle rest mass, and $c$ is the speed of light. 
When applied to a cold stream, this equation can be reduced to a quasi-one-dimensional problem along the cold-stream axis, with $n_j=n_j(E,r,t)$ for galacto-centric radius $r$. 

The terms in Eq.~\ref{eq:transport} represent, from left to right, time evolution, spatial diffusion, effective advection along the stream, continuous energy changes, source or injection terms, and catastrophic losses. 
We take the characteristic CR transport speed along the stream to be the sum of the bulk stream velocity and the local Alfv\'{e}n speed, $u_{\rm eff}(r)=v_{\stream}(r)+v_{\rm A}(r)$, with both speeds defined as positive in the inward direction 
and specified by the adopted stream and magnetic-field model (see Sec.~\ref{sec:inflow_halo_model}).  
This represents advection with the inflowing cold stream gas plus Alfv\'{e}nic streaming relative to the gas in the self-confinement limit. In our model, $v_{\rm A}$ remains well below the bulk stream velocity, $v_{\stream}$, through most of the cold stream spine. We retain it for consistency with the CR streaming-loss term and to capture the dependence on the local magnetisation.  
The source term \(S_j(E,r)\) describes volumetric particle 
injection rate. The coefficient \(b_j(E,r)\) describes
continuous energy changes; positive values correspond to losses (cooling) 
and negative values to energy gains (e.g. through adiabatic compression). Catastrophic destruction processes,
such as inelastic interactions or particle decay, are encoded by
the loss rate \(\lambda_j(E,r)\). The detailed forms of \(b_j\),
\(S_j\), and \(\lambda_j\) are specified separately for
protons and electrons. These are described in Appendix~\ref{sec:CR_processes}, where the corresponding timescales over which they operate are also provided.  

CR transport is followed along the stream axis, while allowing the local gas density, magnetic-field strength, bulk stream velocity, and stream cross-section to vary with galacto-centric radius.  
Here, we follow the standard simplified treatment for CR diffusion in galaxy and CGM environments, which is characterized by the coefficient 
\begin{equation}
D(E,r)=D_{0}
\left[
\frac{r_{\rm L}(E,B)}{r_{\rm L}(E_{\rm ref},B_{\rm ref})}
\right]^{1/2},
\end{equation}
where $r_{\rm L}$ is the Larmor radius, $E_{\rm ref}=1\,{\rm GeV}$, and $B_{\rm ref}=50\,\mu{\rm G}$. 
The index of $1/2$ corresponds to a Kraichnan-like rigidity scaling~\citep[e.g.][]{Berezinskii1990acr, Strong2007ARNPS}. 
We adopt 
$D_{0}=3\times10^{29}\,{\rm cm^2\,s^{-1}}$ as the normalization for both CR protons and electrons.\footnote{Galaxy halo-scale CR transport calculations calibrated to reproduce the observed $\gamma$-ray luminosities of nearby dwarf and $L_\star$ galaxies favour diffusion coefficients around $3\times10^{29}\,{\rm cm^2\,s^{-1}}$~\citep[e.g.][]{Chan2019MNRAS,Ji2020MNRAS}. 
This value is larger than the typical choice of $3\times10^{28}\,{\rm cm^2\,s^{-1}}$ invoked for Galactic interstellar medium conditions for CR protons of a few GeV~\citep{Strong2007ARNPS}, but is a reasonable effective choice for lower-density halo and circumgalactic environments, 
and smaller than the upper limit of $5\times10^{30}\,{\rm cm^2\,s^{-1}}$ suggested by $\gamma$-ray emission studies of galaxy haloes~\citep{Recchia2021ApJ}. Although beyond the scope of the current work, we note that studies of CR transport in dense molecular clumps have reported indications for diffusion coefficients suppressed by up to two orders of magnitude relative to the canonical ISM value in denser gas~\citep[e.g.][]{Yang2023NatAs, Ng2026PASJ}, which would suggest that $D_0$ could vary through the stream spine and the mixing layer. 
The exact choice of this parameter does not have a strong impact on our conclusions, but can have a noticeable impact on the resulting CR heating patterns throughout a stream (see Appendix~\ref{sec:diffusion_study}, which demonstrates how variation of 
this parameter choice affects our results).} 

The CR energy density at the outer boundary follows from Sec.~\ref{sec:boundary_condition} (Eq.~\ref{eq:cr_compression_factor}). It is partitioned between protons and electrons using $K_{p/e}\equiv U_p/U_e=100$, and the imposed boundary spectra are power laws with index $q=2.2$. 
As a baseline, we include CR kinetic energies from $E_{\rm min}=10$ MeV to $E_{\rm max}=10^4$ GeV.  
The lower bound is chosen to retain the sub-GeV proton population that dominates Coulomb and ionisation heating in cold-stream gas~\citep[e.g.][]{Padovani2009A&A}, while avoiding an extrapolation to even lower energies where the external spectrum, transport approximation, and local thermalisation become increasingly uncertain.\footnote{The exact choice of $E_{\rm min}$, if reasonable, does not have a strong bearing on our results. Values substantially below 10 MeV marginally boost CR heating near the virial radius, but do not affect our conclusions. Higher values of $E_{\rm min}$ begin to weaken the impact of Coulomb and ionisation heating in cold-stream gas, reducing the impact of the CRs.} 
Such low-energy CRs can be strongly attenuated by Coulomb and ionisation losses when transported through dense gas. They may therefore be depleted in feedback-dominated environments where CRs must first pass through dense interstellar gas before reaching the halo~\citep[e.g.][]{Werhahn2021MNRAS}, while remaining more abundant in sufficiently tenuous cosmic-web environments. The low-energy content of the externally supplied CR population considered here may therefore differ from that of CRs supplied by local galactic feedback. The adopted low-energy extension should be regarded as part of the imposed external boundary model rather than a prediction of the CR supply process.  
For CR electrons, we adopt an exponential cut-off at $E_{\rm cut, e} = 10$ GeV to represent cooling prior to entrainment. The exact choice of this energy cut-off in the electron spectrum, and the overall choice of $E_{\rm max}$ in our calculations do not have a noticeable impact on our results (for reasonable choices). 
The values adopted for our representative CR boundary model are summarised in Table~\ref{tab:fiducial_cr_stream}, while the full boundary spectrum is described in Appendix~\ref{app:cr_profile}. 

\begin{table*}
\caption{Fiducial CR boundary spectrum properties as it enters a halo of $M_{\mathrm{h}} = 10^{12}\,\Msun$ at 
$z = 2$. These choices are selected for our fiducial calculations, but the compressed boundary CR energy density $U_{\rm CR,ext}$ is computed self-consistently when different halo masses or redshifts are considered. Further discussion of these parameter choices is presented in Sec.~\ref{sec:cr_transport} and Appendix~\ref{app:cr_profile}.}
\label{tab:fiducial_cr_stream}
\centering

\begin{tabular}{llr}
\hline\hline
Quantity & Symbol & Value \\
\hline
Background CR energy density & $U_{\rm CR,bg}$ & $10^{-4} \;{\rm eV \; cm^{-3}}$ \\
CR energy density at virial radius & $U_{\rm CR,ext}$ & $0.4 \; {\rm eV \; cm^{-3}}$ \\
CR proton-to-electron energy-density ratio & $K_{p/e}$ & 100 \\
Spectral index (protons and electrons) & $q$ & 2.2 \\
Electron spectral cut-off energy & $E_{\rm cut,e}$ & 10 GeV \\
Minimum CR kinetic energy & $E_{\rm min}$ & $10^{-2}$ GeV \\
Maximum CR kinetic energy & $E_{\rm max}$ & 10$^4$ GeV \\
CR diffusion coefficient normalisation & $D_0$ & $3\times10^{29} \; {\rm cm^2 \; s^{-1}}$ \\
\hline
\end{tabular}
\end{table*}

The microphysical cooling and interaction timescales that regulate CR energy deposition are generally shorter than the global evolution time of the cold stream in the regions where CR effects could be important.
We therefore assume that the CR population reaches a quasi-stationary configuration along the stream, and solve Eq.~\ref{eq:transport} in the time-independent limit. 
The solution is obtained from an outer boundary at the virial radius $R_{\rm v}$ to an inner radius, $r_{\rm in}=0.1 R_{\rm v}$. This allows us to follow CR energy deposition through the circumgalactic part of the stream, while avoiding the inner-galaxy region where the cold-stream model is no longer appropriate. 
We solve the steady-state transport equation on a fixed logarithmic energy grid and a one-dimensional radial grid, using an implicit finite-difference discretisation of radial diffusion/advection and a sign-aware upwind discretisation of the continuous energy-change term.   

\subsection{Cold stream, halo and magnetic field model}
\label{sec:inflow_halo_model}

We describe the host halo and stream properties at the virial radius using the analytic framework of \citet{Dekel2013}, later extended by \citet{Mandelker2020b}, for which we consider a flat Lambda Cold Dark Matter
($\Lambda$CDM) cosmological model with parameters given by Planck Collaboration VI \citep{Planck2020_Cosmo_param}\footnote{Here, we use the TT,TE,EE+lowE+lensing+BAO parameters.} such that $\left(\Omega_\mathrm{m},\Omega_\mathrm{b}, \Omega_\Lambda,H_0\right) = \left(0.3111,0.0490,0.6889,67.66\, \mathrm{km\,s^{-1}\,Mpc^{-1}}\right)$.
The dark-matter halo is assumed to follow a Navarro-Frenk-White profile, with the concentration parameter taken from the redshift- and mass-dependent fitting formulae of \citet{Correa2015MNRAS}. 
The model is specified primarily by the halo mass $M_{\rm h}$, and the virial density $\rho_\mathrm{v}(z)=\rho_\mathrm{c}(z)\Delta_\mathrm{c}(z)$ where $\rho_\mathrm{c}(z)$ is the critical density of the Universe, and $\Delta_\mathrm{c}(z)$ is the over-density factor taken from \citet{Bryan1998}. We express the variables in terms of the normalised quantities $M_{12}=M_{\rm h}/(10^{12} \,\Msun)$ and $\rho_\mathrm{v,2} =\rho_\mathrm{v}(z)/\rho_\mathrm{v}(z=2)=\rho_\mathrm{v}(z)/(1.34\times10^{-26}\,\mathrm{g\,cm^{-3}})$.
Other parameters are fixed to fiducial values or varied over the ranges stated below as described in \citet{Mandelker2020b} unless specified otherwise. Representative radial profiles for fiducial parameter choices of $M_{\mathrm{h}} = 10^{12}\,\Msun$ at 
$z = 2$ are shown in Appendix~\ref{app:stream_prof}. 

\subsubsection{Halo model}

The halo is defined by its virial radius, $R_{\mathrm{v}}$, a temperature proportional to the virial temperature, and the normalisation of the hot gas density at $R_{\mathrm{v}}$, 
\begin{align}\label{eq:halo_prop}
    R_{\mathrm{v}} = & 106.5 \, M_{12}^{1/3} \rho_\mathrm{v,2}^{-1/3} \,\, \kpc , \\
    T_{\rm h} = & 1.4 \times 10^{6} \, M_{12}^{2/3}\rho_\mathrm{v,2}^{1/3} \left(\frac{\Theta_{\mathrm{h}}}{1}\right) \, \, \rm K,\\
    \rho_{\mathrm{h}}(R_{\mathrm{v}}) = &1.6\times 10^{-28} \, \rho_\mathrm{v,2} \, \left(\frac{\Delta_{\mathrm{Rv}}(c_\mathrm{h})}{0.246}\right)\left(\frac{f_{\mathrm{h}}}{0.3}\right) \, \, \rm g\, cm^{-3},
\end{align}
where $\Theta_{\mathrm{h}} \in [3/8,1]$ is the ratio of the hot halo gas temperature to the virial temperature, $c_\mathrm{h}$ is the halo concentration parameter, and $f_{\mathrm{h}}\in [0.2,0.4]$ is the hot gas fraction. The hot gas temperature also assumes a baryon fraction of $f_\mathrm{b}=\Omega_\mathrm{b}/\Omega_\mathrm{m}\sim 0.157$. The quantity $\Delta_{\mathrm{Rv}}(c_h)$ denotes the density contrast at $R_{\mathrm{v}}$ relative to the mean, for which we adopt a fiducial value $\Delta_{\mathrm{Rv}}(c_\mathrm{h}=5)\simeq 0.246$.
The concentration parameter $c_\mathrm{h}$ is taken from redshift- and mass-dependent fits by \citet[][their Eqs. 19–20]{Correa2015MNRAS}. 

\subsubsection{Cold stream model}
\label{sec:cold_stream_model}

The cold-stream properties at the virial radius follows the empirical analytic model of \citet{Dekel2013}, as extended by \citet{Mandelker2020b} using our choice of recent cosmological parameters.
We idealise cold streams as coherent, cross-section-averaged structures embedded in a hot, lower-density CGM.
In reality, such cold gas can interact with the surrounding halo through shear instabilities, turbulence, mixing layers, thermal conduction, radiative cooling, and condensation.
These processes can affect angular-momentum transport, the development of multiphase halo gas, and the redistribution of magnetic flux around the stream-CGM interface~\citep[e.g.][]{Danovich2015,Mandelker2019,bib_MHD_Berlok_2019b,Mandelker2020a,Ledos2024a,Ledos2024b,Hong2024,Aung2024,Kaul2025,Gronke2026}.
High-resolution numerical simulations further show that cold CGM gas, such as that in cold streams, can fragment into small-scale multiphase structures~\citep{Nelson2020MNRAS,Ramesh2024MNRAS,Bennett2020,Yao2025}.
We do not attempt to resolve this multidimensional structure directly.
Instead, we use a one-dimensional representative stream model, with parameter variations that allow us to explore how the CR heating responds to changes between different stream configurations.   
The cold stream properties are defined by its velocity, temperature, density ratio, and density as follows:
\begin{align}\label{eq:stream_prop}
    v_{\stream}(R_{\mathrm{v}}) = & 201 \, M_{12}^{1/3} \rho_\mathrm{v,2}^{1/6} \, \left(\frac{\eta}{1}\right) \,\, \rm km\, s^{-1} , \\
    T_{\rm cold} = & 1.5 \times 10^{4} \left(\frac{\Theta_{\stream}}{1}\right) \, \, \rm K,\\
    \delta = & \frac{T_{\rm h}}{T_{\rm cold}} = \frac{\rho_{\stream}}{\rho_{\mathrm{h}}} = 96.2 \, \, M_{12}^{2/3}\rho_\mathrm{v,2}^{1/3} \left(\frac{\Theta_{\mathrm{h}}}{\Theta_{\stream}}\right) \\
    \rho_{\stream}(R_{\mathrm{v}}) = & 1.5\times 10^{-26} \, M_{12}^{2/3} \rho_\mathrm{v,2}^{4/3} \, \left(\frac{\Delta_{\mathrm{Rv}}(c_h)}{0.246}\right)\left(\frac{f_{\mathrm{h}}}{0.3}\right)\left(\frac{\Theta_{\mathrm{h}}}{\Theta_{\stream}}\right) \, \, \rm g\, cm^{-3}, 
\end{align}
where quantities are considered to be uniform across the stream cross-section at a given radius, 
and where $\eta\in [0.5,\sqrt{2}]$ is the ratio of the stream velocity to the virial velocity, and $\Theta_{\stream} \in [0.5,2]$ is a dimensionless factor accounting for uncertainties in the stream temperature. Following \citet{Mandelker2020b}, the cold gas accretion rate is derived from the dark matter accretion rate, $\dot{M}_{\mathrm{v}}\propto s M_{12} (z+1)^{5/2}$, where $s\in [0.5,2]$ accounts for halo-to-halo variations (fiducial value $s=1$). This yields a cold gas accretion rate per stream of 
\begin{equation}\label{eq:Mdots0}
 \dot{M}_{\stream}(R_{\mathrm{v}}) = 17.26 \, M_{12}\,a_3^{-5/2} \, \left(\frac{s}{1}\right)\left(\frac{3}{N_{\stream}}\right)\left(\frac{1-f_{\mathrm{h}}}{0.7}\right) \, \, \rm \Msun \, yr^{-1},   
\end{equation}
where $N_{\stream}\in [2,5]$ is the number of streams penetrating the halo, and $a_3= 3/(1+z)$. Compared to \citet{Mandelker2020b}, we include an additional factor $(1-f_{\mathrm{h}})$ representing the cold gas fraction, which leads to slightly lower accretion rates.
From the accretion rate, density, and velocity, the stream radius follows as 
\begin{align}\label{eq:rs0}
 r_{\stream}(R_{\mathrm{v}}) &= \sqrt{\frac{\dot{M}_{\stream}(R_{\mathrm{v}})}{\pi \rho_{\stream}(R_{\mathrm{v}}) v_{\stream}(R_{\mathrm{v}})}} \\ \nonumber
 &=  11.0 \, \, a_3^{-5/4} \rho_\mathrm{v,2}^{9/4} \, \left(\frac{\eta_{\stream}}{0.778}\right)^{1/2}\left(\frac{0.246}{\Delta_{\mathrm{Rv}}(c_h)}\right)^{1/2}  \, \, \rm kpc, 
\end{align}
where $\eta_{\stream}$ absorbs the combined uncertainty factors:
\begin{equation}\label{eq:etas}
 \frac{\eta_{\stream}}{0.778} = \left(\frac{s}{\eta}\right)\left(\frac{\Theta_{\stream}}{\Theta_{\mathrm{h}}}\right)\left(\frac{3}{N_{\stream}}\right)\left(\frac{1-f_{\mathrm{h}}}{0.7}\right)\left(\frac{0.3}{f_{\mathrm{h}}}\right), 
\end{equation}
yielding $\eta_{\stream}\in [0.053,42]$. In practice, the upper bound of this 
parameter range results in a very thick stream ($r_{\stream}(R_{\mathrm{v}}) > R_{\mathrm{v}}$). To focus on narrower streams, we therefore impose $r_{\stream}(R_{\mathrm{v}}) < 0.5 R_{\mathrm{v}}$. 
Stream properties
can still vary substantially at fixed halo mass and redshift, as illustrated by
the range of profiles in Appendix~\ref{app:stream_prof}.  We therefore use the
central parameter values to define a ``fiducial stream'' model, and adopt
representative lower-density and higher-density configurations to bracket the
range of plausible stream properties. We refer to the lower-density,
lower-column case as the ``diffuse stream'' model hereafter.\footnote{These bounds are not sharply defined, especially for low-density streams. Current state-of-the-art simulations do not numerically converge on the minimum size of cold CGM structures~\citep[e.g.][]{Nelson2020MNRAS,Bennett2020}, with increasing resolution continuing to reveal smaller cold structures.}

\subsubsection{Radial profiles}

The quantities above are defined at the virial radius. 
To describe their radial variation, where $r$ denotes the galacto-centric radial position along the stream, we prescribe the ambient CGM structure using the analytic polytropic halo model of \citet{Komatsu2001}.
This accounts for the fact that the baryonic gas does not trace the dark-matter profile directly, owing to radiative cooling and pressure support in the halo gas~\citep{Komatsu2001,Aung2024}.
For a polytropic equation of state, $P_{\cgm}\propto \rho_{\rm h}^{\gamma_{\rm poly}}$, the CGM density profile can be written as \citep[][their Eq. 19]{Komatsu2001}:
\begin{equation}
    y\left(r\right) =\frac{\rho_{\rm h}\left(r\right)}{\rho_{\rm h,0}} = \left[1 +\frac{3}{\sigma_0}\frac{\gamma_{\rm poly}}{\gamma_{\rm poly}-1}\frac{c_h}{m\left(c_h\right)}\left(\frac{\log{\left(x+1\right)}}{x}-1\right) \right]^{\frac{1}{\gamma_{\rm poly}-1}},
\end{equation}
where $r$ is the halo radial coordinate, $x\equiv c_h r/R_{\mathrm{v}}$ with $R_{\mathrm{v}}$ the virial radius, $m\left(c_h\right)=\log{\left(c_h+1\right)} - c_h/\left(c_h+1\right)$, and $\rho_{\rm h,0}=\rho_{\rm h}\left(r=0\right)$.
The constant $\sigma_0$ and the polytropic index $\gamma_{\rm poly}$ are empirical functions of $c_h$ given by \citet[][their Eqs. 25--26]{Komatsu2001}. 
The stream density profile is then given by 
\begin{equation}\label{eq:dens_s}
    \frac{\rho_{\stream}\left(r\right)}{\rho_{\stream}\left(R_{\mathrm{v}}\right)} = \left[\frac{y\left(r\right)}{y\left(R_{\mathrm{v}}\right)}\right]^{\gamma_{\rm poly}}. 
\end{equation}
The stream velocity $v_{\stream}(r)$ is assumed to follow the free-fall velocity along the halo radius \citep[e.g. ][]{Aung2024}. The stream radial thickness then follows from 
mass-flux conservation along the stream, giving 
\begin{equation}\label{eq:rs}
    \frac{r_{\stream}\left(r\right)}{r_{\stream}\left(R_{\mathrm{v}}\right)} = \left[\frac{\rho_{\stream}\left(R_{\mathrm{v}}\right) \; v_{\stream}\left(R_{\mathrm{v}}\right)}{\rho_{\stream}\left(r\right) \; v_{\stream}\left(r\right)}\right]^{1/2} \ . 
\end{equation}

\subsubsection{Cold stream magnetisation}

We model the magnetic field in the stream following \citet{Ledos2024b}. The upstream magnetic field at the virial radius is normalised by adopting a plasma beta parameter $\beta(R_{\mathrm{v}})=10^5$, corresponding to field strengths of order nano-Gauss (nG).
An initial amplification arises from the tangling and stretching of magnetic field lines within the mixing layer surrounding the stream~\citep{Ledos2024b,Kaul2025}.
More generally, turbulent stretching and redistribution of magnetic fields around cold gas structures may also contribute to the early magnetisation inferred in young galaxies~\citep{Geach2023Natur,Chen2024A&A,deRoo2025MNRAS}, although this is not explicitly modelled here. 
The time-averaged amplification is well described by
\begin{equation}\label{eq:Bmodel1}
    \chi_{\mathrm{B}}=\frac{\left\langle B_{\rm{th}}\right\rangle}{B_0} = \frac{1}{2}\left[ \left(\beta \frac{r_{\stream}}{r_{\stream} + v_{\stream}t_{\mathrm{c}}}\right)^2 + 4\left(\beta+1\right)\right]^{1/2} - \frac{1}{2} \beta \frac{r_{\stream}}{r_{\stream} + v_{\stream}t_{\mathrm{c}}},
\end{equation}
where $B_{\rm{th}}$ is the theoretical magnetic field, $B_0$ the initial magnetic field corresponding to $\beta=10^5$, and $t_{\mathrm{c}}\sim R_{\mathrm{v}}/v_{\stream}$ is the characteristic timescale associated with the rapid amplification phase. 
This leads to an average $\sim 40$-fold increase in the field strength~\citep{Ledos2024b}, and operates on timescales shorter than the virial crossing time. 
As shown by \citet{Ledos2024a}, $\langle B_{\rm{th}} \rangle$ captures the time-averaged magnetisation but underestimates the early growth phase, motivating the use of a relatively large $t_{\mathrm{c}}$.
This amplification mechanism is only applicable when the cold stream is embedded in a hot CGM, since it is driven by mixing between the two phases. Shock-stability analyses indicate that a virial shock develops for $M_{\mathrm{h}} \gtrsim 10^{11.6} \Msun$ \citep{Birnboim2003MNRAS,Dekel2006}. We therefore adopt this mass as a threshold below which $\chi_{\mathrm{B}}=1$. 
As the stream propagates inward, it contracts, producing additional magnetic amplification through flux freezing. We approximate this compression as isotropic, so that $B\ell^2=\mathrm{const}$ and $\rho \ell^3=\mathrm{const}$ for a characteristic contraction scale $\ell$, giving $B\propto\rho^{2/3}$. This should be regarded as an effective prescription. A real cold stream may contract anisotropically, and the resulting magnetic amplification depends on the field orientation and deformation geometry. A more detailed treatment of anisotropic compression and magnetic-field evolution is left to future multidimensional calculations. 
The cold stream magnetic-field profile is then given by: 
\begin{equation}\label{eq:B_r}
B_{\stream}(r) = \left\{ \begin{array}{ll}
\chi_{\mathrm{B}}B_0 \left(\frac{\rho_{\stream}\left(r\right)}{\rho_{\stream}\left(R_{\mathrm{v}}\right)}\right)^{2/3} & \text{if} \, \, M_{\mathrm{h}} \geq 10^{11.6}\, \Msun ,\\
B_0 \left(\frac{\rho_{\stream}\left(r\right)}{\rho_{\stream}\left(R_{\mathrm{v}}\right)}\right)^{2/3} & \text{if} \, \, M_{\mathrm{h}} < 10^{11.6}\, \Msun.
\end{array}\right.
\end{equation}

\subsection{Timescale analysis}
\label{sec:timescales}

The characteristic temperature of a cold stream in equilibrium with the ambient UV background is 
$T_{\rm cold}\sim 1-4\times10^4\,{\rm K}$, with its exact temperature dependent on the stream density and the strength of the background radiation.  
In massive hot haloes, mixing between the cold stream and the ambient gas produces a mixing layer with characteristic density and temperature:
\begin{align}
\rho_{\mathrm{mix}} &= \sqrt{\rho_{\stream}\rho_{\mathrm{h}}} \hspace{0.25cm} = \delta^{-1/2}\rho_{\stream} \ ,
\label{eq:rho_mix}
\\
T_{\mathrm{mix}} &= \sqrt{T_{\rm cold}T_{\rm h}} = \delta^{1/2}T_{\rm cold} \ , 
\label{eq:mix}
\end{align}
as derived analytically by \cite{Begelman1990} and \cite{Hillier2019}, and numerically verified in \citet[their Appendix C]{Ledos2024a}. 
This mixing layer is out of thermal equilibrium and cools on a characteristic timescale 
\begin{equation} \label{eq:tcool}
t_{\rm{cool}} = \frac{ n_{\rm{mix}}T_{\rm{mix}} k_{\rm{B}}}{\left(\gamma_{\rm g} -1 \right)\Lambda_{\rm net, mix}},
\end{equation}
where $n_{\rm mix}$ is the total number density of the gas in the mixing layer, $k_{\rm B}$ is the Boltzmann constant, and $\Lambda_{\rm net, mix}$ is the net cooling rate in the mixing layer. Throughout this work, tabulated cooling and heating rates are computed using the photoionisation code {\tt CLOUDY} \citep{Ferland2017RMxAA}, which includes both atomic and metal-line cooling processes. 
The photoheating rates from \citet{Haardt2012ApJ} are adopted to model the UV background from galaxies and quasars. We evaluate this background as a function of redshift $z$ and neglect the local radiation sources in our setup. Examples of the resulting cooling curves are shown in Appendix~\ref{app:cooling_net}. Due to the shape of the cooling curves, the cooling strength is maximum near $10^5\, \mathrm{K}\sim T_\mathrm{mix}$. The cooling time in the mixing layer is therefore an indicator to determine whether the gas in the mixing layer would suffer from thermal instability \citep{Field1965ApJ,Balbus1995ASPC} and rapidly cool down back to the stream temperature.  
 
Observations indicate that the metallicity of cold accretion streams spans a wide range, from $10^{-3.8}$ to $1\, Z_{\odot}$ \citep[e.g.][]{Crighton2013,Martin2019}. To account for metal enrichment arising from mixing with the more metal-rich hot CGM, we adopt a fiducial stream metallicity of $Z=0.1\,Z_{\odot}$. The metallicity of accreting gas remains poorly constrained observationally, but 
its variation over a plausible range of values has little impact on our results.
The cooling time defined above plays a key role in determining the stability of the stream against hydrodynamical instabilities~\citep{Mandelker2020a} and thermal conduction~\citep{Ledos2024a}.  

To compare CR heating with the local dynamical evolution of the stream, we define a dynamical timescale 
\begin{equation} \label{eq:tadv} 
t_{\rm{dyn}} = \frac{\ell_\mathrm{c} }{v_{\stream}}, 
\end{equation}
where $\ell_\mathrm{c}$ is the local compression length.
As the stream falls toward the galaxy, the relevant length scale is the radial distance over which the stream density changes appreciably, since this controls the increase in the cooling rate. 
For CRs that remain coupled to the inflow, this is also the characteristic time over which advective transport carries the CR population across one local density scale height, so we identify this as the CR advection timescale $t_{\rm adv}\simeq t_{\rm dyn}$ for our diagnostic comparisons.  
The compression length derived from Eq.~\ref{eq:dens_s} is 
\begin{equation} \label{eq:lcomp} 
\ell_\mathrm{c}(r) = \left\vert \frac{\rho_{\stream}(r)}{\nabla \rho_{\stream}(r)}\right\vert= \frac{R_\mathrm{v}}{c_\mathrm{h}}\frac{x(x+1)\left((\xi-1)x-\xi\log(x+1)\right)}{g_\mathrm{poly}\xi\left((x+1)\log(x+1)-x \right)} \ ,
\end{equation}
where $g_\mathrm{poly} = \gamma_\mathrm{poly}/(\gamma_\mathrm{poly}-1)$, and $\xi = 3g_\mathrm{poly}c_\mathrm{h}/\sigma_0m(c_\mathrm{h})$. For $R_\mathrm{v}=100\,\kpc$ and $c_\mathrm{h}=5$, $\ell_\mathrm{c}(R_\mathrm{v})\sim 39\,\kpc$ and $\ell_\mathrm{c}(0.1R_\mathrm{v})\sim 1.4\,\kpc$. 

We define heating timescales for each CR heating process $\ell$ described in Sec.~\ref{sec:cr_heating_rates}, for both the stream and the mixing layer, as
\begin{align}
\label{eq:theat}
    t_{{\rm heat},\ell,j,{\rm mix}} &= 
    \frac{n_{\rm mix} k_{\rm B}T_{\rm{mix, heat}}}
    {\left(\gamma_{\rm g} -1 \right)Q_{\ell,j, \rm mix}} \, , \\
    t_{{\rm heat},\ell,j,{\rm s}} &= 
    \frac{n_{\rm s} k_{\rm B}T_{\rm{s, heat}}}
    {\left(\gamma_{\rm g} -1 \right)Q_{\ell,j, \rm s}} \,  ,
\end{align}
where $Q_{\ell,j, \rm mix}$ and $Q_{\ell,j, \rm s}$ are the 
relevant CR heating processes operating in the mixing layer and stream, respectively. These are specified 
in Appendix~\ref{sec:cr_heating_rates}.  
These timescales are then compared with the mixing layer cooling time through 
    $\tau_{\ell,j,{\rm mix}} = {t_{{\rm heat},\ell,j,{\rm mix}}}/{t_{\rm cool}}$ and 
    $\tau_{\ell,j,{\rm s}} = {t_{{\rm heat},\ell,j,{\rm s}}}/{t_{\rm cool}}$. 

\begin{figure}
    \centering  
    \includegraphics[width=1\linewidth]{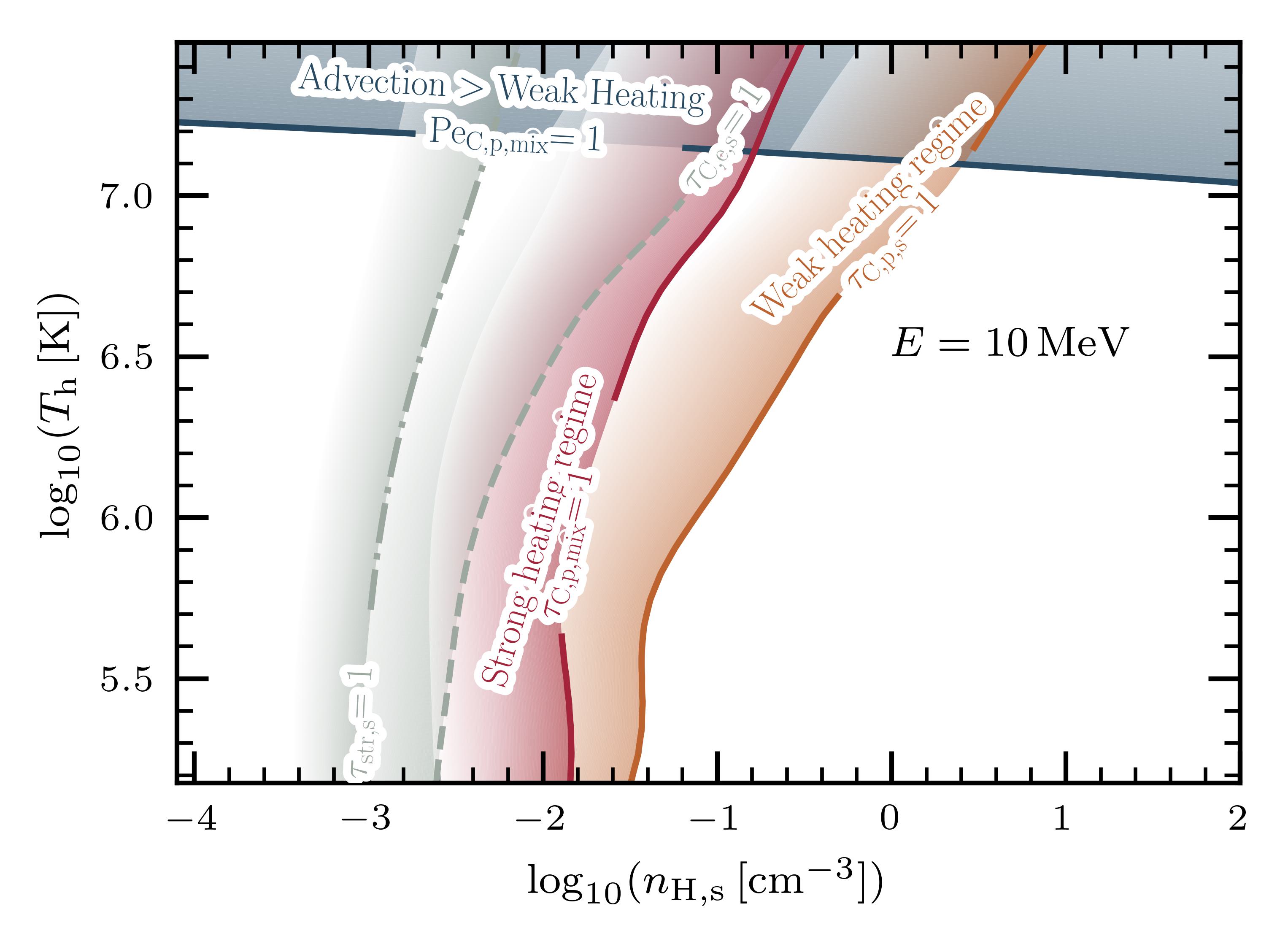}
    \caption{Balance of the timescales regulating CR heating and its spatial distribution, shown over plausible variations in cold-stream gas number density $n_{\rm H, s}$ and hot-halo temperature $T_{\rm h}$. Red and orange contours show the dominant proton Coulomb-heating timescale to gas cooling timescale ratios in the mixing layer, $\tau_{{\rm C,p,mix}}$, and in the stream, $\tau_{{\rm C,p,s}}$, respectively. Light grey contours show sub-dominant heating channels, including 
    electron Coulomb heating, $\tau_{{\rm C,e,s}}$, and streaming heating, $\tau_{{\rm str,s}}$. 
Coloured contours indicate where CR heating is faster than radiative cooling, extending to $\tau_{\ell}=0.1$. 
The dark blue line shows the ratio between the dominant CR heating time and the local stream advection time, while dark blue contours indicate the regime where advection is faster than heating, i.e. the Péclet number $\mathrm{Pe}_\mathrm{C,p,mix}=t_{{\rm heat},C,{\rm p,mix}}/t_{\rm dyn}$ extending to $\mathrm{Pe}_\mathrm{C,p,mix}=10$. 
Results are shown for $E=10\,{\rm MeV}$.  
    }
    \label{fig:timescale_map}
\end{figure}

We illustrate the resulting balance of timescales over plausible variations in the cold-stream gas number density and halo gas temperature in Fig.~\ref{fig:timescale_map}. For this diagnostic calculation, the CR heating timescales are evaluated at a representative CR kinetic energy of 10 MeV, chosen to highlight the low-energy hadronic CRs that dominate Coulomb and ionisation heating. In the full transport calculations shown in later sections, CR heating is spectrally resolved and includes the contribution of the full CR energy distribution. 
Appendix~\ref{app:cr_timescale_energy} 
repeats the same analysis for other representative CR energies, illustrating which parts of the CR spectrum contribute most efficiently to heating.  
In these illustrative calculations, we adopt a stream radius $r_{\rm s}=10\,{\rm kpc}$, and set the plasma beta parameter $\beta=P_{\rm th}/P_{\rm mag}$ to $\beta=1$ in the mixing layer and $\beta=100$ in the stream. 
These choices are consistent with both idealised simulations~\citep{Ledos2024a,Kaul2025} and cosmological zoom-in simulations~\citep{Nelson2020MNRAS}. 
Their exact values do not noticeably affect the dominant heating mechanism found in the stream over the parameter range considered here.

The dominant CR heating channel (proton Coulomb-heating) timescales relative to radiative cooling time $t_{\rm cool}$ in the stream core $\tau_{{\rm C,p,s}}$ and mixing layer $\tau_{{\rm C,p,mix}}$ are shown in orange and red, respectively. These indicate two characteristic CR-heating regimes: 
\begin{itemize}
    \item For $\tau_{{\rm C,p,s}}<1$ in the stream, CR heating is fast enough to raise the stream temperature, even if it does not overcome cooling in the mixing layer. This corresponds to a ``weak heating'' regime. 
    \item For $\tau_{{\rm C,p,mix}}<1$ in the mixing layer, CR heating is effective even where radiative cooling is strongest. This corresponds to a ``strong heating'' regime, where the stream-CGM interface can be heated efficiently and the survival of the cold stream may be affected.
\end{itemize}
For a typical hot gas temperature of $\sim 10^6\,{\rm K}$, these regimes occur for cold-stream number densities of approximately $n_{\rm H,s}\lesssim10^{-2}\,{\rm cm^{-3}}$ and $n_{\rm H,s}\lesssim10^{-2.5}\,{\rm cm^{-3}}$, respectively. 
We also compare the heating and advection timescales, using $t_{{\rm heat},C,{\rm p}}/t_{\rm dyn}$, to determine whether the stream is compressed faster than it is heated. 
For ambient gas temperatures below $\sim 10^{7}\,{\rm K}$, the heating time is shorter than the local dynamical time, indicating that CR heating can act before the stream is substantially compressed during its free-fall into the halo.  

In a cosmological context, increasing halo gas temperature on the $y$-axis in Fig.~\ref{fig:timescale_map} provides an approximate proxy for halo mass. 
The stream density on the $x$-axis can be read in two ways, which are not equivalent. If considering density as a radial sequence within a given halo, it traces the increase in gas density toward smaller galacto-centric radii. In this case 
the CR energy density is set by the boundary supply at $R_{\rm v}$ and its subsequent attenuation, and is not tied to the local density. If reading density instead as a 
redshift sequence, the results are not straightforward. This is because the cosmic-web density also rises as $n_{\rm H, web}(z) \propto (1+z)^3$ (see Sec.~\ref{sec:boundary_condition}), so the entrained CR energy density 
increases together with the stream density. Since Fig.~\ref{fig:timescale_map} is evaluated at a fixed supplied CR energy density, this second effect is not captured by the contours shown. The magnitude of CR heating follows from the heating rates in Appendix~\ref{sec:cr_heating_rates}. Each collisional channel is the product of the CR number density and a loss rate proportional to the ambient gas density (see the Coulomb and ionisation terms in Eq.~\ref{eq:Qi_general}, and the hadronic term of Eq.~\ref{eq:Q_had_p}, for which $t_{\rm pp} \propto n_{\rm H}^{-1}$; Eq.~\ref{eq:had_time}). 

At fixed CR spectral shape this gives a heating rate $\propto n_{\rm H}$, and combining with adiabatic compression scaling $U_{\rm CR}\propto n_{\rm H}^{4/3}$ (Eq.~\ref{eq:cr_compression_factor}) gives a CR heating rate scaling as $\propto n_\mathrm{H}^{7/3}$. The volumetric radiative gas cooling rate scales as $\propto n_{\rm H}^2$, so the ratio of heating to cooling grows weakly with density ($\propto n_{\rm H}^{1/3}$). CR heating is 
therefore favoured at higher redshift, where both the stream and its upstream reservoir are denser. 
This scaling applies to the supplied CR population in the adiabatic-entrainment limit. Along the stream itself, collisional attenuation and transport losses modify the CR energy density independently of the local compression, and full transport calculations will depart from this scaling accordingly (see Sec.~\ref{sec:results}). 
This timescale analysis therefore indicates that CR heating is most significant near the virial radius, in more massive haloes, for smaller streams, and at higher redshift, up to the point where advection becomes too rapid for heating to remain effective.    

\section{Results and discussion}
\label{sec:results}

We have shown that the impact of externally supplied CRs depends sensitively on the local balance between CR transport, energy deposition, radiative cooling, and the physical configuration of the stream.
We now examine the resulting transport solutions to determine how an external CR population entrained from the cosmic web evolves along the flow, where it deposits energy, and which parts of the cold stream are most susceptible to heating.
We first consider a single halo as a fiducial reference model, then explore how the physical behaviour depends on galaxy halo mass, redshift, and the density structure of the cold stream. 

\subsection{CR heating structure in a cold stream}
\label{sec:CR_heating_profile}

We construct a fiducial galaxy model with a halo mass 
$M_{\rm h}=10^{12}\,\Msun$, and located at redshift $z=2$.  This is broadly representative of a massive galaxy growing through gas supplied through cold streams near the cosmic noon. 
This case provides a useful reference point for assessing how the entrained CR population evolves as it is transported from the virial region inwards, and how strongly it can modify the thermal state of the stream. 
The underlying CR spectrum we adopt is discussed in more detail in Appendix~\ref{app:cr_profile} (see  Fig.~\ref{fig:cr_spectrum}). 

\subsubsection{Heating-cooling balance}
\label{sec:heating_cooling_profile}

\begin{figure}
    \centering
    \includegraphics[width=1.\linewidth]{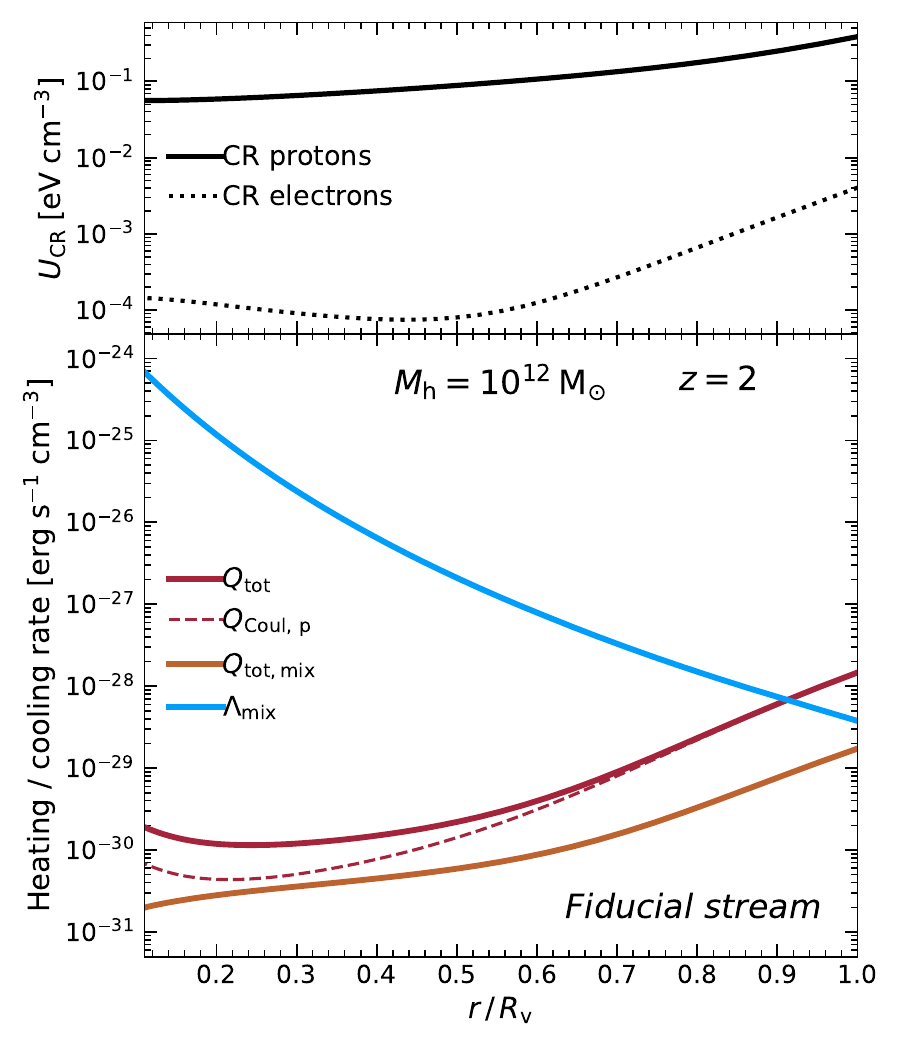}
    \caption{Top: CR distribution along the cold stream in 
    the fiducial ($M_{\rm h}=10^{12}\,\Msun$; $z=2$) model. 
    Bottom: Corresponding radial heating and cooling profiles with radius normalized to the halo virial radius ($R_{\rm v}$), where 
    the blue solid line shows the (volumetric) radiative cooling rate
$\Lambda_{\rm mix}(r)$ in the mixing layer, while the red/orange solid curves show the total CR heating
rate in the stream's spine ($Q_{\rm tot}$) and mixing layer ($Q_{\rm tot, mix}$). The dominant heating channel is via CR proton Coulomb thermalisation, for which the contribution to the total heating is given by the dashed red line ($Q_{\rm Coul, \;p}$). The remainder is from  
collisional heating by CR electrons, 
streaming heating of CR protons, and direct hadronic 
heating. The results for a diffuse stream are shown in Fig.~\ref{fig:heating_profile_diffuse}.} 
    \label{fig:heating_profile}
\end{figure}

\begin{figure}
    \centering
    \includegraphics[width=1.\linewidth]{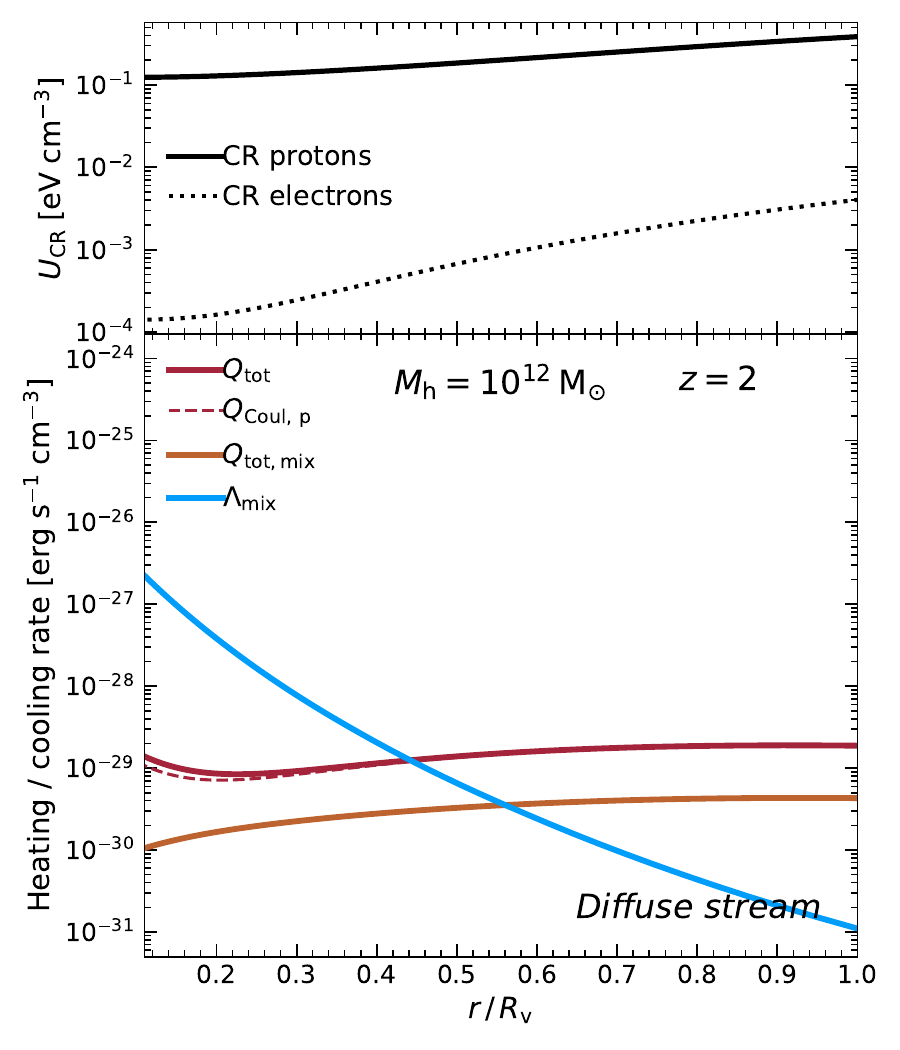}
    \caption{Same as Fig.~\ref{fig:heating_profile}, but for a diffuse stream.}
    \label{fig:heating_profile_diffuse}
\end{figure}

The heating and cooling profiles for the fiducial and diffuse stream models are
shown in Figs.~\ref{fig:heating_profile} and~\ref{fig:heating_profile_diffuse},
respectively, where the upper panels show the CR distribution underlying the
heating effect. In the fiducial case (Fig.~\ref{fig:heating_profile}), radiative cooling (blue solid line) dominates through most of the stream, but the total CR heating rate (dark red solid line) rises toward the outer halo and becomes comparable to, or slightly exceeds, the cooling rate near the virial radius. The total heating is dominated by Coulomb losses of CR protons, mainly driven by the low-energy CR population. Direct thermalisation through hadronic interactions becomes more important only in denser regions, where heating by secondary electrons also begin to contribute after the primary electron component injected near the virial radius has declined. 
Heating by CR streaming is generally sub-dominant and weakens toward smaller galacto-centric radii.  This occurs because streaming heating is tied to the surviving CR proton pressure gradient, while the CR population becomes increasingly attenuated by pp losses as it traverses larger gas columns. 
For smaller or lower-column streams, as shown in Fig.~\ref{fig:heating_profile_diffuse}, attenuation is less severe, allowing CRs to penetrate more deeply into the flow and modify its thermal structure more noticeably.  

Taken together, these results indicate that CR heating develops most efficiently in the outer, sufficiently magnetised parts of the stream, where magnetic coupling and streaming losses remain relevant, while the dominant thermal input is supplied by a long-lived hadronic CR component entering from the cosmic web. 
Radiative cooling becomes progressively stronger toward smaller radii as the stream gas becomes denser, and this inward increase is not offset by CR heating. 
The heating--cooling balance is reversed only in the outermost part of the stream. 
In the fiducial model, CR heating should therefore be interpreted as an outer-halo boundary effect that regulates the thermal state of cold streams as they enter the halo, rather than as a process that globally reheats the full stream. 

\subsubsection{Thermal response of the cold stream}
\label{sec:temperature_profile}

\begin{figure}
    \centering
    \includegraphics[width=1.\linewidth]{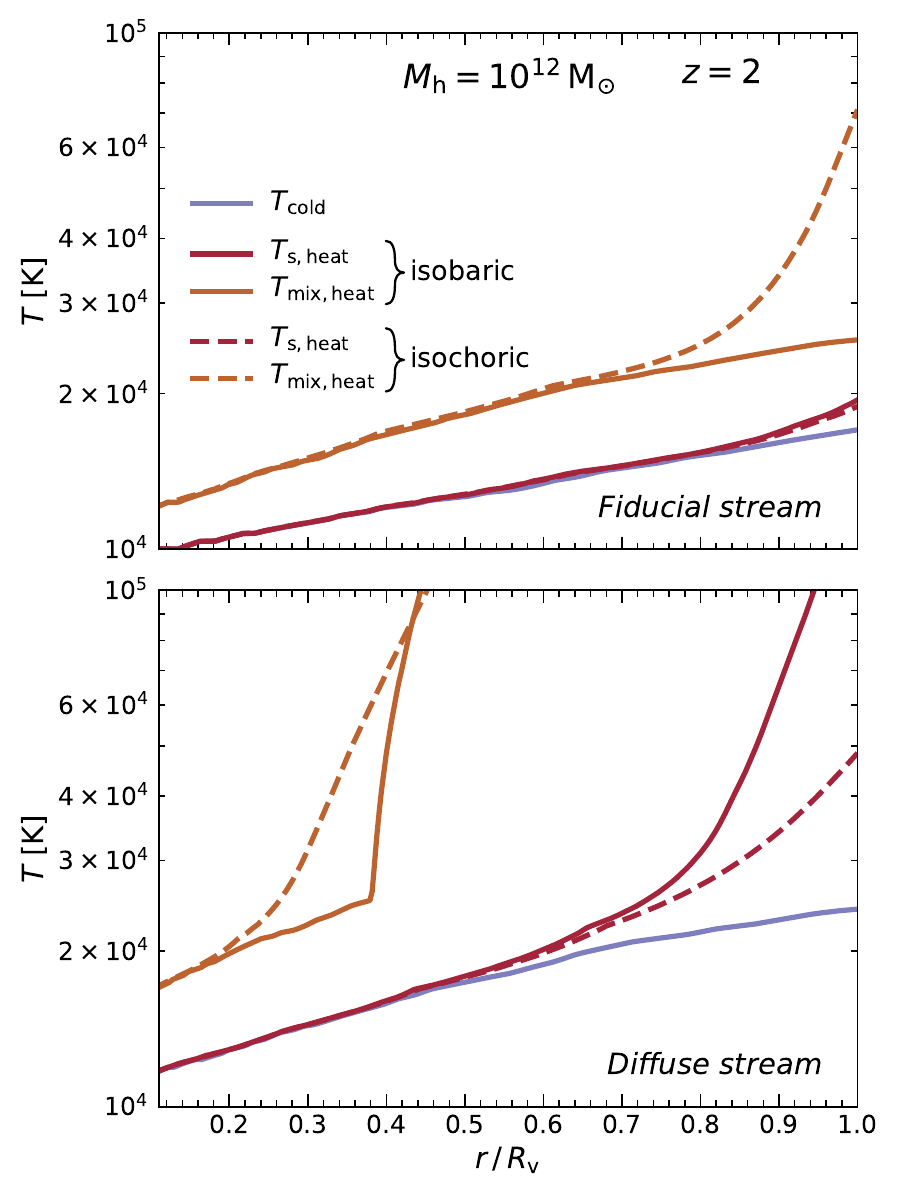}
    \caption{Top: Radial equilibrium temperature profile of the stream gas in the fiducial model 
($M_{\rm h}=10^{12}\,\Msun$; $z=2$). The faded solid dark-blue curve shows the background cold-stream temperature, 
$T_{\rm cold}(r)$, from the cold stream model. 
Solid curves show the isobaric heating-cooling balance temperature and dashed curves show the isochoric balance temperature. Results for the stream spine, $T_{\rm s,heat}(r)$, are in dark red, while the mixing layer, $T_{\rm mix,heat}(r)$, are in orange. The two thermodynamic limits bracket the true response. 
Bottom: Same as top panel, but for a representative diffuse (lower density-bound) stream, where the impact of CR heating is more severe.}
    \label{fig:temperature_profile}
\end{figure}
 Fig.~\ref{fig:temperature_profile} compares the temperatures reached under local CR heating in the stream ($T_{\rm s,heat}$) and its mixing layer ($T_{\rm mix,heat}$), with the corresponding cold-stream temperature profile in the absence of CR heating, $T_{\rm cold}(r)$. The temperatures are obtained by numerically integrating the  
energy equation  
\begin{equation}
    n \;\! c_{\rm X} \frac{\partial T}{\partial t} = Q -\Lambda_\mathrm{net} \ , 
\end{equation}
where $Q=\sum_j\sum_lQ_{j,l}\left(r,\rho,T\right)$ is the total CR heating rate, with heating mechanism $j$ for species $l$ described in Appendix~\ref{sec:cr_heating_rates}, $\Lambda_\mathrm{net}$ is the local net cooling rate (see Appendix~\ref{app:cooling_net}), and $c_{\rm X}$ is the heat capacity per
particle appropriate to the thermodynamic response. 
We solve this equation in
two limiting cases: an isobaric limit, for which $c_{\rm X}=c_{\rm P} \equiv k_{\rm B} \gamma_{\rm g}/(\gamma_{\rm g} - 1)$, and an isochoric limit, for which $c_{\rm X} = c_{\rm V} \equiv k_{\rm B}/(\gamma_{\rm g} - 1)$.  
The isobaric limit is appropriate when the gas pressure can adjust itself faster than it heats, i.e., the sound-crossing time in the cold gas is lower than the heating time.  This leads to the density evolving as $n\propto T^{-1}$. The work of this expansion is 
already accounted for through $c_{\rm P}$, so we do not include a separate gas adiabatic-cooling term. 
The isochoric limit is appropriate when heating proceeds faster than the sound-crossing time, so the gas cannot expand before it heats, and the density, $n$, is held fixed at its initial value. To ensure the heating does not compete with the cold gas contraction as it is advected deeper within the halo, we integrate the energy equation in both limits over one local dynamical time, $t=t_\mathrm{dyn}$. Results are shown for our fiducial parameter choices, $M_{\rm h}=10^{12}\,\Msun$ and $z=2$ for both a fiducial and diffuse stream. 

The isobaric and isochoric calculations bracket the expected thermal response of the gas.  
In the isobaric limit, the gas expands as it heats and maintains fixed pressure, whereas in the isochoric limit its density is held fixed.  These two limits therefore make explicit how the outcome depends on the local thermodynamic response. 
For the fiducial stream, the two solutions remain similar, indicating that CR heating is relatively weak. 
For the diffuse stream, however, they separate at large galacto-centric radii, with the isobaric solution entering the strong-heating regime more readily.  
This occurs because expansion in the isobaric case lowers the gas density and suppresses radiative cooling, whereas the fixed-density isochoric case retains a larger cooling rate. 
Since the diffuse outer stream is initially cold, and may not expand efficiently before being heated, the isochoric solution is likely the more conservative estimate in this region. 

Changing the stream density modifies both the radiative cooling rate and the efficiency of CR heating. 
This dependence can be understood from the timescale analysis in Fig.~\ref{fig:timescale_map}, 
which shows where the dominant CR heating time becomes shorter than the local cooling time. 
Streams denser than our fiducial model generally remain on the cooling-dominated side of this transition and experience a negligible thermal response to the entrained CR population, while lower-density, lower-column streams lie closer to the heating-dominated regime and are affected more strongly. 

\begin{figure*}[t]
    \centering
    \includegraphics[width=\textwidth]{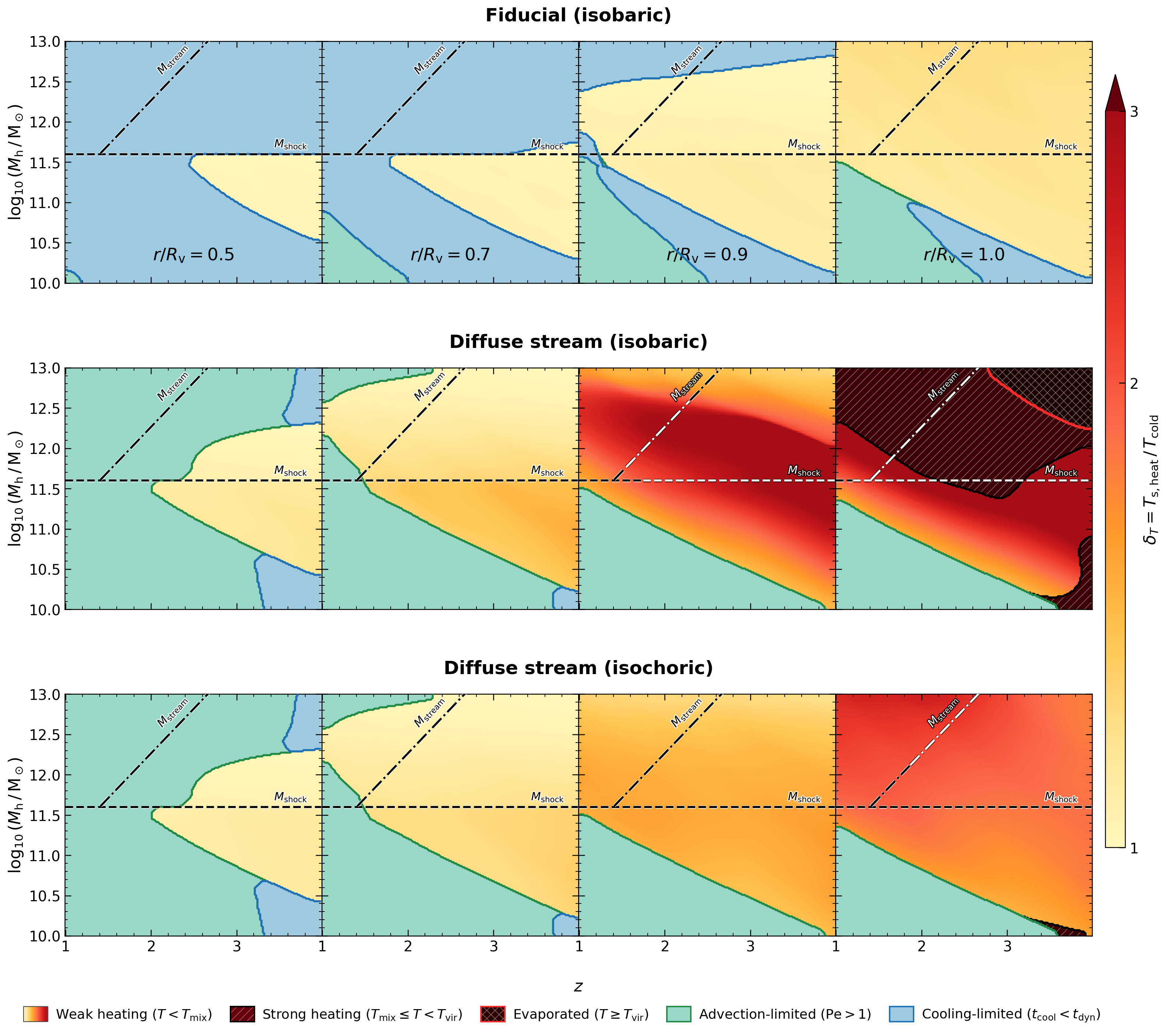}
    \caption{Temperature enhancement factor for gas in cold streams subject to CR heating, 
$\delta_T \equiv T_{\rm s,heat}/T_{\rm cold}$, over halo mass, redshift, and radius, 
for the isobaric and isochoric thermodynamic limits (rows) and fiducial/diffuse stream density as labelled.  
Columns correspond to $r/R_{\rm v}=0.5, 0.7, 0.9,$ and $1.0$. 
Colour indicates the local heating regime, where the light yellow/orange gradient indicates weak heating ($T<T_{\rm mix}$); 
dark brown hatching marks strong heating where the gas is driven to or above the mixing-layer temperature ($T_{\rm mix} \leq T < T_{\rm vir}$); black cross-hatching with a red outline marks evaporation ($T \geq T_{\rm vir}$); light green marks the advection-limited regime, where advection is faster than heating (Péclet number  ${\rm Pe} = t_{\rm heat}/t_{\rm dyn} > 1$); and light blue indicates the cooling-limited regime, where radiative cooling is the fastest process ($t_{\rm cool} < t_{\rm dyn}$) such that CR heating is offset by cooling. Both the advection-limited and cooling-limited regimes correspond to $\delta_{\rm T} \approx 1$, but distinguish whether advection or radiative cooling overcomes CR heating. 
Features around $10^{11.6}\,\Msun$ reflect the absence of a hot halo and associated shear-driven magnetic amplification. Black dashed and dash-dotted lines indicate the approximate shock-stability thresholds $M_{\rm shock}$ and $M_{\rm stream}$, motivated by the cold/hot accretion transition~\citep{Dekel2006}; these are intended as qualitative guides rather than firm boundaries, given the uncertainty in this transition~\citep{Daddi2022ApJ}.} 
    \label{fig:heating_redshift_mH_evolution}
\end{figure*} 
For the fiducial case, $T_{\rm s,heat}(r)$ remains close to $T_{\rm cold}(r)$ through most of the stream, indicating that the entrained CR population does not strongly alter the thermal state of the inner stream. 
However, the outer stream near the virial region can experience non-negligible CR heating, reaching temperatures of a few $\times 10^4\,{\rm K}$. 
This may modify the thermal state of the gas supplied downstream, but the gas remains far below the virial temperature. 
The fiducial stream is therefore heated but not destroyed, corresponding to the weak-heating regime. 
This outcome is not universal, and depends on the stream density and halo conditions. 
For example, the dense-stream case does not significantly respond to CR heating. 
Conversely, the diffuse-stream case is more susceptible (see Fig.~\ref{fig:temperature_profile}, bottom panel), with its temperature noticeably elevated for $r\gtrsim0.4R_{\rm v}$. 
The heating is strong enough that radiative cooling can no longer balance it, placing the flow in the strong-heating regime identified in Sec.~\ref{sec:timescales}. 
In this case, the stream can be heated near or above the temperature of the mixing layer, likely affecting its survival against hydrodynamic instabilities. 

\subsection{Dependence on halo mass, redshift, and stream structure} 
\label{sec:cosmological_evo}

The properties of cold accretion flows vary systematically with halo mass and redshift, modifying their gas density, cooling efficiency, size, and susceptibility to CR heating. Here, we examine how these factors regulate the thermal impact of externally supplied CRs and, consequently, the role of CR heating in cold accretion across the galaxy population. 

\subsubsection{Heating efficiency across the halo population}
\label{sec:heating_efficiency_population}

Figure~\ref{fig:heating_redshift_mH_evolution} shows the temperature increase produced by CR heating,
$\delta_T \equiv T_{\rm s,heat}/T_{\rm cold}$, at several fixed radii as a function of halo mass, $M_{\rm h}$, and redshift, $z$, spanning the regime where cold streams are expected to contribute significantly to galaxy growth \citep[e.g.][]{Dekel2009,Waterval2025}. 
The upper panels show our fiducial stream model, while the lower panels correspond to diffuse streams. Dense streams are omitted because they remain largely unaffected by CR heating, even though some CR loss channels, such as pp interactions, proceed more rapidly in denser gas. Values of $\delta_T \simeq 1$ indicate little modification of the stream thermal equilibrium, whereas larger values correspond to progressively stronger heating within the weak-heating regime. Dark brown hatched regions identify cases where CR heating overwhelms radiative cooling before thermal equilibrium is reached below the mixing-layer temperature. In these regions, the gas can be heated above the mixing-layer temperature, potentially promoting strong stream erosion or disruption. 

The overall trends closely follow the timescale analysis of Sec.~\ref{sec:timescales}. Larger galactocentric radii correspond to lower gas densities, allowing CR heating to compete more effectively with radiative cooling. Higher redshifts increase the ambient CR energy density and therefore enhance CR heating. Increasing halo mass produces a hotter CGM, yielding a hotter, more diffuse mixing layer (Eq.~\ref{eq:mix}) with lower radiative cooling efficiency and thus greater susceptibility to CR heating.
We note that some cosmological hydrodynamical simulations indicate that feedback-driven outflows can preferentially expand through the low-density regions between cold accretion filaments, while the denser streams continue to penetrate the halo~\citep{Powell2011MNRAS, Waterval2025}. In particular,~\citet{Waterval2025} found that hot outflows occupy the regions between and around cold filaments at $z\approx2-4$ and extend beyond the virial radius, with relatively little interaction between the cold inflowing and hot outflowing gas. Strong feedback can also raise the temperature of the hot halo component above that expected from the halo virial temperature alone~\citep{Dubois2013MNRAS}. The ambient gas surrounding a cold stream in some regions may therefore be hotter than assumed by our fiducial halo model.

Fig.~\ref{fig:timescale_map} treats $T_{\rm h}$ directly, and can 
therefore also be read as an approximate indication of this effect. At fixed 
$T_{\rm cold}$, increasing $T_{\rm h}$ raises the temperature contrast between the stream and the ambient gas. 
For an initially pressure-balanced mixing layer, $T_{\rm mix} \propto T_{\rm h}^{1/2}$ and $n_{\rm H, mix} \propto T_{\rm h}^{-1/2}$. At a fixed CR irradiation, the collisional CR heating rate scales approximately with $n_{\rm H, mix}$, while the radiative cooling rate scales as $n_{\rm H, mix}^2 \Lambda(T_{\rm mix})$. A hotter ambient medium therefore generally makes the initial mixing layer more susceptible to CR heating, particularly once $T_{\rm mix}$ moves away from the peak of the cooling curve. However, the corresponding heating time also increases relative to the stream dynamical time. The enhancement therefore persists only until the flow enters the advection-limited regime at $T_{\rm h} \sim 10^{7}$ K.  
Whether produced by increasing halo mass or by additional heating of the ambient gas, this trend persists only until rapid stream advection and compression increase the gas density faster than CR heating can act, producing the advection-limited region (light green) of Fig.~\ref{fig:heating_redshift_mH_evolution}. Within this parameter space where CR heating remains inefficient (${\rm Pe}>1$), we further distinguish the cooling-limited regions ($t_{\rm cool}<t_{\rm dyn}$; light blue), where cooling balances CR heating. 

For the fiducial stream model (upper panels of Fig.~\ref{fig:heating_redshift_mH_evolution}), streams generally remain close to $\delta_T\simeq1$, particularly inside $r/R_{\rm v}\lesssim0.9$. Once a stream reaches the inner halo without substantial CR heating, it is therefore unlikely to be significantly affected at smaller radii. The strongest thermal impact is confined to the outer halo, where the CR population has undergone less attenuation and radiative cooling remains comparatively inefficient. This behaviour is observed across all redshifts. In more massive haloes, the hotter, more diffuse mixing layer initially favours CR heating, until increasing stream velocities render advection and compression dominant. Consequently, significant CR heating is largely restricted to haloes with masses near or above $10^{12}\,\rm M_\odot$, while lower-mass systems remain comparatively resilient throughout cosmic time. 

Our results reflect the competition between the supplied CR density and spectrum, CR survival, and local gas cooling. This is evident in Fig.~\ref{fig:timescale_map}, which illustrates that CR heating is most effective when low-energy particles can deposit energy on a timescale shorter than radiative cooling and advection. The response of a stream to CR heating is therefore controlled by local gas conditions, transport physics, and the supplied CR energy density and spectrum. The spectral dependence provides a further distinction between the scenario considered here and CRs produced by galactic feedback. Appendix~\ref{app:cr_timescale_energy} shows that when the representative proton energy is increased to 1 GeV, advection becomes faster than heating across almost the entire stream parameter space. A feedback-supplied population that has been depleted at lower energies may therefore remain dynamically important through its non-thermal pressure, while producing substantially less collisional heating on the stream transport timescale. Our heating results therefore apply most directly to an externally supplied population that retains a substantial sub-GeV component. 

This sensitivity to transport and local gas conditions is qualitatively consistent with previous simulations of feedback-driven CR haloes~\citep[e.g.][]{Salem2016MNRAS,Butsky2020ApJ,Butsky2022ApJ}. However, 
the main physical effect considered in those studies differs from the collisional heating studied here. Feedback-associated CR populations are generally represented by the $\sim$ GeV particles that carry most of the non-thermal pressure, while the heating found in our calculations is dominated by sub-GeV CRs. This distinction can be seen in~\cite{Roy2025arXiv}, which used a GeV CR-fluid treatment to study the survival of cold gas stripped from satellite galaxies. They found that CR pressure causes the cold clouds to expand, increasing their surface area and promoting radiative cooling in the surrounding mixing layer. This allows the clouds to grow and survive for longer. In our externally supplied CR scenario, lower-energy CRs instead deposit energy collisionally into diffuse or partially mixed stream material. This can move the interface from a cooling-dominated to a heating-dominated regime, reducing the amount of gas that remains in the cold phase. These results therefore describe complementary CR regimes: a pressure-dominated GeV population can promote cold-gas survival through its effect on cloud structure and mixing, while a lower-energy external population can weaken the cold interface through collisional heating. The response of cold halo gas to CRs therefore depends on the CR spectrum and dominant coupling mechanism, as well as on the local gas conditions and transport physics. In the externally supplied scenario considered here, the retention of a sub-GeV CR population favours collisional heating in diffuse or partially mixed stream gas, particularly in massive haloes ($M_{\rm h}\gtrsim10^{12}\,{\rm M_\odot}$) and at higher redshift, where
the supplied CR energy density is larger.  

\subsubsection{Beyond stream survival}
\label{sec:implications_stream_survival}

\begin{figure}
    \centering
    \includegraphics[width=\columnwidth]{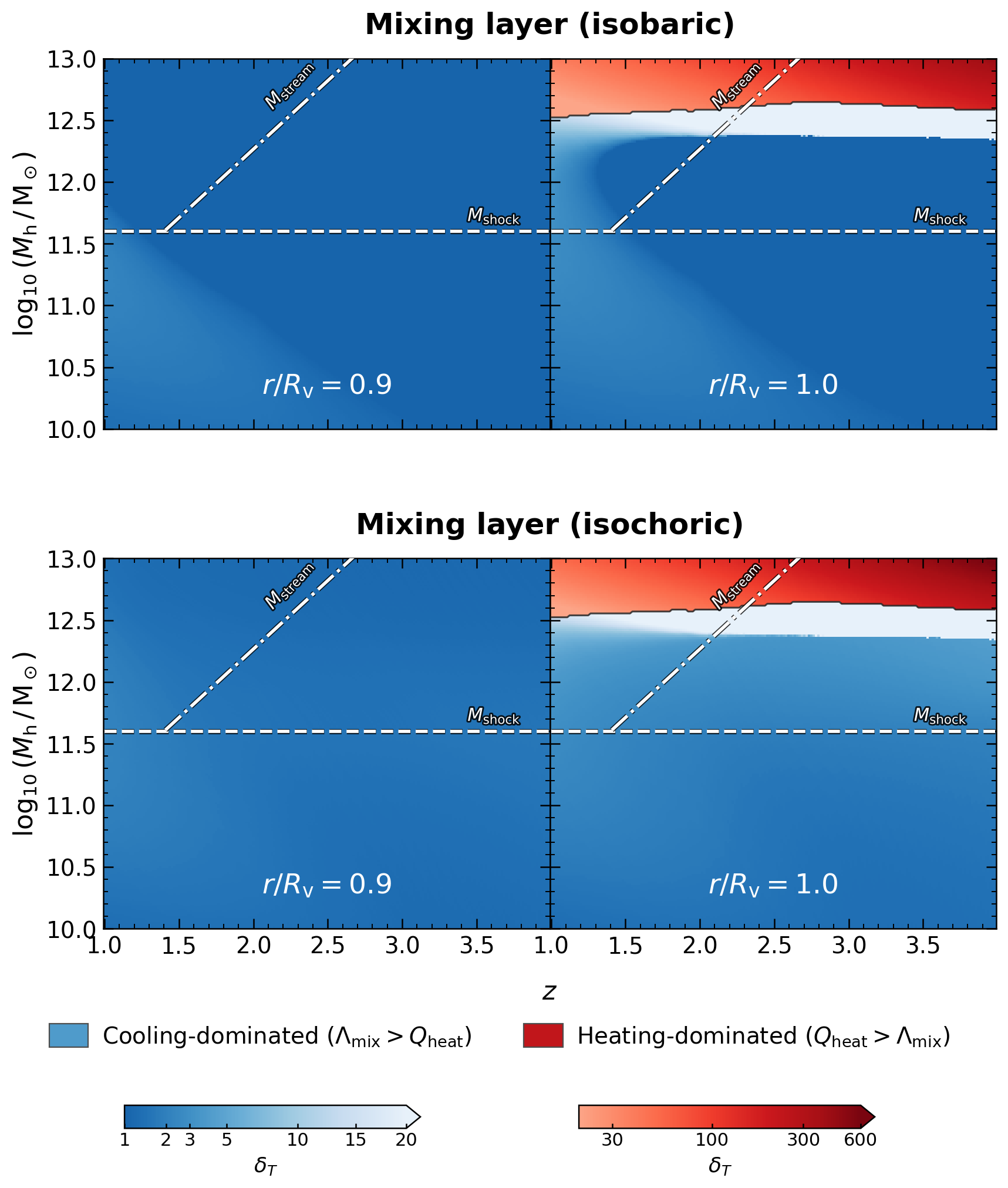}
    \caption{Local heating/cooling balance of the mixing-layer gas for a fiducial stream, over halo mass and redshift, in the isobaric (top) and isochoric (bottom) thermodynamic limits. Columns correspond to results at $r/R_{\rm v} = 0.9$ and 1.0 near the outer region of the halo where CR heating has the greatest impact. The two coloured regions are set by the local heating-cooling rate balance, with the heating-dominated regions indicating where $Q_{\rm heat} > \Lambda_{\rm mix}$ and the cooling-dominated regions where $\Lambda_{\rm mix} > Q_{\rm heat}$. The mixing layer is a transient, phase‑exchanging interface rather than a coherent advecting channel, so its thermal fate is set by whether CR heating dominates over radiative cooling locally. The colour intensity in the two regions shows the CR‑heated gas temperature relative to the cold stream temperature, $\delta_T \equiv T_{\rm mix,heat}/T_{\rm cold}$. Note that colour bars corresponding to the two regions are on different scales for visual clarity, and the values we obtain for $\delta_T$ in the mixing layer are generally substantially higher than in the stream spine shown in Fig.~\ref{fig:heating_redshift_mH_evolution}. Dashed and dash‑dotted lines mark the approximate shock-stability thresholds $M_{\rm shock}$ and $M_{\rm stream}$, as in Fig.~\ref{fig:heating_redshift_mH_evolution}.}
    \label{fig:mixing_param_checks}
\end{figure}

The physical implication of Fig.~\ref{fig:heating_redshift_mH_evolution} is that CR heating is unlikely to affect all cold streams in the same way. 
Dense, coherent stream cores remain comparatively robust, while more diffuse streams are much more vulnerable to CR-driven heating. 
In a less idealised stream configuration, this contrast is likely to become even more pronounced. 
The mixing layer is not expected to remain a smooth, static interface. 
It is a turbulent envelope~\citep[see, e.g.][]{Mandelker2020b} where gas is exchanged between the hot CGM and cold stream. Its interfacial gas may shatter or fragment into smaller cold structures~\citep[e.g.][]{Yao2025}. 
Such fragments can seed the halo with cold substructures and contribute to a highly multiphase CGM. 
Because this interfacial and fragmented material is generally more diffuse, less shielded, and more closely coupled to the ambient halo gas than the central stream spine, it is much more prone to thermal disruption by CRs. 
Thus, even if the dense core of a cold stream remains largely unaffected by CR heating, its surrounding mixing layer and associated cold substructures may not be.

Fig.~\ref{fig:mixing_param_checks} shows the temperature enhancement of the mixing layer for fiducial-like streams. 
Because this gas is less dense than the stream core and has already been partially heated by mixing with the ambient CGM, it is more thermally fragile (i.e. the mixed gas has a shorter cooling time, lower heat capacity per unit volume, and lies closer to the temperature range where the radiative cooling curve can promote isobaric thermal instability;~\citealt{Field1965ApJ, Balbus1995ASPC}). 
Rather than producing the modest temperature enhancement seen in the dense stream gas in Fig.~\ref{fig:heating_redshift_mH_evolution}, even moderate CR heating can push the mixing layer from a cooling/condensation-dominated regime into a heating/evaporation-dominated regime. 
The dense stream spine can absorb substantially more energy before approaching the same threshold, while the mixing layer has a lower heat capacity per unit volume and a reduced ability to maintain radiative balance. 
This thermal fragility suggests two possible outcomes: 
\begin{itemize}
    \item When heating acts on a timescale shorter than $t_\mathrm{dyn}$ (red region in Fig.~\ref{fig:mixing_param_checks}), and radiative cooling cannot compensate the deposited CR energy, the mixed interface gas would become over-pressurised and responds hydrodynamically through expansion or evaporation-like mass exchange with the surrounding CGM.\footnote{This is conceptually analogous to chromospheric evaporation in solar 
    flares, where the energy deposited into dense, radiatively-efficient gas exceeds what 
    that gas can radiate away. The excess drives an over-pressure and an upward 
    enthalpy flux into the overlying tenuous medium rather than being lost locally 
    \citep[for reviews of the observational picture and the underlying MHD 
    description, see][respectively]{Benz2017LRSP,Shibata2011LRSP}. In the  case of solar flares, 
    the energy is supplied by flare-accelerated electron beams or by thermal 
    conduction along reconnected field lines. Here, it is supplied by 
     CR interactions in the stream-CGM interface, but 
    the failure of local radiative losses to balance the deposited energy is common to both cases.} 
    In this case, part of the injected CR energy would be converted into an enthalpy flux rather than being radiated locally.  
    This would tend to erode the cool mixed gas at the stream-CGM interface and smooth the transition between the cold stream and the hot CGM, similar in outcome to the smooth interfacial configurations reported for stream dissipation driven by thermal conduction~\citep{Ledos2024a}, viscosity~\citep{Marin2025}, or magnetic tension~\citep{bib_MHD_Berlok_2019b,Ledos2024a,Das2024MNRAS,Kaul2025}. 
    If the interface is eroded in this way, the dense stream core may become temporarily less prone to shattering and hydrodynamical disruption over part of its trajectory, reducing the formation of clumpy cold CGM substructure. 
    \item When heating acts on a timescale longer than $t_\mathrm{dyn}$ and $t_\mathrm{cool}$, it would instead behave as a persistent effective reduction of the net cooling capacity of the interface gas, as in Eq.~\ref{eq:tcool}. 
    In this case, CR heating could make the interface more susceptible to other dissipative processes, including thermal conduction and thermally driven stream disruption. 
\end{itemize}

Taken together, Figs.~\ref{fig:heating_redshift_mH_evolution} and~\ref{fig:mixing_param_checks} suggest that CR heating introduces a new mode of selectivity into cold-gas accretion onto massive galaxies. 
Rather than acting uniformly on all inflowing gas, an external CR population preferentially affects high redshift, partially mixed, or already fragile stream material within massive halos. 
CR heating may therefore counteract cooling and condensation in the stream-CGM interface, and could contribute an additional physical selection mechanism relevant to the survival, fragmentation, and minimum scale of cold gas in the multiphase CGM~\citep{Nelson2020MNRAS,Bennett2020}. 

\subsection{Astrophysical and observational implications}
\label{sec:implications}

Our results suggest that externally supplied CRs are expected to act first on the mixed envelope of cold streams, introducing an additional selection effect in the survival and observability of marginal cold streams. 
This picture may be considered in the broader context of the transition from cold-stream feeding to hot-mode accretion. 
In theoretical models, stable virial shocks and hot haloes become increasingly important above a characteristic halo mass of order $M_{\rm h}\sim10^{11.6}$--$10^{12}\,\Msun$ at $z\sim2$ \citep{Birnboim2003MNRAS,Dekel2006,Hong2024,Waterval2025,Medlock2026ApJ}. 
Around this regime, galaxy growth depends on whether dense cold streams can survive within an increasingly hot halo environment. 
Observationally, extended cold-gas emission around high-redshift galaxies has been interpreted as a possible signature of cold-stream feeding \citep{Emonts2023Sci,Zhang2023Sci}. 
More massive systems are expected to host hotter CGM gas and appear to show reduced cool-gas covering fractions in their inner CGM, although they are not necessarily devoid of cool gas \citep{Anand2022MNRAS,Barone2024CmPhy,Chang2025arXiv}. 
While these observational interpretations remain subject to uncertainties from  radiative-transfer effects, AGN or star-formation contamination, and the uncertain geometry of the emitting gas, they point to a regime where the persistence of cold gas is likely to be sensitive to additional heating and mixing processes~\citep[e.g.][]{Mandelker2020a, Ledos2024a}. 
Our results suggest that CR heating could contribute to this sensitivity by shifting the effective survival boundary of marginal streams by preferential heating of their outer and partially mixed layers. 
This is consistent with a scenario in which cold-gas delivery declines gradually as haloes become more massive, rather than being abruptly terminated at a single threshold mass \citep{Daddi2022A&A}. 

A possible observational consequence is increased scatter in cold-gas tracers at fixed halo mass and inferred accretion rate. 
Extended Ly$\alpha$ emission is most efficient when a substantial fraction of the inflowing or interface gas remains in a cool, dense, $\sim10^4$ K phase. If CR heating raises part of this gas toward the mixing-layer or hot-CGM temperature, then the same total baryonic inflow rate would contain a smaller fraction of gas in the Ly$\alpha$-emitting phase. 
In this case, the extended Ly$\alpha$ luminosity per unit baryonic accretion rate could decline as a smaller fraction of the stream is held in the cold Ly$\alpha$-emitting phase~\citep{Daddi2022ApJ}. CR-rich environments could therefore show weaker or less extended Ly$\alpha$ emission than otherwise similar systems.   
This should be regarded as an indirect diagnostic, since Ly$\alpha$ morphology and luminosity depend on radiative transfer, illumination, gas geometry, and CGM thermodynamics \citep[see][]{Cantalupo2014,Ouchi2020}. 
However, combining Ly$\alpha$ structures or very metal-poor Lyman-limit systems as candidate signatures of cold accretion \citep{Fumagalli2016MNRAS} with independent indicators of non-thermal activity in the surrounding cosmic web, such as diffuse radio synchrotron emission \citep[e.g.][]{Vernstrom2021MNRAS,Vazza2025A&A},\footnote{Note that we find any non-thermal synchrotron emission associated directly with the CR electron population in the cold stream would be weak and far below plausible current or future detection thresholds.} could test whether CR-rich environments show reduced cold-flow signatures or more halo-to-halo scatter.

 \subsection{Limitations}
\label{sec:limitations}

Our calculations invoke a number of simplifying assumptions and approximations. The most important of these is in our treatment 
of the dynamical interaction between the flow and the CRs. 
The cold-stream structure is taken from an analytic halo/stream model, while CR transport is solved on a fixed background. 
The gas density, velocity, radius, and magnetic-field strength therefore do not respond dynamically to CR heating or CR pressure. This fixed-background approximation is most relevant near the virial region, where CR  heating and pressure can become dynamically important. Our model can identify when the flow becomes susceptible to CR heating, but it does not follow the subsequent expansion, re-mixing, or magnetic reorganisation of the gas. Since the thermal response depends on whether the heated gas evolves isobarically or isochorically, a fully self-consistent treatment would follow the transition between them as the gas heats. This lies beyond our fixed-background approach, but we bracket the possible impacts by showing both limits. This would be particularly relevant to the mixing layer, where CR heating could be particularly effective locally, without being offset by increased cooling.  

Our calculations also use a reduced geometrical description of the stream. We follow CR transport along the stream axis and, within the adopted one-dimensional geometry, we evolve the CR population consistently using the diffusion-advection-loss equation (Eq.~\ref{eq:transport}). This treatment captures axial transport, attenuation, energy losses, secondary production, and CR heating within the stream. 
However, it does not resolve the full cylindrical structure of the flow. Transverse gradients between the dense spine, turbulent envelope, and mixing layer are therefore not evolved self-consistently. This limitation means that the lateral exchange of CRs between the stream and the surrounding CGM is neglected. This includes leakage from the stream, re-entry from the ambient medium, field-line wandering, and transport effects associated with magnetic-field evolution driven by turbulence or mixing. If efficient, such lateral transport could reduce the CR residence time in the stream by making the stream-CGM boundary more permeable to CRs. Substantial CR leakage through the stream boundary would therefore lower the CR heating efficiency compared to our one-dimensional results. Firmly quantifying the magnitude of this effect would require a multidimensional treatment of the ratio of perpendicular to parallel CR diffusion in the local magnetic-field geometry, accounting for the turbulence properties of the stream and its mixing layer. Test-particle and particle-tracing calculations in turbulent magnetic fields often find perpendicular diffusion to be substantially slower than parallel diffusion~\citep[e.g.][]{Casse2001PhRvD, Candia2004JCAP, Desiati2014ApJ}. Although this ratio can depend strongly on the turbulence amplitude, field-line wandering, magnetic-field topology, and transport regime~\citep[e.g.][]{Yan2008ApJ, Xu2013ApJ,Shalchi2020SSRv,Lazarian2023FrASS}, these earlier studies suggest that lateral diffusive leakage is not likely to have a dominant impact in most regimes relevant to our results. This is still an important uncertainty that should be tested with future multidimensional CR-MHD and particle-tracing calculations tailored to cold-stream geometries. 

A further limitation concerns the halo regime where the stream model should be interpreted. 
In this work, we do not model the formation, stability or geometry of a virial shock, and we do not determine where a given halo lies on the cold-flow or hot-mode side of the cold-flow/hot-halo transition~\citep{Birnboim2003MNRAS,Dekel2006,Dekel2009}. Our calculations 
should therefore be interpreted as conditional on a configuration where a hot CGM is present but cold streams can still penetrate it. This is most relevant for massive high-redshift haloes near the cold-flow/hot-halo transition regime. They should not be interpreted as a model of very low-mass haloes without a well-developed hot atmosphere, nor as a prediction for very massive group-scale haloes, $M_{\rm h}\gtrsim10^{13} \; \Msun$, where cold stream penetration may be suppressed or qualitatively different. The lowest- and highest-mass limits of our model parameter survey should therefore be considered as controlled extrapolations of the adopted analytic stream model. 

The adopted halo model also restricts the gas temperature to $T_{\rm h} = \Theta_{\rm h} T_{\rm vir}$, with $\Theta_{\rm h} \in [3/8, 1]$. It therefore does not explicitly represent regions where feedback heats the ambient CGM above the virial-temperature expectation~\citep[e.g.][]{Dubois2013MNRAS}. The independent variation of $T_{\rm h}$ in Fig.~\ref{fig:timescale_map} provides an 
approximate indication of how such additional heating could affect the local heating-cooling balance. 
However, this interpretation retains the pressure-equilibrium scaling of Eqs.~\ref{eq:rho_mix}-~\ref{eq:mix}. 
Feedback-heated gas may instead be spatially intermittent, over-pressurised, or metal-enriched, modifying both the density and cooling efficiency of the mixing layer. The quantitative mapping of feedback-heated channels onto our stream model should therefore not be over-interpreted.  

The externally supplied CR population is also imposed as a fixed boundary condition at $R_{\rm v}$, with a power-law spectrum and prescribed proton-to-electron energy-density ratio. 
This provides a controlled representation of an ambient CR reservoir, rather than a self-consistent model of its origin, entrainment, or time evolution. The adopted upstream CR energy density should therefore be interpreted as a phenomenological normalization for the externally supplied component. The uncertainties associated with this supply are not determined by our model, since we do not self-consistently follow the entrainment process, its efficiency, or temporal/spatial variability in the upstream CR supply. These assumptions are appropriate for an exploratory first calculation, but imply that the precise location of the heating threshold, and the absolute strength of the heating effect, should not be over-interpreted. 
Additional local CR sources within the stream environment are also not included. For example, turbulent re-acceleration or second-order Fermi acceleration could operate at the stream-CGM interface, where turbulence~\citep{Ptuskin1988SvAL, Bustard2022ApJ} and condensation/compression~\citep{Habegger2025arXiv} may energize pre-existing particles. Depending on the magnetic-field configuration and mixing efficiency, these processes could enhance the CR population entering the stream, seed CRs into the surrounding CGM, or modify the spectrum incident on the dense stream spine. These effects require a detailed treatment of turbulence, magnetic-field topology, and CR scattering beyond the present one-dimensional approach, and are better suited to future multidimensional CR-MHD or particle-tracing calculations.  

Finally, we adopt a simplified effective CR transport model, invoking a scalar diffusion coefficient with a fixed Kraichnan-like energy scaling, and approximating the characteristic CR advection speed as the sum of the bulk stream velocity and the local Alfv\'{e}n speed. These are standard assumptions, but they correspond to resolving only the mean magnetic structure of the flow, rather than the detailed transport physics associated with a fully evolving magnetic geometry~\citep{Hanasz2021LRCA, Sampson2023MNRAS}, self-confinement through CR-generated waves~\citep{Zweibel2017PhPl}, their damping~\citep{Lazarian2022FrP, Schroer2025PhRvL}, and the full dissipation physics of CR-driven perturbations and phase-dependent transport effects in a multiphase medium~\citep[e.g.][]{Bustard2021ApJ, Tsung2023MNRAS, Weber2025A&A}. This limitation is important because CR transport in galaxy haloes and the CGM remains theoretically uncertain, and galaxy-scale predictions are known to depend sensitively on the assumed diffusion and streaming prescriptions~\citep[e.g.][]{Hopkins2020MNRAS, Hopkins2021MNRAS}. The advantage of our simplified transport treatment is that it allows us to retain a detailed, energy-dependent description of CR interactions and heating. In particular, we follow spectrally resolved CR protons and electrons through the evolving density and magnetic-field structure of the stream, and explicitly calculate the associated hadronic, Coulomb, secondary electron, and streaming-related heating channels. Our approach therefore provides a complementary limit to full galaxy-scale CR-MHD simulations, which can follow the coupled dynamical response of gas, magnetic fields, turbulence, and CR pressure on galactic and circumgalactic scales~\citep[e.g.][]{Pakmor2016ApJ, Chan2019MNRAS, Hopkins2020MNRAS, Farcy2022MNRAS}, but often invoke energy-averaged or otherwise effective CR-fluid treatments.\footnote{Spectrally resolved CR-MHD methods have recently provided an important step beyond this approximation; see, e.g., \citealt{Girichidis2020MNRAS, Girichidis2022MNRAS, Girichidis2024MNRAS}.} Our results are therefore useful to determine the conditions where strong CR heating and thermo-mechanical disruption may arise in a cold stream. Predicting the subsequent dynamical response will require future CR-MHD simulations that treat CR interactions, thermal gas and magnetic fields self-consistently, while resolving the turbulent, partially mixed stream--CGM interface.  

\section{Conclusions}
\label{sec:summary_conclusions}

In this work, we investigated whether a long-lived CR reservoir associated with the cosmic web could influence cold-stream accretion onto massive galaxies during the cosmic noon.
We constructed a model for the transport and heating of externally supplied CRs entrained into magnetised cold streams, coupling a spectrally resolved CR transport calculation to a redshift-dependent cold-stream model.
This allowed us to assess whether CRs advected from the surrounding filamentary environment can survive within the flow and modify the thermal state of the inflowing gas before it reaches the central galaxy. We found:
\begin{enumerate}
    \item Externally supplied CRs can heat cold streams. This is concentrated mainly in the stream’s outer regions near the virial radius, and is driven mainly by CR protons, which can be advected in from the cosmic web environment (see Sec.~\ref{sec:heating_cooling_profile}; Figs.~\ref{fig:heating_profile} and~\ref{fig:heating_profile_diffuse}). 
    \item Dense streams remain largely resilient to CR heating, but the impact of CR heating in diffuse streams and mixing layers depends sensitively on local thermodynamic conditions, halo mass, redshift, and stream density structure, through the competition between CR transport, losses, advection, and radiative cooling. At $T_{\rm h}\sim10^6 \; {\rm K}$, CR heating becomes faster than local cooling in a stream spine for $n_{\rm H,s}\lesssim10^{-2} \; {\rm cm^{-3}}$, corresponding to a weak-heating regime, while CR heating of the mixing layer becomes efficient for $n_{\rm H,s}\lesssim10^{-2.5} \; {\rm cm^{-3}}$, corresponding to a strong-heating regime (see Sec.~\ref{sec:timescales}; Fig.~\ref{fig:timescale_map}). These structured CR heating regimes are captured by the spectrally resolved, species-dependent, and density-dependent CR interaction treatment developed in this work.     
    \item This phase-dependence implies that externally supplied CRs are likely to act first on partially mixed gas and the stream-CGM interface, rather than on the dense central spine of a cold stream. The mixing layer can remain in an efficient CR-heating regime up to densities $\sim 3$ times higher than in the spine at $T_{\rm h}\sim10^6 \; {\rm K}$ (see Fig.~\ref{fig:timescale_map}). This makes the mixed envelope of a stream more susceptible to thermal modification, while the cold spine can remain comparatively intact (see Sec.~\ref{sec:heating_efficiency_population}; Fig.~\ref{fig:heating_redshift_mH_evolution}; Sec.~\ref{sec:implications_stream_survival}; Fig.~\ref{fig:mixing_param_checks}).
    \item CR heating may therefore reduce the cold, efficiently cooling cross-section of a stream without fully destroying its dense core. By preferentially heating diffuse or partially mixed gas at the stream-CGM interface, externally supplied CRs could erode the cold envelope, suppress cold substructure, or make the surviving cold component appear thinner and more sharply confined (see Sec.~\ref{sec:cosmological_evo}; Figs.~\ref{fig:heating_redshift_mH_evolution} and~\ref{fig:mixing_param_checks}). This provides a possible mechanism for increasing halo-to-halo scatter in cold-gas survival near the transition between cold-stream feeding and hot-mode accretion (see Sec.~\ref{sec:implications}).  
\end{enumerate}
While our calculations are subject to substantial limitations (see Sec.~\ref{sec:limitations}), 
they provide a first indication that CRs supplied externally from the cosmic web can modify cold-gas accretion onto young massive galaxies. 
Rather than acting uniformly on all inflowing gas, this CR population preferentially affects diffuse, partially mixed, or marginally surviving stream material. 
Externally supplied CRs may therefore represent an additional non-thermal selection mechanism in cold-stream accretion, capable of weakening fragile inflows, eroding cold envelopes, and modifying the gas supply available for massive-galaxy growth. 

\begin{acknowledgement}
E.R.O. acknowledges support from the RIKEN Special Postdoctoral Researcher Program for junior scientists, the 
RIKEN Incentive Research Project ("Diffusion and beyond: probing anomalous cosmic-ray transport in magnetized
molecular clouds") 
and the Postdoctoral Fellowship of the Japan Society for the Promotion of Science, supported by KAKENHI Grant Number JP22F22327 and hosted by the University of Osaka, where this work was initiated.  
N.L. acknowledges support from the European Research Council (ERC) under the European Union’s Horizon 2020 research and innovation program grant agreement No 864361. K.N. acknowledges support from JSPS KAKENHI grant 20H00180, 24H00002, 24H00241, JP25K01032, and the JSPS International Leading Research (ILR) project, JP22K21349. K.N. also acknowledges support from the Kavli IPMU, the World Premier Research Center Initiative (WPI), UTIAS, and the University of Tokyo. ChatGPT-5 (OpenAI 2025) and Claude-Opus-4.7 (Anthropic 2026) were used to refine the wording of some sections of the text, and to format the Figures. All content where these tools were used was carefully reviewed, edited, and verified for accuracy by the authors, who assume full responsibility for the outcome.  
\end{acknowledgement}
\vspace{-0.8cm}

\bibliographystyle{bibtex/aa} 
\bibliography{references.bib} 

@article{Ackermann2012ApJ,
  adsurl =
                  {https://ui.adsabs.harvard.edu/abs/2012ApJ...755..164A},
  archiveprefix ={arXiv},
  author =	 {{Ackermann}, M. and {Ajello}, M. and {Allafort},
                  A. and {Baldini}, L. and {Ballet}, J. and
                  {Bastieri}, D. and {Bechtol}, K. and {Bellazzini},
                  R. and {Berenji}, B. and {Bloom}, E.~D. and
                  {Bonamente}, E. and {Borgland}, A.~W. and {Bouvier},
                  A. and {Bregeon}, J. and {Brigida}, M. and {Bruel},
                  P. and {Buehler}, R. and {Buson}, S. and
                  {Caliandro}, G.~A. and {Cameron}, R.~A. and
                  {Caraveo}, P.~A. and {Casandjian}, J.~M. and
                  {Cecchi}, C. and {Charles}, E. and {Chekhtman},
                  A. and {Cheung}, C.~C. and {Chiang}, J. and
                  {Cillis}, A.~N. and {Ciprini}, S. and {Claus},
                  R. and {Cohen-Tanugi}, J. and {Conrad}, J. and
                  {Cutini}, S. and {de Palma}, F. and {Dermer},
                  C.~D. and {Digel}, S.~W. and {Silva}, E. do Couto
                  e. and {Drell}, P.~S. and {Drlica-Wagner}, A. and
                  {Favuzzi}, C. and {Fegan}, S.~J. and {Fortin},
                  P. and {Fukazawa}, Y. and {Funk}, S. and {Fusco},
                  P. and {Gargano}, F. and {Gasparrini}, D. and
                  {Germani}, S. and {Giglietto}, N. and {Giordano},
                  F. and {Glanzman}, T. and {Godfrey}, G. and
                  {Grenier}, I.~A. and {Guiriec}, S. and {Gustafsson},
                  M. and {Hadasch}, D. and {Hayashida}, M. and {Hays},
                  E. and {Hughes}, R.~E. and {J{\'o}hannesson}, G. and
                  {Johnson}, A.~S. and {Kamae}, T. and {Katagiri},
                  H. and {Kataoka}, J. and {Kn{\"o}dlseder}, J. and
                  {Kuss}, M. and {Lande}, J. and {Longo}, F. and
                  {Loparco}, F. and {Lott}, B. and {Lovellette},
                  M.~N. and {Lubrano}, P. and {Madejski}, G.~M. and
                  {Martin}, P. and {Mazziotta}, M.~N. and {McEnery},
                  J.~E. and {Michelson}, P.~F. and {Mizuno}, T. and
                  {Monte}, C. and {Monzani}, M.~E. and {Morselli},
                  A. and {Moskalenko}, I.~V. and {Murgia}, S. and
                  {Nishino}, S. and {Norris}, J.~P. and {Nuss}, E. and
                  {Ohno}, M. and {Ohsugi}, T. and {Okumura}, A. and
                  {Omodei}, N. and {Orlando}, E. and {Ozaki}, M. and
                  {Parent}, D. and {Persic}, M. and {Pesce-Rollins},
                  M. and {Petrosian}, V. and {Pierbattista}, M. and
                  {Piron}, F. and {Pivato}, G. and {Porter}, T.~A. and
                  {Rain{\`o}}, S. and {Rando}, R. and {Razzano},
                  M. and {Reimer}, A. and {Reimer}, O. and {Ritz},
                  S. and {Roth}, M. and {Sbarra}, C. and {Sgr{\`o}},
                  C. and {Siskind}, E.~J. and {Spandre}, G. and
                  {Spinelli}, P. and {Stawarz}, {\L}ukasz and
                  {Strong}, A.~W. and {Takahashi}, H. and {Tanaka},
                  T. and {Thayer}, J.~B. and {Tibaldo}, L. and
                  {Tinivella}, M. and {Torres}, D.~F. and {Tosti},
                  G. and {Troja}, E. and {Uchiyama}, Y. and
                  {Vandenbroucke}, J. and {Vianello}, G. and {Vitale},
                  V. and {Waite}, A.~P. and {Wood}, M. and {Yang}, Z.},
  doi =		 {10.1088/0004-637X/755/2/164},
  eid =		 164,
  eprint =	 {1206.1346},
  journal =	 {\apj},
  month =	 aug,
  number =	 2,
  pages =	 164,
  primaryclass = {astro-ph.HE},
  title =	 {{GeV Observations of Star-forming Galaxies with the
                  Fermi Large Area Telescope}},
  volume =	 755,
  year =	 2012
}

@article{Anand2022MNRAS,
  adsurl =
                  {https://ui.adsabs.harvard.edu/abs/2022MNRAS.513.3210A},
  archiveprefix ={arXiv},
  author =	 {{Anand}, Abhijeet and {Kauffmann}, Guinevere and
                  {Nelson}, Dylan},
  doi =		 {10.1093/mnras/stac928},
  eprint =	 {2201.07811},
  journal =	 {\mnras},
  month =	 jul,
  number =	 3,
  pages =	 {3210-3227},
  primaryclass = {astro-ph.GA},
  title =	 {{Cool circumgalactic gas in galaxy clusters:
                  connecting the DESI legacy imaging survey and SDSS
                  DR16 Mg II absorbers}},
  volume =	 513,
  year =	 2022
}

@article{Aung2024,
  author =	 {{Aung}, Han and {Mandelker}, Nir and {Dekel},
                  Avishai and {Nagai}, Daisuke and {Semenov}, Vadim
                  and {van den Bosch}, Frank C.},
  doi =		 {10.1093/mnras/stae1673},
  journal =	 {\mnras},
  month =	 aug,
  number =	 3,
  pages =	 {2965-2987},
  title =	 {{Entrainment of hot gas into cold streams: the
                  origin of excessive star formation rates at cosmic
                  noon}},
  volume =	 532,
  year =	 2024
}

@inproceedings{Balbus1995ASPC,
  adsurl =
                  {https://ui.adsabs.harvard.edu/abs/1995ASPC...80..328B},
  author =	 {{Balbus}, S.~A.},
  booktitle =	 {The Physics of the Interstellar Medium and
                  Intergalactic Medium},
  editor =	 {{Ferrara}, A. and {McKee}, C.~F. and {Heiles},
                  C. and {Shapiro}, P.~R.},
  month =	 jan,
  pages =	 328,
  series =	 {Astronomical Society of the Pacific Conference
                  Series},
  title =	 {{Thermal Instability}},
  volume =	 80,
  year =	 1995
}

@article{Barone2024CmPhy,
  adsurl =
                  {https://ui.adsabs.harvard.edu/abs/2024CmPhy...7..286B},
  archiveprefix ={arXiv},
  author =	 {{Barone}, Tania M. and {Kacprzak}, Glenn G. and
                  {Nightingale}, James W. and {Nielsen}, Nikole M. and
                  {Glazebrook}, Karl and {Tran}, Kim-Vy H. and
                  {Jones}, Tucker and {Nateghi}, Hasti and {Vasan
                  Gopala Chandrasekaran}, Keerthi and {Sahu}, Nandini
                  and {Nanayakkara}, Themiya and {Skobe}, Hannah and
                  {van de Sande}, Jesse and {Lopez}, Sebastian and
                  {Lewis}, Geraint F.},
  doi =		 {10.1038/s42005-024-01778-4},
  eid =		 286,
  eprint =	 {2408.07984},
  journal =	 {Communications Physics},
  month =	 dec,
  number =	 1,
  pages =	 286,
  primaryclass = {astro-ph.GA},
  title =	 {{Gravitational lensing reveals cool gas within 10-20
                  kpc around a quiescent galaxy}},
  volume =	 7,
  year =	 2024
}

@article{Begelman1990,
  author =	 {Begelman, Mitchell C.},
  journal =	 {MNRAS},
  pages =	 {26-29},
  title =	 {{Turbulent mixing layers in the interstellar and
                  intracluster medium}},
  volume =	 000,
  year =	 1990
}

@article{Bennett2020,
  author =	 {Bennett, Jake S. and Sijacki, Debora},
  doi =		 {10.1093/mnras/staa2835},
  eprint =	 {2006.10058},
  issn =	 13652966,
  journal =	 {MNRAS},
  number =	 1,
  pages =	 {597--615},
  publisher =	 {Oxford University Press},
  title =	 {{Resolving shocks and filaments in galaxy formation
                  simulations: Effects on gas properties and star
                  formation in the circumgalactic medium}},
  volume =	 499,
  year =	 2020
}

@article{Benz2017LRSP,
  adsurl =
                  {https://ui.adsabs.harvard.edu/abs/2017LRSP...14....2B},
  author =	 {{Benz}, Arnold O.},
  doi =		 {10.1007/s41116-016-0004-3},
  eid =		 2,
  journal =	 {Living Reviews in Solar Physics},
  month =	 dec,
  number =	 1,
  pages =	 2,
  title =	 {{Flare Observations}},
  volume =	 14,
  year =	 2017
}

@book{Berezinskii1990acr,
  adsurl =
                  {https://ui.adsabs.harvard.edu/abs/1990acr..book.....B},
  author =	 {{Berezinskii}, V.~S. and {Bulanov}, S.~V. and
                  {Dogiel}, V.~A. and {Ptuskin}, V.~S.},
  title =	 {{Astrophysics of cosmic rays}},
  year =	 1990
}

@article{Berezinsky1997ApJ,
  adsurl =
                  {https://ui.adsabs.harvard.edu/abs/1997ApJ...487..529B},
  archiveprefix ={arXiv},
  author =	 {{Berezinsky}, V.~S. and {Blasi}, P. and {Ptuskin},
                  V.~S.},
  doi =		 {10.1086/304622},
  eprint =	 {astro-ph/9609048},
  journal =	 {\apj},
  month =	 oct,
  number =	 2,
  pages =	 {529-535},
  primaryclass = {astro-ph},
  title =	 {{Clusters of Galaxies as Storage Room for Cosmic
                  Rays}},
  volume =	 487,
  year =	 1997
}

@article{Birnboim2003MNRAS,
  adsurl =
                  {https://ui.adsabs.harvard.edu/abs/2003MNRAS.345..349B},
  archiveprefix ={arXiv},
  author =	 {{Birnboim}, Yuval and {Dekel}, Avishai},
  doi =		 {10.1046/j.1365-8711.2003.06955.x},
  eprint =	 {astro-ph/0302161},
  journal =	 {\mnras},
  month =	 oct,
  number =	 1,
  pages =	 {349-364},
  primaryclass = {astro-ph},
  title =	 {{Virial shocks in galactic haloes?}},
  volume =	 345,
  year =	 2003
}

@article{Blasi2013A&ARv,
  adsurl =
                  {https://ui.adsabs.harvard.edu/abs/2013A&ARv..21...70B},
  archiveprefix ={arXiv},
  author =	 {{Blasi}, Pasquale},
  doi =		 {10.1007/s00159-013-0070-7},
  eid =		 70,
  eprint =	 {1311.7346},
  journal =	 {\aapr},
  month =	 nov,
  pages =	 70,
  primaryclass = {astro-ph.HE},
  title =	 {{The origin of galactic cosmic rays}},
  volume =	 21,
  year =	 2013
}

@article{Brunetti2014IJMPD,
  adsurl =
                  {https://ui.adsabs.harvard.edu/abs/2014IJMPD..2330007B},
  archiveprefix ={arXiv},
  author =	 {{Brunetti}, Gianfranco and {Jones}, Thomas W.},
  doi =		 {10.1142/S0218271814300079},
  eid =		 {1430007-98},
  eprint =	 {1401.7519},
  journal =	 {International Journal of Modern Physics D},
  month =	 mar,
  number =	 4,
  pages =	 {1430007-98},
  primaryclass = {astro-ph.CO},
  title =	 {{Cosmic Rays in Galaxy Clusters and Their Nonthermal
                  Emission}},
  volume =	 23,
  year =	 2014
}

@article{Bryan1998,
  adsurl =
                  {https://ui.adsabs.harvard.edu/abs/1998ApJ...495...80B},
  archiveprefix ={arXiv},
  author =	 {{Bryan}, Greg L. and {Norman}, Michael L.},
  doi =		 {10.1086/305262},
  eprint =	 {astro-ph/9710107},
  journal =	 {\apj},
  month =	 mar,
  number =	 1,
  pages =	 {80-99},
  primaryclass = {astro-ph},
  title =	 {{Statistical Properties of X-Ray Clusters: Analytic
                  and Numerical Comparisons}},
  volume =	 495,
  year =	 1998
}

@article{Bustard2021ApJ,
  adsurl =
                  {https://ui.adsabs.harvard.edu/abs/2021ApJ...913..106B},
  archiveprefix ={arXiv},
  author =	 {{Bustard}, Chad and {Zweibel}, Ellen G.},
  doi =		 {10.3847/1538-4357/abf64c},
  eid =		 106,
  eprint =	 {2012.06585},
  journal =	 {\apj},
  month =	 jun,
  number =	 2,
  pages =	 106,
  primaryclass = {astro-ph.HE},
  title =	 {{Cosmic-Ray Transport, Energy Loss, and Influence in
                  the Multiphase Interstellar Medium}},
  volume =	 913,
  year =	 2021
}

@article{Bustard2022ApJ,
  adsurl =
                  {https://ui.adsabs.harvard.edu/abs/2022ApJ...941...65B},
  archiveprefix ={arXiv},
  author =	 {{Bustard}, Chad and {Oh}, S. Peng},
  doi =		 {10.3847/1538-4357/aca021},
  eid =		 65,
  eprint =	 {2208.02261},
  journal =	 {\apj},
  month =	 dec,
  number =	 1,
  pages =	 65,
  primaryclass = {astro-ph.HE},
  title =	 {{Turbulent Reacceleration of Streaming Cosmic Rays}},
  volume =	 941,
  year =	 2022
}

@article{Butsky2020ApJ,
  adsurl =
                  {https://ui.adsabs.harvard.edu/abs/2020ApJ...903...77B},
  archiveprefix ={arXiv},
  author =	 {{Butsky}, Iryna S. and {Fielding}, Drummond B. and
                  {Hayward}, Christopher C. and {Hummels}, Cameron
                  B. and {Quinn}, Thomas R. and {Werk}, Jessica K.},
  doi =		 {10.3847/1538-4357/abbad2},
  eid =		 77,
  eprint =	 {2008.04915},
  journal =	 {\apj},
  month =	 nov,
  number =	 2,
  pages =	 77,
  primaryclass = {astro-ph.GA},
  title =	 {{The Impact of Cosmic Rays on Thermal Instability in
                  the Circumgalactic Medium}},
  volume =	 903,
  year =	 2020
}

@article{Butsky2022ApJ,
  adsurl =
                  {https://ui.adsabs.harvard.edu/abs/2022ApJ...935...69B},
  archiveprefix ={arXiv},
  author =	 {{Butsky}, Iryna S. and {Werk}, Jessica K. and
                  {Tchernyshyov}, Kirill and {Fielding}, Drummond
                  B. and {Breneman}, Joseph and {Piacitelli}, Daniel
                  R. and {Quinn}, Thomas R. and {Sanchez}, N. Nicole
                  and {Cruz}, Akaxia and {Hummels}, Cameron B. and
                  {Burchett}, Joseph N. and {Tremmel}, Michael},
  doi =		 {10.3847/1538-4357/ac7ebd},
  eid =		 69,
  eprint =	 {2106.14889},
  journal =	 {\apj},
  month =	 aug,
  number =	 2,
  pages =	 69,
  primaryclass = {astro-ph.GA},
  title =	 {{The Impact of Cosmic Rays on the Kinematics of the
                  Circumgalactic Medium}},
  volume =	 935,
  year =	 2022
}

@article{Candia2004JCAP,
  adsurl =
                  {https://ui.adsabs.harvard.edu/abs/2004JCAP...10..007C},
  archiveprefix ={arXiv},
  author =	 {{Candia}, Juli{\'a}n and {Roulet}, Esteban},
  doi =		 {10.1088/1475-7516/2004/10/007},
  eid =		 007,
  eprint =	 {astro-ph/0408054},
  journal =	 {\jcap},
  month =	 oct,
  number =	 10,
  pages =	 007,
  primaryclass = {astro-ph},
  title =	 {{Diffusion and drift of cosmic rays in highly
                  turbulent magnetic fields}},
  volume =	 2004,
  year =	 2004
}

@article{Cantalupo2014,
  adsurl =
                  {https://ui.adsabs.harvard.edu/abs/2014Natur.506...63C},
  archiveprefix ={arXiv},
  author =	 {{Cantalupo}, Sebastiano and {Arrigoni-Battaia},
                  Fabrizio and {Prochaska}, J. Xavier and {Hennawi},
                  Joseph F. and {Madau}, Piero},
  doi =		 {10.1038/nature12898},
  eprint =	 {1401.4469},
  journal =	 {\nat},
  month =	 feb,
  number =	 7486,
  pages =	 {63-66},
  primaryclass = {astro-ph.CO},
  title =	 {{A cosmic web filament revealed in
                  Lyman-{\ensuremath{\alpha}} emission around a
                  luminous high-redshift quasar}},
  volume =	 506,
  year =	 2014
}

@inproceedings{Cantalupo2017ASSL,
  adsurl =
                  {https://ui.adsabs.harvard.edu/abs/2017ASSL..430..195C},
  archiveprefix ={arXiv},
  author =	 {{Cantalupo}, Sebastiano},
  booktitle =	 {Gas Accretion onto Galaxies},
  doi =		 {10.1007/978-3-319-52512-9_9},
  editor =	 {{Fox}, Andrew and {Dav{\'e}}, Romeel},
  eprint =	 {1612.00491},
  month =	 jan,
  pages =	 195,
  primaryclass = {astro-ph.GA},
  series =	 {Astrophysics and Space Science Library},
  title =	 {{Gas Accretion and Giant Ly{\ensuremath{\alpha}}
                  Nebulae}},
  volume =	 430,
  year =	 2017
}

@article{Casse2001PhRvD,
  adsurl =
                  {https://ui.adsabs.harvard.edu/abs/2001PhRvD..65b3002C},
  archiveprefix ={arXiv},
  author =	 {{Casse}, Fabien and {Lemoine}, Martin and
                  {Pelletier}, Guy},
  doi =		 {10.1103/PhysRevD.65.023002},
  eid =		 023002,
  eprint =	 {astro-ph/0109223},
  journal =	 {\prd},
  month =	 dec,
  number =	 2,
  pages =	 023002,
  primaryclass = {astro-ph},
  title =	 {{Transport of cosmic rays in chaotic magnetic
                  fields}},
  volume =	 65,
  year =	 2001
}

@article{Chan2019MNRAS,
  adsurl =
                  {https://ui.adsabs.harvard.edu/abs/2019MNRAS.488.3716C},
  archiveprefix ={arXiv},
  author =	 {{Chan}, T.~K. and {Kere{\v{s}}}, D. and {Hopkins},
                  P.~F. and {Quataert}, E. and {Su}, K.-Y. and
                  {Hayward}, C.~C. and {Faucher-Gigu{\`e}re}, C.-A.},
  doi =		 {10.1093/mnras/stz1895},
  eprint =	 {1812.10496},
  journal =	 {\mnras},
  month =	 sep,
  number =	 3,
  pages =	 {3716-3744},
  primaryclass = {astro-ph.GA},
  title =	 {{Cosmic ray feedback in the FIRE simulations:
                  constraining cosmic ray propagation with GeV
                  {\ensuremath{\gamma}}-ray emission}},
  volume =	 488,
  year =	 2019
}

@article{Chang2025arXiv,
  adsurl =
                  {https://ui.adsabs.harvard.edu/abs/2025arXiv251203845C},
  archiveprefix ={arXiv},
  author =	 {{Chang}, Yu-Ling and {Lan}, Ting-Wen and
                  {Prochaska}, J. Xavier and {Siudek}, Malgorzata and
                  {Aguilar}, J. and {Ahlen}, S. and {Anand}, A. and
                  {Bianchi}, D. and {Brooks}, D. and {Castander},
                  F.~J. and {Claybaugh}, T. and {de la Macorra},
                  A. and {Doel}, P. and {Ferraro}, S. and
                  {Font-Ribera}, A. and {Forero-Romero}, J.~E. and
                  {Gaztanaga}, E. and {Gontcho}, S. Gontcho A and
                  {Gutierrez}, G. and {Guy}, J. and {Honscheid},
                  K. and {Joyce}, R. and {Juneau}, S. and {Kremin},
                  A. and {Lahav}, O. and {Lamman}, C. and {Landriau},
                  M. and {Le Guillou}, L. and {Levi}, M.~E. and
                  {Manera}, M. and {Meisner}, A. and {Miquel}, R. and
                  {Nadathur}, S. and {Newman}, J.~A. and {Percival},
                  W.~J. and {Poppett}, C. and {Prada}, F. and
                  {Perez-Rafols}, I. and {Rossi}, G. and {Sanchez},
                  E. and {Schlegel}, D. and {Schubnell}, M. and
                  {Sprayberry}, D. and {Tarle}, G. and {Weaver},
                  B.~A. and {Zhou}, R. and {Zou}, H.},
  doi =		 {10.48550/arXiv.2512.03845},
  eid =		 {arXiv:2512.03845},
  eprint =	 {2512.03845},
  journal =	 {arXiv e-prints},
  month =	 dec,
  pages =	 {arXiv:2512.03845},
  primaryclass = {astro-ph.GA},
  title =	 {{Tracing the Cosmic Evolution of the Cool
                  Circumgalactic Medium of Luminous Red Galaxies with
                  DESI Year 1 Data}},
  year =	 2025
}

@article{Chen2024A&A,
  adsurl =
                  {https://ui.adsabs.harvard.edu/abs/2024A&A...692A..34C},
  archiveprefix ={arXiv},
  author =	 {{Chen}, Jianhang and {Lopez-Rodriguez}, Enrique and
                  {Ivison}, R.~J. and {Geach}, James E. and {Dye},
                  Simon and {Liu}, Xiaohui and {Bendo}, George},
  doi =		 {10.1051/0004-6361/202450969},
  eid =		 {A34},
  eprint =	 {2407.14596},
  journal =	 {\aap},
  month =	 dec,
  pages =	 {A34},
  primaryclass = {astro-ph.GA},
  title =	 {{A kiloparsec-scale ordered magnetic field in a
                  galaxy at z = 5.6}},
  volume =	 692,
  year =	 2024
}

@article{Correa2015MNRAS,
  adsurl =
                  {https://ui.adsabs.harvard.edu/abs/2015MNRAS.452.1217C},
  archiveprefix ={arXiv},
  author =	 {{Correa}, Camila A. and {Wyithe}, J. Stuart B. and
                  {Schaye}, Joop and {Duffy}, Alan R.},
  doi =		 {10.1093/mnras/stv1363},
  eprint =	 {1502.00391},
  journal =	 {\mnras},
  month =	 sep,
  number =	 2,
  pages =	 {1217-1232},
  primaryclass = {astro-ph.CO},
  title =	 {{The accretion history of dark matter haloes -
                  III. A physical model for the concentration-mass
                  relation}},
  volume =	 452,
  year =	 2015
}

@article{Crighton2013,
  adsurl =
                  {https://ui.adsabs.harvard.edu/abs/2013ApJ...776L..18C},
  archiveprefix ={arXiv},
  author =	 {{Crighton}, Neil H.~M. and {Hennawi}, Joseph F. and
                  {Prochaska}, J. Xavier},
  doi =		 {10.1088/2041-8205/776/2/L18},
  eid =		 {L18},
  eprint =	 {1307.6588},
  journal =	 {\apjl},
  month =	 oct,
  number =	 2,
  pages =	 {L18},
  primaryclass = {astro-ph.CO},
  title =	 {{Metal-poor, Cool Gas in the Circumgalactic Medium
                  of a z = 2.4 Star-forming Galaxy: Direct Evidence
                  for Cold Accretion?}},
  volume =	 776,
  year =	 2013
}

@article{Daddi2022A&A,
  adsurl =
                  {https://ui.adsabs.harvard.edu/abs/2022A&A...661L...7D},
  archiveprefix ={arXiv},
  author =	 {{Daddi}, E. and {Delvecchio}, I. and {Dimauro},
                  P. and {Magnelli}, B. and {Gomez-Guijarro}, C. and
                  {Coogan}, R. and {Elbaz}, D. and {Kalita}, B.~S. and
                  {Le Bail}, A. and {Rich}, R.~M. and {Tan}, Q.},
  doi =		 {10.1051/0004-6361/202243574},
  eid =		 {L7},
  eprint =	 {2203.10880},
  journal =	 {\aap},
  month =	 may,
  pages =	 {L7},
  primaryclass = {astro-ph.CO},
  title =	 {{The bending of the star-forming main sequence
                  traces the cold- to hot-accretion transition mass
                  over 0 < z < 4}},
  volume =	 661,
  year =	 2022
}

@article{Daddi2022ApJ,
  adsurl =
                  {https://ui.adsabs.harvard.edu/abs/2022ApJ...926L..21D},
  archiveprefix ={arXiv},
  author =	 {{Daddi}, E. and {Rich}, R.~M. and {Valentino},
                  F. and {Jin}, S. and {Delvecchio}, I. and {Liu},
                  D. and {Strazzullo}, V. and {Neill}, J. and {Gobat},
                  R. and {Finoguenov}, A. and {Bournaud}, F. and
                  {Elbaz}, D. and {Kalita}, B.~S. and {O'Sullivan},
                  D. and {Wang}, T.},
  doi =		 {10.3847/2041-8213/ac531f},
  eid =		 {L21},
  eprint =	 {2202.03715},
  journal =	 {\apjl},
  month =	 feb,
  number =	 2,
  pages =	 {L21},
  primaryclass = {astro-ph.CO},
  title =	 {{Evidence for Cold-stream to Hot-accretion
                  Transition as Traced by Ly{\ensuremath{\alpha}}
                  Emission from Groups and Clusters at 2 < z < 3.3}},
  volume =	 926,
  year =	 2022
}

@article{Danovich2015,
  author =	 {Danovich, Mark and Dekel, Avishai and Hahn, Oliver
                  and Ceverino, Daniel and Primack, Joel},
  doi =		 {10.1093/mnras/stv270},
  eprint =	 {1407.7129},
  issn =	 13652966,
  journal =	 {MNRAS},
  number =	 2,
  pages =	 {2087--2111},
  title =	 {{Four phases of angular-momentum buildup in high-z
                  galaxies: From cosmic-web streams through an
                  extended ring to disc and bulge}},
  volume =	 449,
  year =	 2015
}

@article{Das2024MNRAS,
  adsurl =
                  {https://ui.adsabs.harvard.edu/abs/2024MNRAS.527..991D},
  archiveprefix ={arXiv},
  author =	 {{Das}, Hitesh Kishore and {Gronke}, Max},
  doi =		 {10.1093/mnras/stad3125},
  eprint =	 {2307.06411},
  journal =	 {\mnras},
  month =	 jan,
  number =	 1,
  pages =	 {991-1013},
  primaryclass = {astro-ph.GA},
  title =	 {{Magnetic fields in multiphase turbulence: impact on
                  dynamics and structure}},
  volume =	 527,
  year =	 2024
}

@article{Dekel2006,
  author =	 {Dekel, Avishai and Birnboim, Yuval},
  doi =		 {10.1111/j.1365-2966.2006.10145.x},
  issn =	 00358711,
  journal =	 {MNRAS},
  number =	 1,
  pages =	 {2--20},
  title =	 {{Galaxy bimodality due to cold flows and shock
                  heating}},
  volume =	 368,
  year =	 2006
}

@article{Dekel2009,
  author =	 {Dekel, A. and Birnboim, Y. and Engel, G. and
                  Freundlich, J. and Goerdt, T. and Mumcuoglu, M. and
                  Neistein, E. and Pichon, C. and Teyssier, R. and
                  Zinger, E.},
  doi =		 {10.1038/nature07648},
  eprint =	 {0808.0553},
  issn =	 00280836,
  journal =	 {\nat},
  number =	 7228,
  pages =	 {451--454},
  publisher =	 {Nature Publishing Group},
  title =	 {{Cold streams in early massive hot haloes as the
                  main mode of galaxy formation}},
  volume =	 457,
  year =	 2009
}

@article{Dekel2013,
  author =	 {Dekel, Avishai and Zolotov, A and Tweed, D and
                  Cacciato, M and Ceverino, D and Primack, J R},
  doi =		 {10.1093/mnras/stt1338},
  journal =	 {MNRAS},
  pages =	 {999--1019},
  title =	 {{Toy models for galaxy formation versus
                  simulations}},
  volume =	 435,
  year =	 2013
}

@article{Desiati2014ApJ,
  adsurl =
                  {https://ui.adsabs.harvard.edu/abs/2014ApJ...791...51D},
  archiveprefix ={arXiv},
  author =	 {{Desiati}, Paolo and {Zweibel}, Ellen G.},
  doi =		 {10.1088/0004-637X/791/1/51},
  eid =		 51,
  eprint =	 {1402.1475},
  journal =	 {\apj},
  month =	 aug,
  number =	 1,
  pages =	 51,
  primaryclass = {astro-ph.HE},
  title =	 {{The Transport of Cosmic Rays Across Magnetic
                  Fieldlines}},
  volume =	 791,
  year =	 2014
}

@article{Dubois2013MNRAS,
  adsurl =
                  {https://ui.adsabs.harvard.edu/abs/2013MNRAS.428.2885D},
  archiveprefix ={arXiv},
  author =	 {{Dubois}, Yohan and {Pichon}, Christophe and
                  {Devriendt}, Julien and {Silk}, Joseph and
                  {Haehnelt}, Martin and {Kimm}, Taysun and {Slyz},
                  Adrianne},
  doi =		 {10.1093/mnras/sts224},
  eprint =	 {1206.5838},
  journal =	 {\mnras},
  month =	 feb,
  number =	 4,
  pages =	 {2885-2900},
  primaryclass = {astro-ph.CO},
  title =	 {{Blowing cold flows away: the impact of early AGN
                  activity on the formation of a brightest cluster
                  galaxy progenitor}},
  volume =	 428,
  year =	 2013
}

@article{Emonts2023Sci,
  adsurl =
                  {https://ui.adsabs.harvard.edu/abs/2023Sci...379.1323E},
  archiveprefix ={arXiv},
  author =	 {{Emonts}, Bjorn H.~C. and {Lehnert}, Matthew D. and
                  {Yoon}, Ilsang and {Mandelker}, Nir and
                  {Villar-Mart{\'\i}n}, Montserrat and {Miley}, George
                  K. and {De Breuck}, Carlos and {P{\'e}rez-Torres},
                  Miguel A. and {Hatch}, Nina A. and {Guillard},
                  Pierre},
  doi =		 {10.1126/science.abh2150},
  eprint =	 {2303.17484},
  journal =	 {Science},
  month =	 mar,
  number =	 6639,
  pages =	 {1323-1326},
  primaryclass = {astro-ph.GA},
  title =	 {{A cosmic stream of atomic carbon gas connected to a
                  massive radio galaxy at redshift 3.8}},
  volume =	 379,
  year =	 2023
}

@article{Farcy2022MNRAS,
  adsurl =
                  {https://ui.adsabs.harvard.edu/abs/2022MNRAS.513.5000F},
  archiveprefix ={arXiv},
  author =	 {{Farcy}, Marion and {Rosdahl}, Joakim and {Dubois},
                  Yohan and {Blaizot}, J{\'e}r{\'e}my and
                  {Martin-Alvarez}, Sergio},
  doi =		 {10.1093/mnras/stac1196},
  eprint =	 {2202.01245},
  journal =	 {\mnras},
  month =	 jul,
  number =	 4,
  pages =	 {5000-5019},
  primaryclass = {astro-ph.GA},
  title =	 {{Radiation-magnetohydrodynamics simulations of
                  cosmic ray feedback in disc galaxies}},
  volume =	 513,
  year =	 2022
}

@article{Fardal2001,
  author =	 {Fardal, Mark A. and Katz, Neal and Gardner, Jeffrey
                  P. and Hernquist, Lars and Weinberg, David H. and
                  Dave, Romeel},
  doi =		 {10.1086/323519},
  issn =	 {0004-637X},
  journal =	 {ApJ},
  number =	 2,
  pages =	 {605--617},
  title =	 {{Cooling Radiation and the Ly$\alpha$ Luminosity of
                  Forming Galaxies}},
  volume =	 562,
  year =	 2001
}

@article{Ferland2017RMxAA,
  adsurl =
                  {https://ui.adsabs.harvard.edu/abs/2017RMxAA..53..385F},
  archiveprefix ={arXiv},
  author =	 {{Ferland}, G.~J. and {Chatzikos}, M. and
                  {Guzm{\'a}n}, F. and {Lykins}, M.~L. and {van Hoof},
                  P.~A.~M. and {Williams}, R.~J.~R. and {Abel},
                  N.~P. and {Badnell}, N.~R. and {Keenan}, F.~P. and
                  {Porter}, R.~L. and {Stancil}, P.~C.},
  doi =		 {10.48550/arXiv.1705.10877},
  eprint =	 {1705.10877},
  journal =	 {\rmxaa},
  month =	 oct,
  pages =	 {385-438},
  primaryclass = {astro-ph.GA},
  title =	 {{The 2017 Release Cloudy}},
  volume =	 53,
  year =	 2017
}

@article{Field1965ApJ,
  adsurl =
                  {https://ui.adsabs.harvard.edu/abs/1965ApJ...142..531F},
  author =	 {{Field}, George B.},
  doi =		 {10.1086/148317},
  journal =	 {\apj},
  month =	 aug,
  pages =	 531,
  title =	 {{Thermal Instability.}},
  volume =	 142,
  year =	 1965
}

@article{Fumagalli2016MNRAS,
  adsurl =
                  {https://ui.adsabs.harvard.edu/abs/2016MNRAS.462.1978F},
  archiveprefix ={arXiv},
  author =	 {{Fumagalli}, Michele and {Cantalupo}, Sebastiano and
                  {Dekel}, Avishai and {Morris}, Simon L. and
                  {O'Meara}, John M. and {Prochaska}, J. Xavier and
                  {Theuns}, Tom},
  doi =		 {10.1093/mnras/stw1782},
  eprint =	 {1607.03893},
  journal =	 {\mnras},
  month =	 oct,
  number =	 2,
  pages =	 {1978-1988},
  primaryclass = {astro-ph.GA},
  title =	 {{MUSE searches for galaxies near very metal-poor gas
                  clouds at z {\ensuremath{\sim}} 3: new constraints
                  for cold accretion models}},
  volume =	 462,
  year =	 2016
}

@article{Gabici2019IJMPD,
  adsurl =
                  {https://ui.adsabs.harvard.edu/abs/2019IJMPD..2830022G},
  archiveprefix ={arXiv},
  author =	 {{Gabici}, Stefano and {Evoli}, Carmelo and
                  {Gaggero}, Daniele and {Lipari}, Paolo and
                  {Mertsch}, Philipp and {Orlando}, Elena and
                  {Strong}, Andrew and {Vittino}, Andrea},
  doi =		 {10.1142/S0218271819300222},
  eid =		 {1930022-339},
  eprint =	 {1903.11584},
  journal =	 {International Journal of Modern Physics D},
  month =	 jan,
  number =	 15,
  pages =	 {1930022-339},
  primaryclass = {astro-ph.HE},
  title =	 {{The origin of Galactic cosmic rays: Challenges to
                  the standard paradigm}},
  volume =	 28,
  year =	 2019
}

@article{GalarragaEpinosa2021A&A,
  adsurl =
                  {https://ui.adsabs.harvard.edu/abs/2021A&A...649A.117G},
  archiveprefix ={arXiv},
  author =	 {{Gal{\'a}rraga-Espinosa}, Daniela and {Aghanim},
                  Nabila and {Langer}, Mathieu and {Tanimura}, Hideki},
  doi =		 {10.1051/0004-6361/202039781},
  eid =		 {A117},
  eprint =	 {2010.15139},
  journal =	 {\aap},
  month =	 may,
  pages =	 {A117},
  primaryclass = {astro-ph.CO},
  title =	 {{Properties of gas phases around cosmic filaments at
                  z = 0 in the IllustrisTNG simulation}},
  volume =	 649,
  year =	 2021
}

@article{Geach2023Natur,
  adsurl =
                  {https://ui.adsabs.harvard.edu/abs/2023Natur.621..483G},
  archiveprefix ={arXiv},
  author =	 {{Geach}, J.~E. and {Lopez-Rodriguez}, E. and
                  {Doherty}, M.~J. and {Chen}, Jianhang and {Ivison},
                  R.~J. and {Bendo}, G.~J. and {Dye}, S. and {Coppin},
                  K.~E.~K.},
  doi =		 {10.1038/s41586-023-06346-4},
  eprint =	 {2309.02034},
  journal =	 {\nat},
  month =	 sep,
  number =	 7979,
  pages =	 {483-486},
  primaryclass = {astro-ph.GA},
  title =	 {{Polarized thermal emission from dust in a galaxy at
                  redshift 2.6}},
  volume =	 621,
  year =	 2023
}

@article{Girichidis2020MNRAS,
  adsurl =
                  {https://ui.adsabs.harvard.edu/abs/2020MNRAS.491..993G},
  archiveprefix ={arXiv},
  author =	 {{Girichidis}, Philipp and {Pfrommer}, Christoph and
                  {Hanasz}, Micha{\l} and {Naab}, Thorsten},
  doi =		 {10.1093/mnras/stz2961},
  eprint =	 {1909.12840},
  journal =	 {\mnras},
  month =	 jan,
  number =	 1,
  pages =	 {993-1007},
  primaryclass = {astro-ph.HE},
  title =	 {{Spectrally resolved cosmic ray hydrodynamics -
                  I. Spectral scheme}},
  volume =	 491,
  year =	 2020
}

@article{Girichidis2022MNRAS,
  adsurl =
                  {https://ui.adsabs.harvard.edu/abs/2022MNRAS.510.3917G},
  archiveprefix ={arXiv},
  author =	 {{Girichidis}, Philipp and {Pfrommer}, Christoph and
                  {Pakmor}, R{\"u}diger and {Springel}, Volker},
  doi =		 {10.1093/mnras/stab3462},
  eprint =	 {2109.13250},
  journal =	 {\mnras},
  month =	 mar,
  number =	 3,
  pages =	 {3917-3938},
  primaryclass = {astro-ph.GA},
  title =	 {{Spectrally resolved cosmic rays -
                  II. Momentum-dependent cosmic ray diffusion drives
                  powerful galactic winds}},
  volume =	 510,
  year =	 2022
}

@article{Girichidis2024MNRAS,
  adsurl =
                  {https://ui.adsabs.harvard.edu/abs/2024MNRAS.52710897G},
  archiveprefix ={arXiv},
  author =	 {{Girichidis}, Philipp and {Werhahn}, Maria and
                  {Pfrommer}, Christoph and {Pakmor}, R{\"u}diger and
                  {Springel}, Volker},
  doi =		 {10.1093/mnras/stad3628},
  eprint =	 {2303.03417},
  journal =	 {\mnras},
  month =	 feb,
  number =	 4,
  pages =	 {10897-10920},
  primaryclass = {astro-ph.GA},
  title =	 {{Spectrally resolved cosmic rays - III. Dynamical
                  impact and properties of the circumgalactic medium}},
  volume =	 527,
  year =	 2024
}

@article{Globus2025ARA&A,
  adsurl =
                  {https://ui.adsabs.harvard.edu/abs/2025ARA&A..63..339G},
  archiveprefix ={arXiv},
  author =	 {{Globus}, No{\'e}mie and {Blandford}, Roger D.},
  doi =		 {10.1146/annurev-astro-052622-033150},
  eprint =	 {2505.21846},
  journal =	 {\araa},
  month =	 aug,
  number =	 1,
  pages =	 {339-377},
  primaryclass = {astro-ph.HE},
  title =	 {{Ultrahigh-Energy Cosmic Rays}},
  volume =	 63,
  year =	 2025
}

@article{Gronke2026,
  adsurl =
                  {https://ui.adsabs.harvard.edu/abs/2026arXiv260116566G},
  archiveprefix ={arXiv},
  author =	 {{Gronke}, Max and {Schneider}, Evan},
  doi =		 {10.48550/arXiv.2601.16566},
  eid =		 {arXiv:2601.16566},
  eprint =	 {2601.16566},
  journal =	 {arXiv e-prints},
  month =	 jan,
  pages =	 {arXiv:2601.16566},
  primaryclass = {astro-ph.GA},
  title =	 {{Simulations of multi-phase gas in and around
                  galaxies}},
  year =	 2026
}

@article{Guo2008MNRAS,
  adsurl =
                  {https://ui.adsabs.harvard.edu/abs/2008MNRAS.384..251G},
  archiveprefix ={arXiv},
  author =	 {{Guo}, Fulai and {Oh}, S. Peng},
  doi =		 {10.1111/j.1365-2966.2007.12692.x},
  eprint =	 {0706.1274},
  journal =	 {\mnras},
  month =	 feb,
  number =	 1,
  pages =	 {251-266},
  primaryclass = {astro-ph},
  title =	 {{Feedback heating by cosmic rays in clusters of
                  galaxies}},
  volume =	 384,
  year =	 2008
}

@article{Gupta2024ApJ,
  adsurl =
                  {https://ui.adsabs.harvard.edu/abs/2024ApJ...976...10G},
  archiveprefix ={arXiv},
  author =	 {{Gupta}, Siddhartha and {Caprioli}, Damiano and
                  {Spitkovsky}, Anatoly},
  doi =		 {10.3847/1538-4357/ad7c4c},
  eid =		 10,
  eprint =	 {2408.16071},
  journal =	 {\apj},
  month =	 nov,
  number =	 1,
  pages =	 10,
  primaryclass = {astro-ph.HE},
  title =	 {{Electron Acceleration at Quasi-parallel
                  Nonrelativistic Shocks: A 1D Kinetic Survey}},
  volume =	 976,
  year =	 2024
}

@article{Haardt2012ApJ,
  adsurl =
                  {https://ui.adsabs.harvard.edu/abs/2012ApJ...746..125H},
  author =	 {{Haardt}, Francesco and {Madau}, Piero},
  doi =		 {10.1088/0004-637X/746/2/125},
  eid =		 125,
  journal =	 {\apj},
  month =	 feb,
  number =	 2,
  pages =	 125,
  title =	 {{Radiative Transfer in a Clumpy Universe. IV. New
                  Synthesis Models of the Cosmic UV/X-Ray Background}},
  volume =	 746,
  year =	 2012
}

@article{Habegger2025arXiv,
  adsurl =
                  {https://ui.adsabs.harvard.edu/abs/2025arXiv251024622H},
  archiveprefix ={arXiv},
  author =	 {{Habegger}, Roark and {Ruszkowski}, Mateusz and
                  {Zweibel}, Ellen G.},
  doi =		 {10.48550/arXiv.2510.24622},
  eid =		 {arXiv:2510.24622},
  eprint =	 {2510.24622},
  journal =	 {arXiv e-prints},
  month =	 oct,
  pages =	 {arXiv:2510.24622},
  primaryclass = {astro-ph.GA},
  title =	 {{The Impact of Cosmic Ray Transport on the $γ$-Ray
                  Luminosity of Diffuse Gas}},
  year =	 2025
}

@article{Hanasz2021LRCA,
  adsurl =
                  {https://ui.adsabs.harvard.edu/abs/2021LRCA....7....2H},
  archiveprefix ={arXiv},
  author =	 {{Hanasz}, Micha{\l} and {Strong}, Andrew W. and
                  {Girichidis}, Philipp},
  doi =		 {10.1007/s41115-021-00011-1},
  eid =		 2,
  eprint =	 {2106.08426},
  journal =	 {Living Reviews in Computational Astrophysics},
  month =	 dec,
  number =	 1,
  pages =	 2,
  primaryclass = {astro-ph.HE},
  title =	 {{Simulations of cosmic ray propagation}},
  volume =	 7,
  year =	 2021
}

@article{Hillier2019,
  author =	 {Hillier, Andrew and Arregui, I{\~{n}}igo},
  doi =		 {10.3847/1538-4357/ab4795},
  eprint =	 {1909.11351},
  issn =	 15384357,
  journal =	 {ApJ},
  number =	 2,
  pages =	 101,
  title =	 {{Coronal Cooling as a Result of Mixing by the
                  Nonlinear Kelvin--Helmholtz Instability}},
  volume =	 885,
  year =	 2019
}

@article{Hong2024,
  author =	 {Hong, Wen-Sheng and Zhu, Weishan and Wang, Tian-Rui
                  and Yang, Xiaohu and Feng, Long-Long},
  doi =		 {10.1093/mnras/stae777},
  eprint =
                  {https://academic.oup.com/mnras/article-pdf/529/4/4262/57127551/stae777.pdf},
  issn =	 {0035-8711},
  journal =	 {\mnras},
  month =	 03,
  number =	 4,
  pages =	 {4262-4286},
  title =	 {{Evolution of cold streams in hot gaseous haloes}},
  url =		 {https://doi.org/10.1093/mnras/stae777},
  volume =	 529,
  year =	 2024
}

@article{Hopkins2020MNRAS,
  adsurl =
                  {https://ui.adsabs.harvard.edu/abs/2020MNRAS.492.3465H},
  archiveprefix ={arXiv},
  author =	 {{Hopkins}, Philip F. and {Chan}, T.~K. and
                  {Garrison-Kimmel}, Shea and {Ji}, Suoqing and {Su},
                  Kung-Yi and {Hummels}, Cameron B. and {Kere{\v{s}}},
                  Du{\v{s}}an and {Quataert}, Eliot and
                  {Faucher-Gigu{\`e}re}, Claude-Andr{\'e}},
  doi =		 {10.1093/mnras/stz3321},
  eprint =	 {1905.04321},
  journal =	 {\mnras},
  month =	 mar,
  number =	 3,
  pages =	 {3465-3498},
  primaryclass = {astro-ph.GA},
  title =	 {{But what about...: cosmic rays, magnetic fields,
                  conduction, and viscosity in galaxy formation}},
  volume =	 492,
  year =	 2020
}

@article{Hopkins2021MNRAS,
  adsurl =
                  {https://ui.adsabs.harvard.edu/abs/2021MNRAS.501.3663H},
  archiveprefix ={arXiv},
  author =	 {{Hopkins}, Philip F. and {Chan}, T.~K. and {Squire},
                  Jonathan and {Quataert}, Eliot and {Ji}, Suoqing and
                  {Kere{\v{s}}}, Du{\v{s}}an and
                  {Faucher-Gigu{\`e}re}, Claude-Andr{\'e}},
  doi =		 {10.1093/mnras/staa3692},
  eprint =	 {2004.02897},
  journal =	 {\mnras},
  month =	 mar,
  number =	 3,
  pages =	 {3663-3669},
  primaryclass = {astro-ph.GA},
  title =	 {{Effects of different cosmic ray transport models on
                  galaxy formation}},
  volume =	 501,
  year =	 2021
}

@article{Ji2020MNRAS,
  adsurl =
                  {https://ui.adsabs.harvard.edu/abs/2020MNRAS.496.4221J},
  archiveprefix ={arXiv},
  author =	 {{Ji}, Suoqing and {Chan}, T.~K. and {Hummels},
                  Cameron B. and {Hopkins}, Philip F. and {Stern},
                  Jonathan and {Kere{\v{s}}}, Du{\v{s}}an and
                  {Quataert}, Eliot and {Faucher-Gigu{\`e}re},
                  Claude-Andr{\'e} and {Murray}, Norman},
  doi =		 {10.1093/mnras/staa1849},
  eprint =	 {1909.00003},
  journal =	 {\mnras},
  month =	 aug,
  number =	 4,
  pages =	 {4221-4238},
  primaryclass = {astro-ph.GA},
  title =	 {{Properties of the circumgalactic medium in cosmic
                  ray-dominated galaxy haloes}},
  volume =	 496,
  year =	 2020
}

@article{Jubelgas2008A&A,
  adsurl =
                  {https://ui.adsabs.harvard.edu/abs/2008A&A...481...33J},
  archiveprefix ={arXiv},
  author =	 {{Jubelgas}, M. and {Springel}, V. and {En{\ss}lin},
                  T. and {Pfrommer}, C.},
  doi =		 {10.1051/0004-6361:20065295},
  eprint =	 {astro-ph/0603485},
  journal =	 {\aap},
  month =	 apr,
  number =	 1,
  pages =	 {33-63},
  primaryclass = {astro-ph},
  title =	 {{Cosmic ray feedback in hydrodynamical simulations
                  of galaxy formation}},
  volume =	 481,
  year =	 2008
}

@article{Kachelriess2015ApJ,
  adsurl =
                  {https://ui.adsabs.harvard.edu/abs/2015ApJ...803...54K},
  archiveprefix ={arXiv},
  author =	 {{Kachelrie{\ss}}, Michael and {Moskalenko}, Igor
                  V. and {Ostapchenko}, Sergey S.},
  doi =		 {10.1088/0004-637X/803/2/54},
  eid =		 54,
  eprint =	 {1502.04158},
  journal =	 {\apj},
  month =	 apr,
  number =	 2,
  pages =	 54,
  primaryclass = {astro-ph.HE},
  title =	 {{New Calculation of Antiproton Production by Cosmic
                  Ray Protons and Nuclei}},
  volume =	 803,
  year =	 2015
}

@article{Kachelriess2019CoPhC,
  adsurl =
                  {https://ui.adsabs.harvard.edu/abs/2019CoPhC.24506846K},
  archiveprefix ={arXiv},
  author =	 {{Kachelrie{\ss}}, M. and {Moskalenko}, I.~V. and
                  {Ostapchenko}, S.},
  doi =		 {10.1016/j.cpc.2019.08.001},
  eid =		 106846,
  eprint =	 {1904.05129},
  journal =	 {Computer Physics Communications},
  month =	 dec,
  pages =	 106846,
  primaryclass = {hep-ph},
  title =	 {{AAfrag: Interpolation routines for Monte Carlo
                  results on secondary production in proton-proton,
                  proton-nucleus and nucleus-nucleus interactions}},
  volume =	 245,
  year =	 2019
}

@article{Kafexhiu2014PhRvD,
  adsurl =
                  {https://ui.adsabs.harvard.edu/abs/2014PhRvD..90l3014K},
  archiveprefix ={arXiv},
  author =	 {{Kafexhiu}, Ervin and {Aharonian}, Felix and
                  {Taylor}, Andrew M. and {Vila}, Gabriela S.},
  doi =		 {10.1103/PhysRevD.90.123014},
  eid =		 123014,
  eprint =	 {1406.7369},
  journal =	 {\prd},
  month =	 dec,
  number =	 12,
  pages =	 123014,
  primaryclass = {astro-ph.HE},
  title =	 {{Parametrization of gamma-ray production cross
                  sections for p p interactions in a broad proton
                  energy range from the kinematic threshold to PeV
                  energies}},
  volume =	 90,
  year =	 2014
}

@article{Kamae2006ApJ,
  adsurl =
                  {https://ui.adsabs.harvard.edu/abs/2006ApJ...647..692K},
  archiveprefix ={arXiv},
  author =	 {{Kamae}, Tuneyoshi and {Karlsson}, Niklas and
                  {Mizuno}, Tsunefumi and {Abe}, Toshinori and {Koi},
                  Tatsumi},
  doi =		 {10.1086/505189},
  eprint =	 {astro-ph/0605581},
  journal =	 {\apj},
  month =	 aug,
  number =	 1,
  pages =	 {692-708},
  primaryclass = {astro-ph},
  title =	 {{Parameterization of {\ensuremath{\gamma}},
                  e$^{+/-}$, and Neutrino Spectra Produced by p-p
                  Interaction in Astronomical Environments}},
  volume =	 647,
  year =	 2006
}

@article{Kaul2025,
  adsurl =
                  {https://ui.adsabs.harvard.edu/abs/2025MNRAS.539.3669K},
  archiveprefix ={arXiv},
  author =	 {{Kaul}, Ish and {Tan}, Brent and {Oh}, S. Peng and
                  {Mandelker}, Nir},
  doi =		 {10.1093/mnras/staf706},
  eprint =	 {2502.17549},
  journal =	 {\mnras},
  month =	 jun,
  number =	 4,
  pages =	 {3669-3696},
  primaryclass = {astro-ph.GA},
  title =	 {{Tales of tension: magnetized infalling cold clouds
                  and streams in the CGM}},
  volume =	 539,
  year =	 2025
}

@article{Keres2005,
  author =	 {Kere{\v{s}}, Du{\v{s}}an and Katz, Neal and
                  Weinberg, David H. and Dav{\'{e}}, Romeel},
  doi =		 {10.1111/j.1365-2966.2005.09451.x},
  eprint =	 0407095,
  issn =	 00358711,
  journal =	 {MNRAS},
  number =	 1,
  pages =	 {2--28},
  title =	 {{How do galaxies get their gas?}},
  volume =	 363,
  year =	 2005
}

@article{Keres2009MNRAS,
  adsurl =
                  {https://ui.adsabs.harvard.edu/abs/2009MNRAS.395..160K},
  archiveprefix ={arXiv},
  author =	 {{Kere{\v{s}}}, Du{\v{s}}an and {Katz}, Neal and
                  {Fardal}, Mark and {Dav{\'e}}, Romeel and
                  {Weinberg}, David H.},
  doi =		 {10.1111/j.1365-2966.2009.14541.x},
  eprint =	 {0809.1430},
  journal =	 {\mnras},
  month =	 may,
  number =	 1,
  pages =	 {160-179},
  primaryclass = {astro-ph},
  title =	 {{Galaxies in a simulated {\ensuremath{\Lambda}}CDM
                  Universe - I. Cold mode and hot cores}},
  volume =	 395,
  year =	 2009
}

@article{Komatsu2001,
  adsurl =
                  {https://ui.adsabs.harvard.edu/abs/2001MNRAS.327.1353K},
  author =	 {{Komatsu}, E. and {Seljak}, U.},
  doi =		 {10.1046/j.1365-8711.2001.04838.x},
  eprint =	 {astro-ph/0106151},
  journal =	 {\mnras},
  month =	 nov,
  number =	 4,
  pages =	 {1353-1366},
  primaryclass = {astro-ph},
  title =	 {{Universal gas density and temperature profile}},
  volume =	 327,
  year =	 2001
}

@article{Lazarian2022FrP,
  adsurl =
                  {https://ui.adsabs.harvard.edu/abs/2022FrP....10.2799L},
  archiveprefix ={arXiv},
  author =	 {{Lazarian}, Alex and {Xu}, Siyao},
  doi =		 {10.3389/fphy.2022.702799},
  eid =		 702799,
  eprint =	 {2201.05168},
  journal =	 {Frontiers in Physics},
  month =	 may,
  pages =	 702799,
  primaryclass = {astro-ph.GA},
  title =	 {{Damping of Alfv{\'e}n Waves in MHD Turbulence and
                  Implications for Cosmic Ray Streaming Instability
                  and Galactic Winds}},
  volume =	 10,
  year =	 2022
}

@article{Lazarian2023FrASS,
  adsurl =
                  {https://ui.adsabs.harvard.edu/abs/2023FrASS..1054760L},
  archiveprefix ={arXiv},
  author =	 {{Lazarian}, Alex and {Xu}, Siyao and {Hu}, Yue},
  doi =		 {10.3389/fspas.2023.1154760},
  eid =		 1154760,
  eprint =	 {2304.02684},
  journal =	 {Frontiers in Astronomy and Space Sciences},
  month =	 may,
  pages =	 1154760,
  primaryclass = {astro-ph.GA},
  title =	 {{Cosmic ray propagation in turbulent magnetic
                  fields}},
  volume =	 10,
  year =	 2023
}

@article{Ledos2024a,
  adsurl =
                  {https://ui.adsabs.harvard.edu/abs/2024MNRAS.52711304L},
  archiveprefix ={arXiv},
  author =	 {{Ledos}, Nicolas and {Takasao}, Shinsuke and
                  {Nagamine}, Kentaro},
  doi =		 {10.1093/mnras/stad3814},
  eprint =	 {2308.05412},
  journal =	 {\mnras},
  month =	 feb,
  number =	 4,
  pages =	 {11304-11326},
  primaryclass = {astro-ph.GA},
  title =	 {{Stability and Ly {\ensuremath{\alpha}} emission of
                  Cold Stream in the Circumgalactic Medium: impact of
                  magnetic fields and thermal conduction}},
  volume =	 527,
  year =	 2024
}

@article{Ledos2024b,
  adsurl =
                  {https://ui.adsabs.harvard.edu/abs/2024A&A...691A.280L},
  archiveprefix ={arXiv},
  author =	 {{Ledos}, Nicolas and {Ntormousi}, Evangelia and
                  {Takasao}, Shinsuke and {Nagamine}, Kentaro},
  doi =		 {10.1051/0004-6361/202451139},
  eid =		 {A280},
  eprint =	 {2408.17438},
  journal =	 {\aap},
  month =	 nov,
  pages =	 {A280},
  primaryclass = {astro-ph.GA},
  title =	 {{Magnetising galaxies with cold inflows}},
  volume =	 691,
  year =	 2024
}

@article{Leite2017MNRAS,
  adsurl =
                  {https://ui.adsabs.harvard.edu/abs/2017MNRAS.469..416L},
  archiveprefix ={arXiv},
  author =	 {{Leite}, N. and {Evoli}, C. and {D'Angelo}, M. and
                  {Ciardi}, B. and {Sigl}, G. and {Ferrara}, A.},
  doi =		 {10.1093/mnras/stx805},
  eprint =	 {1703.09337},
  journal =	 {\mnras},
  month =	 jul,
  number =	 1,
  pages =	 {416-424},
  primaryclass = {astro-ph.CO},
  title =	 {{Do cosmic rays heat the early intergalactic
                  medium?}},
  volume =	 469,
  year =	 2017
}

@article{Lin2023MNRAS,
  adsurl =
                  {https://ui.adsabs.harvard.edu/abs/2023MNRAS.520..963L},
  archiveprefix ={arXiv},
  author =	 {{Lin}, Yen-Hsing and {Yang}, H.-Y. Karen and {Owen},
                  Ellis R.},
  doi =		 {10.1093/mnras/stad185},
  eprint =	 {2301.06025},
  journal =	 {\mnras},
  month =	 mar,
  number =	 1,
  pages =	 {963-975},
  primaryclass = {astro-ph.HE},
  title =	 {{Evolution and feedback of AGN jets of different
                  cosmic ray composition}},
  volume =	 520,
  year =	 2023
}

@article{Loeb2000Natur,
  adsurl =
                  {https://ui.adsabs.harvard.edu/abs/2000Natur.405..156L},
  archiveprefix ={arXiv},
  author =	 {{Loeb}, Abraham and {Waxman}, Eli},
  doi =		 {10.1038/35012018},
  eprint =	 {astro-ph/0003447},
  journal =	 {\nat},
  month =	 may,
  number =	 6783,
  pages =	 {156-158},
  primaryclass = {astro-ph},
  title =	 {{Cosmic {\ensuremath{\gamma}}-ray background from
                  structure formation in the intergalactic medium}},
  volume =	 405,
  year =	 2000
}

@article{Mandelker2019,
  archiveprefix ={arXiv},
  arxivid =	 {1806.05677},
  author =	 {Mandelker, Nir and Nagai, Daisuke and Aung, Han and
                  Dekel, Avishai and Padnos, Dan and Birnboim, Yuval},
  doi =		 {10.1093/mnras/stz012},
  eprint =	 {1806.05677},
  issn =	 13652966,
  journal =	 {MNRAS},
  number =	 1,
  pages =	 {1100--1132},
  publisher =	 {Oxford University Press},
  title =	 {{Instability of supersonic cold streams feeding
                  Galaxies - III. Kelvin-Helmholtz instability in
                  three dimensions}},
  volume =	 484,
  year =	 2019
}

@article{Mandelker2020a,
  archiveprefix ={arXiv},
  arxivid =	 {1910.05344},
  author =	 {Mandelker, Nir and Nagai, Daisuke and Aung, Han and
                  Dekel, Avishai and Birnboim, Yuval and {Van Den
                  Bosch}, Frank C.},
  doi =		 {10.1093/MNRAS/STAA812},
  eprint =	 {1910.05344},
  issn =	 13652966,
  journal =	 {MNRAS},
  number =	 2,
  pages =	 {2641--2663},
  publisher =	 {Oxford University Press},
  title =	 {{Instability of supersonic cold streams feeding
                  galaxies - IV. Survival of radiatively cooling
                  streams}},
  volume =	 494,
  year =	 2020
}

@article{Mandelker2020b,
  adsurl =
                  {https://ui.adsabs.harvard.edu/abs/2020MNRAS.498.2415M},
  archiveprefix ={arXiv},
  author =	 {{Mandelker}, Nir and {van den Bosch}, Frank C. and
                  {Nagai}, Daisuke and {Dekel}, Avishai and
                  {Birnboim}, Yuval and {Aung}, Han},
  doi =		 {10.1093/mnras/staa2421},
  eprint =	 {2003.01724},
  journal =	 {\mnras},
  month =	 oct,
  number =	 2,
  pages =	 {2415-2427},
  primaryclass = {astro-ph.CO},
  title =	 {{Ly {\ensuremath{\alpha}} blobs from cold streams
                  undergoing Kelvin-Helmholtz instabilities}},
  volume =	 498,
  year =	 2020
}

@article{Mannheim1994A&A,
  adsurl =
                  {https://ui.adsabs.harvard.edu/abs/1994A&A...286..983M},
  archiveprefix ={arXiv},
  author =	 {{Mannheim}, K. and {Schlickeiser}, R.},
  doi =		 {10.48550/arXiv.astro-ph/9402042},
  eprint =	 {astro-ph/9402042},
  journal =	 {\aap},
  month =	 jun,
  pages =	 {983-996},
  primaryclass = {astro-ph},
  title =	 {{Interactions of cosmic ray nuclei}},
  volume =	 286,
  year =	 1994
}

@article{Marin2025,
  adsurl =
                  {https://ui.adsabs.harvard.edu/abs/2025arXiv250415345M},
  archiveprefix ={arXiv},
  author =	 {{Marin-Gilabert}, Tirso and {Gronke}, Max and {Oh},
                  S. Peng},
  doi =		 {10.48550/arXiv.2504.15345},
  eid =		 {arXiv:2504.15345},
  eprint =	 {2504.15345},
  journal =	 {arXiv e-prints},
  month =	 apr,
  pages =	 {arXiv:2504.15345},
  primaryclass = {astro-ph.GA},
  title =	 {{The (Limited) Effect of Viscosity in Multiphase
                  Turbulent Mixing}},
  year =	 2025
}

@article{Martin2019,
  adsurl =
                  {https://ui.adsabs.harvard.edu/abs/2019MNRAS.485..796M},
  archiveprefix ={arXiv},
  author =	 {{Martin}, G. and {Kaviraj}, S. and {Laigle}, C. and
                  {Devriendt}, J.~E.~G. and {Jackson}, R.~A. and
                  {Peirani}, S. and {Dubois}, Y. and {Pichon}, C. and
                  {Slyz}, A.},
  doi =		 {10.1093/mnras/stz356},
  eprint =	 {1902.04580},
  journal =	 {\mnras},
  month =	 may,
  number =	 1,
  pages =	 {796-818},
  primaryclass = {astro-ph.GA},
  title =	 {{The formation and evolution of
                  low-surface-brightness galaxies}},
  volume =	 485,
  year =	 2019
}

@article{Medlock2026ApJ,
  adsurl =
                  {https://ui.adsabs.harvard.edu/abs/2026ApJ..1000..222M},
  archiveprefix ={arXiv},
  author =	 {{Medlock}, Isabel and {Nagai}, Daisuke and
                  {Mandelker}, Nir and {Springel}, Volker and {van den
                  Bosch}, Frank C. and {Zinger}, Elad and {Chiang},
                  Barry T.},
  doi =		 {10.3847/1538-4357/ae4c42},
  eid =		 222,
  eprint =	 {2511.21814},
  journal =	 {\apj},
  month =	 apr,
  number =	 2,
  pages =	 222,
  primaryclass = {astro-ph.GA},
  title =	 {{Statistical Properties of Cold Streams in Massive
                  Star-forming Halos in TNG50}},
  volume =	 1000,
  year =	 2026
}

@article{Migkas2025A&A,
  adsurl =
                  {https://ui.adsabs.harvard.edu/abs/2025A&A...698A.270M},
  archiveprefix ={arXiv},
  author =	 {{Migkas}, K. and {Pacaud}, F. and {Tuominen}, T. and
                  {Aghanim}, N.},
  doi =		 {10.1051/0004-6361/202554944},
  eid =		 {A270},
  eprint =	 {2506.14917},
  journal =	 {\aap},
  month =	 jun,
  pages =	 {A270},
  primaryclass = {astro-ph.CO},
  title =	 {{Detection of pure warm-hot intergalactic medium
                  emission from a 7.2 Mpc long filament in the Shapley
                  supercluster using X-ray spectroscopy}},
  volume =	 698,
  year =	 2025
}

@article{Miniati2000ApJ,
  adsurl =
                  {https://ui.adsabs.harvard.edu/abs/2000ApJ...542..608M},
  archiveprefix ={arXiv},
  author =	 {{Miniati}, Francesco and {Ryu}, Dongsu and {Kang},
                  Hyesung and {Jones}, T.~W. and {Cen}, Renyue and
                  {Ostriker}, Jeremiah P.},
  doi =		 {10.1086/317027},
  eprint =	 {astro-ph/0005444},
  journal =	 {\apj},
  month =	 oct,
  number =	 2,
  pages =	 {608-621},
  primaryclass = {astro-ph},
  title =	 {{Properties of Cosmic Shock Waves in Large-Scale
                  Structure Formation}},
  volume =	 542,
  year =	 2000
}

@article{Miniati2001ApJ,
  adsurl =
                  {https://ui.adsabs.harvard.edu/abs/2001ApJ...559...59M},
  archiveprefix ={arXiv},
  author =	 {{Miniati}, Francesco and {Ryu}, Dongsu and {Kang},
                  Hyesung and {Jones}, T.~W.},
  doi =		 {10.1086/322375},
  eprint =	 {astro-ph/0105465},
  journal =	 {\apj},
  month =	 sep,
  number =	 1,
  pages =	 {59-69},
  primaryclass = {astro-ph},
  title =	 {{Cosmic-Ray Protons Accelerated at Cosmological
                  Shocks and Their Impact on Groups and Clusters of
                  Galaxies}},
  volume =	 559,
  year =	 2001
}

@article{Nelson2020MNRAS,
  adsurl =
                  {https://ui.adsabs.harvard.edu/abs/2020MNRAS.498.2391N},
  archiveprefix ={arXiv},
  author =	 {{Nelson}, Dylan and {Sharma}, Prateek and
                  {Pillepich}, Annalisa and {Springel}, Volker and
                  {Pakmor}, R{\"u}diger and {Weinberger}, Rainer and
                  {Vogelsberger}, Mark and {Marinacci}, Federico and
                  {Hernquist}, Lars},
  doi =		 {10.1093/mnras/staa2419},
  eprint =	 {2005.09654},
  journal =	 {\mnras},
  month =	 oct,
  number =	 2,
  pages =	 {2391-2414},
  primaryclass = {astro-ph.GA},
  title =	 {{Resolving small-scale cold circumgalactic gas in
                  TNG50}},
  volume =	 498,
  year =	 2020
}

@article{Ng2026PASJ,
  adsurl =
                  {https://ui.adsabs.harvard.edu/abs/2026PASJ..tmp..170N},
  archiveprefix ={arXiv},
  author =	 {{Ng}, Hayden P.~H. and {Owen}, Ellis R. and {Tsuji},
                  Naomi and {Chen}, Szu-Ting},
  doi =		 {10.1093/pasj/psag070},
  eprint =	 {2605.16698},
  journal =	 {\pasj},
  month =	 jun,
  primaryclass = {astro-ph.HE},
  title =	 {{Multiwavelength probes of cosmic ray transport in
                  molecular cloud structures}},
  year =	 2026
}

@article{Ostapchenko2011PhRvD,
  adsurl =
                  {https://ui.adsabs.harvard.edu/abs/2011PhRvD..83a4018O},
  archiveprefix ={arXiv},
  author =	 {{Ostapchenko}, S.},
  doi =		 {10.1103/PhysRevD.83.014018},
  eid =		 014018,
  eprint =	 {1010.1869},
  journal =	 {\prd},
  month =	 jan,
  number =	 1,
  pages =	 014018,
  primaryclass = {hep-ph},
  title =	 {{Monte Carlo treatment of hadronic interactions in
                  enhanced Pomeron scheme: QGSJET-II model}},
  volume =	 83,
  year =	 2011
}

@inproceedings{Ostapchenko2013EPJWC,
  adsurl =
                  {https://ui.adsabs.harvard.edu/abs/2013EPJWC..5202001O},
  author =	 {{Ostapchenko}, S.},
  booktitle =	 {European Physical Journal Web of Conferences},
  doi =		 {10.1051/epjconf/20125202001},
  eid =		 02001,
  month =	 jun,
  pages =	 02001,
  series =	 {European Physical Journal Web of Conferences},
  title =	 {{QGSJET-II: physics, recent improvements, and
                  results for air showers}},
  volume =	 52,
  year =	 2013
}

@article{Ouchi2020,
  author =	 {Ouchi, Masami and Ono, Yoshiaki and Shibuya,
                  Takatoshi},
  doi =		 {https://doi.org/10.1146/annurev-astro-032620-021859},
  issn =	 {1545-4282},
  journal =	 {Annual Review of Astronomy and Astrophysics},
  number =	 {Volume 58, 2020},
  pages =	 {617-659},
  publisher =	 {Annual Reviews},
  title =	 {Observations of the Lyman-α Universe},
  type =	 {Journal Article},
  url =
                  {https://www.annualreviews.org/content/journals/10.1146/annurev-astro-032620-021859},
  volume =	 58,
  year =	 2020
}

@article{Owen2023Galax,
  adsurl =
                  {https://ui.adsabs.harvard.edu/abs/2023Galax..11...86O},
  archiveprefix ={arXiv},
  author =	 {{Owen}, Ellis R. and {Wu}, Kinwah and {Inoue},
                  Yoshiyuki and {Yang}, H. -Y. Karen and {Mitchell},
                  Alison M.~W.},
  doi =		 {10.3390/galaxies11040086},
  eid =		 86,
  eprint =	 {2306.09924},
  journal =	 {Galaxies},
  month =	 jul,
  number =	 4,
  pages =	 86,
  primaryclass = {astro-ph.GA},
  title =	 {{Cosmic Ray Processes in Galactic Ecosystems}},
  volume =	 11,
  year =	 2023
}

@article{Padovani2009A&A,
  adsurl =
                  {https://ui.adsabs.harvard.edu/abs/2009A&A...501..619P},
  archiveprefix ={arXiv},
  author =	 {{Padovani}, M. and {Galli}, D. and {Glassgold},
                  A.~E.},
  doi =		 {10.1051/0004-6361/200911794},
  eprint =	 {0904.4149},
  journal =	 {\aap},
  month =	 jul,
  number =	 2,
  pages =	 {619-631},
  primaryclass = {astro-ph.SR},
  title =	 {{Cosmic-ray ionization of molecular clouds}},
  volume =	 501,
  year =	 2009
}

@article{Pakmor2016ApJ,
  adsurl =
                  {https://ui.adsabs.harvard.edu/abs/2016ApJ...824L..30P},
  archiveprefix ={arXiv},
  author =	 {{Pakmor}, R. and {Pfrommer}, C. and {Simpson},
                  C.~M. and {Springel}, V.},
  doi =		 {10.3847/2041-8205/824/2/L30},
  eid =		 {L30},
  eprint =	 {1605.00643},
  journal =	 {\apjl},
  month =	 jun,
  number =	 2,
  pages =	 {L30},
  primaryclass = {astro-ph.GA},
  title =	 {{Galactic Winds Driven by Isotropic and Anisotropic
                  Cosmic-Ray Diffusion in Disk Galaxies}},
  volume =	 824,
  year =	 2016
}

@article{Park2015PhRvL,
  adsurl =
                  {https://ui.adsabs.harvard.edu/abs/2015PhRvL.114h5003P},
  archiveprefix ={arXiv},
  author =	 {{Park}, Jaehong and {Caprioli}, Damiano and
                  {Spitkovsky}, Anatoly},
  doi =		 {10.1103/PhysRevLett.114.085003},
  eid =		 085003,
  eprint =	 {1412.0672},
  journal =	 {\prl},
  month =	 feb,
  number =	 8,
  pages =	 085003,
  primaryclass = {astro-ph.HE},
  title =	 {{Simultaneous Acceleration of Protons and Electrons
                  at Nonrelativistic Quasiparallel Collisionless
                  Shocks}},
  volume =	 114,
  year =	 2015
}

@article{Persic2014A&A,
  adsurl =
                  {https://ui.adsabs.harvard.edu/abs/2014A&A...567A.101P},
  archiveprefix ={arXiv},
  author =	 {{Persic}, Massimo and {Rephaeli}, Yoel},
  doi =		 {10.1051/0004-6361/201322664},
  eid =		 {A101},
  eprint =	 {1201.0369},
  journal =	 {\aap},
  month =	 jul,
  pages =	 {A101},
  primaryclass = {astro-ph.HE},
  title =	 {{Estimates of relativistic electron and proton
                  energy densities in starburst galactic nuclei from
                  radio measurements}},
  volume =	 567,
  year =	 2014
}

@inproceedings{Persic2015mgm,
  adsurl =
                  {https://ui.adsabs.harvard.edu/abs/2015mgm..conf.1036P},
  archiveprefix ={arXiv},
  author =	 {{Persic}, Massimo and {Rephaeli}, Yoel},
  booktitle =	 {Thirteenth Marcel Grossmann Meeting: On Recent
                  Developments in Theoretical and Experimental General
                  Relativity, Astrophysics and Relativistic Field
                  Theories},
  doi =		 {10.1142/9789814623995_0075},
  editor =	 {{Rosquist}, Kjell},
  eprint =	 {1405.3107},
  month =	 jan,
  pages =	 {1036-1038},
  primaryclass = {astro-ph.HE},
  title =	 {{Cosmic-Ray Proton to Electron Ratios}},
  year =	 2015
}

@article{Planck2020_Cosmo_param,
  adsurl =
                  {https://ui.adsabs.harvard.edu/abs/2020A&A...641A...6P},
  archiveprefix ={arXiv},
  author =	 {{Planck Collaboration} and {Aghanim}, N. and
                  {Akrami}, Y. and {Ashdown}, M. and {Aumont}, J. and
                  {Baccigalupi}, C. and {Ballardini}, M. and {Banday},
                  A.~J. and {Barreiro}, R.~B. and {Bartolo}, N. and
                  {Basak}, S. and {Battye}, R. and {Benabed}, K. and
                  {Bernard}, J.-P. and {Bersanelli}, M. and
                  {Bielewicz}, P. and {Bock}, J.~J. and {Bond},
                  J.~R. and {Borrill}, J. and {Bouchet}, F.~R. and
                  {Boulanger}, F. and {Bucher}, M. and {Burigana},
                  C. and {Butler}, R.~C. and {Calabrese}, E. and
                  {Cardoso}, J.-F. and {Carron}, J. and {Challinor},
                  A. and {Chiang}, H.~C. and {Chluba}, J. and
                  {Colombo}, L.~P.~L. and {Combet}, C. and
                  {Contreras}, D. and {Crill}, B.~P. and {Cuttaia},
                  F. and {de Bernardis}, P. and {de Zotti}, G. and
                  {Delabrouille}, J. and {Delouis}, J.-M. and {Di
                  Valentino}, E. and {Diego}, J.~M. and {Dor{\'e}},
                  O. and {Douspis}, M. and {Ducout}, A. and {Dupac},
                  X. and {Dusini}, S. and {Efstathiou}, G. and
                  {Elsner}, F. and {En{\ss}lin}, T.~A. and {Eriksen},
                  H.~K. and {Fantaye}, Y. and {Farhang}, M. and
                  {Fergusson}, J. and {Fernandez-Cobos}, R. and
                  {Finelli}, F. and {Forastieri}, F. and {Frailis},
                  M. and {Fraisse}, A.~A. and {Franceschi}, E. and
                  {Frolov}, A. and {Galeotta}, S. and {Galli}, S. and
                  {Ganga}, K. and {G{\'e}nova-Santos}, R.~T. and
                  {Gerbino}, M. and {Ghosh}, T. and
                  {Gonz{\'a}lez-Nuevo}, J. and {G{\'o}rski}, K.~M. and
                  {Gratton}, S. and {Gruppuso}, A. and {Gudmundsson},
                  J.~E. and {Hamann}, J. and {Handley}, W. and
                  {Hansen}, F.~K. and {Herranz}, D. and {Hildebrandt},
                  S.~R. and {Hivon}, E. and {Huang}, Z. and {Jaffe},
                  A.~H. and {Jones}, W.~C. and {Karakci}, A. and
                  {Keih{\"a}nen}, E. and {Keskitalo}, R. and
                  {Kiiveri}, K. and {Kim}, J. and {Kisner}, T.~S. and
                  {Knox}, L. and {Krachmalnicoff}, N. and {Kunz},
                  M. and {Kurki-Suonio}, H. and {Lagache}, G. and
                  {Lamarre}, J.-M. and {Lasenby}, A. and {Lattanzi},
                  M. and {Lawrence}, C.~R. and {Le Jeune}, M. and
                  {Lemos}, P. and {Lesgourgues}, J. and {Levrier},
                  F. and {Lewis}, A. and {Liguori}, M. and {Lilje},
                  P.~B. and {Lilley}, M. and {Lindholm}, V. and
                  {L{\'o}pez-Caniego}, M. and {Lubin}, P.~M. and {Ma},
                  Y.-Z. and {Mac{\'\i}as-P{\'e}rez}, J.~F. and
                  {Maggio}, G. and {Maino}, D. and {Mandolesi}, N. and
                  {Mangilli}, A. and {Marcos-Caballero}, A. and
                  {Maris}, M. and {Martin}, P.~G. and {Martinelli},
                  M. and {Mart{\'\i}nez-Gonz{\'a}lez}, E. and
                  {Matarrese}, S. and {Mauri}, N. and {McEwen},
                  J.~D. and {Meinhold}, P.~R. and {Melchiorri}, A. and
                  {Mennella}, A. and {Migliaccio}, M. and {Millea},
                  M. and {Mitra}, S. and {Miville-Desch{\^e}nes},
                  M.-A. and {Molinari}, D. and {Montier}, L. and
                  {Morgante}, G. and {Moss}, A. and {Natoli}, P. and
                  {N{\o}rgaard-Nielsen}, H.~U. and {Pagano}, L. and
                  {Paoletti}, D. and {Partridge}, B. and {Patanchon},
                  G. and {Peiris}, H.~V. and {Perrotta}, F. and
                  {Pettorino}, V. and {Piacentini}, F. and {Polastri},
                  L. and {Polenta}, G. and {Puget}, J.-L. and
                  {Rachen}, J.~P. and {Reinecke}, M. and
                  {Remazeilles}, M. and {Renzi}, A. and {Rocha},
                  G. and {Rosset}, C. and {Roudier}, G. and
                  {Rubi{\~n}o-Mart{\'\i}n}, J.~A. and {Ruiz-Granados},
                  B. and {Salvati}, L. and {Sandri}, M. and
                  {Savelainen}, M. and {Scott}, D. and {Shellard},
                  E.~P.~S. and {Sirignano}, C. and {Sirri}, G. and
                  {Spencer}, L.~D. and {Sunyaev}, R. and {Suur-Uski},
                  A.-S. and {Tauber}, J.~A. and {Tavagnacco}, D. and
                  {Tenti}, M. and {Toffolatti}, L. and {Tomasi},
                  M. and {Trombetti}, T. and {Valenziano}, L. and
                  {Valiviita}, J. and {Van Tent}, B. and {Vibert},
                  L. and {Vielva}, P. and {Villa}, F. and {Vittorio},
                  N. and {Wandelt}, B.~D. and {Wehus}, I.~K. and
                  {White}, M. and {White}, S.~D.~M. and {Zacchei},
                  A. and {Zonca}, A.},
  doi =		 {10.1051/0004-6361/201833910},
  eid =		 {A6},
  eprint =	 {1807.06209},
  journal =	 {\aap},
  month =	 sep,
  pages =	 {A6},
  primaryclass = {astro-ph.CO},
  title =	 {{Planck 2018 results. VI. Cosmological parameters}},
  volume =	 641,
  year =	 2020
}

@article{Ponnada2026ApJ,
  adsurl =
                  {https://ui.adsabs.harvard.edu/abs/2026ApJ...997L..13P},
  archiveprefix ={arXiv},
  author =	 {{Ponnada}, Sam B. and {Hopkins}, Philip F. and {Lu},
                  Yue Samuel and {Silich}, Emily M. and {Butsky},
                  Iryna S. and {Kere{\v{s}}}, Du{\v{s}}an},
  doi =		 {10.3847/2041-8213/ae2fd9},
  eid =		 {L13},
  eprint =	 {2510.13959},
  journal =	 {\apjl},
  month =	 jan,
  number =	 1,
  pages =	 {L13},
  primaryclass = {astro-ph.GA},
  title =	 {{Strong Evidence for Cosmic-Ray-supported
                  {\ensuremath{\sim}}L* Galaxy Halos via X-Ray and tSZ
                  Constraints}},
  volume =	 997,
  year =	 2026
}

@article{Powell2011MNRAS,
  adsurl =
                  {https://ui.adsabs.harvard.edu/abs/2011MNRAS.414.3671P},
  archiveprefix ={arXiv},
  author =	 {{Powell}, Leila C. and {Slyz}, Adrianne and
                  {Devriendt}, Julien},
  doi =		 {10.1111/j.1365-2966.2011.18668.x},
  eprint =	 {1012.2839},
  journal =	 {\mnras},
  month =	 jul,
  number =	 4,
  pages =	 {3671-3689},
  primaryclass = {astro-ph.CO},
  title =	 {{The impact of supernova-driven winds on stream-fed
                  protogalaxies}},
  volume =	 414,
  year =	 2011
}

@article{Ptuskin1988SvAL,
  adsurl =
                  {https://ui.adsabs.harvard.edu/abs/1988SvAL...14..255P},
  author =	 {{Ptuskin}, V.~S.},
  journal =	 {Soviet Astronomy Letters},
  month =	 mar,
  pages =	 255,
  title =	 {{Cosmic-Ray Acceleration by Long-Wave Turbulence}},
  volume =	 14,
  year =	 1988
}

@article{Quataert2025OJAp,
  adsurl =
                  {https://ui.adsabs.harvard.edu/abs/2025OJAp....8E..66Q},
  archiveprefix ={arXiv},
  author =	 {{Quataert}, Eliot and {Hopkins}, Philip F.},
  doi =		 {10.33232/001c.138772},
  eid =		 66,
  eprint =	 {2502.01753},
  journal =	 {The Open Journal of Astrophysics},
  month =	 may,
  pages =	 66,
  primaryclass = {astro-ph.CO},
  title =	 {{Cosmic Ray Feedback in Massive Halos: Implications
                  for the Distribution of Baryons}},
  volume =	 8,
  year =	 2025
}

@article{Rajpurohit2022ApJ,
  adsurl =
                  {https://ui.adsabs.harvard.edu/abs/2022ApJ...927...80R},
  archiveprefix ={arXiv},
  author =	 {{Rajpurohit}, K. and {van Weeren}, R.~J. and
                  {Hoeft}, M. and {Vazza}, F. and {Brienza}, M. and
                  {Forman}, W. and {Wittor}, D. and
                  {Dom{\'\i}nguez-Fern{\'a}ndez}, P. and {Rajpurohit},
                  S. and {Riseley}, C.~J. and {Botteon}, A. and
                  {Osinga}, E. and {Brunetti}, G. and {Bonnassieux},
                  E. and {Bonafede}, A. and {Rajpurohit}, A.~S. and
                  {Stuardi}, C. and {Drabent}, A. and {Br{\"u}ggen},
                  M. and {Dallacasa}, D. and {Shimwell}, T.~W. and
                  {R{\"o}ttgering}, H.~J.~A. and {de Gasperin}, F. and
                  {Miley}, G.~K. and {Rossetti}, M.},
  doi =		 {10.3847/1538-4357/ac4708},
  eid =		 80,
  eprint =	 {2111.04449},
  journal =	 {\apj},
  month =	 mar,
  number =	 1,
  pages =	 80,
  primaryclass = {astro-ph.CO},
  title =	 {{Deep Low-frequency Radio Observations of
                  A2256. I. The Filamentary Radio Relic}},
  volume =	 927,
  year =	 2022
}

@article{Ramesh2024A&A,
  adsurl =
                  {https://ui.adsabs.harvard.edu/abs/2024A&A...684L..16R},
  archiveprefix ={arXiv},
  author =	 {{Ramesh}, Rahul and {Nelson}, Dylan and {Fielding},
                  Drummond and {Br{\"u}ggen}, Marcus},
  doi =		 {10.1051/0004-6361/202348786},
  eid =		 {L16},
  eprint =	 {2404.01370},
  journal =	 {\aap},
  month =	 apr,
  pages =	 {L16},
  primaryclass = {astro-ph.GA},
  title =	 {{Zooming in on the circumgalactic medium with
                  GIBLE. The topology and draping of magnetic fields
                  around cold clouds}},
  volume =	 684,
  year =	 2024
}

@article{Ramesh2024MNRAS,
  adsurl =
                  {https://ui.adsabs.harvard.edu/abs/2024MNRAS.528.3320R},
  archiveprefix ={arXiv},
  author =	 {{Ramesh}, Rahul and {Nelson}, Dylan},
  doi =		 {10.1093/mnras/stae237},
  eprint =	 {2307.11143},
  journal =	 {\mnras},
  month =	 feb,
  number =	 2,
  pages =	 {3320-3339},
  primaryclass = {astro-ph.GA},
  title =	 {{Zooming in on the circumgalactic medium with GIBLE:
                  Resolving small-scale gas structure in cosmological
                  simulations}},
  volume =	 528,
  year =	 2024
}

@article{Recchia2021ApJ,
  adsurl =
                  {https://ui.adsabs.harvard.edu/abs/2021ApJ...914..135R},
  archiveprefix ={arXiv},
  author =	 {{Recchia}, S. and {Gabici}, S. and {Aharonian},
                  F.~A. and {Niro}, V.},
  doi =		 {10.3847/1538-4357/abfda4},
  eid =		 135,
  eprint =	 {2101.05016},
  journal =	 {\apj},
  month =	 jun,
  number =	 2,
  pages =	 135,
  primaryclass = {astro-ph.HE},
  title =	 {{Giant Cosmic-Ray Halos around M31 and the Milky
                  Way}},
  volume =	 914,
  year =	 2021
}

@article{Romano2025A&A,
  adsurl =
                  {https://ui.adsabs.harvard.edu/abs/2025A&A...701L...5R},
  archiveprefix ={arXiv},
  author =	 {{Romano}, Leonard E.~C. and {Owen}, Ellis R. and
                  {Nagamine}, Kentaro},
  doi =		 {10.1051/0004-6361/202554590},
  eid =		 {L5},
  eprint =	 {2503.13261},
  journal =	 {\aap},
  month =	 sep,
  pages =	 {L5},
  primaryclass = {astro-ph.GA},
  title =	 {{Starburst-driven galactic outflows: Unveiling the
                  suppressive role of cosmic ray halos}},
  volume =	 701,
  year =	 2025
}

@article{Roy2025arXiv,
  adsurl =
                  {https://ui.adsabs.harvard.edu/abs/2025arXiv251021699R},
  archiveprefix ={arXiv},
  author =	 {{Roy}, Manami and {Su}, Kung-Yi and {Tonnesen},
                  Stephanie and {Lu}, Yue Samuel and {Hummels},
                  Cameron and {Ponnada}, Sam B.},
  doi =		 {10.48550/arXiv.2510.21699},
  eid =		 {arXiv:2510.21699},
  eprint =	 {2510.21699},
  journal =	 {arXiv e-prints},
  month =	 oct,
  pages =	 {arXiv:2510.21699},
  primaryclass = {astro-ph.GA},
  title =	 {{To Survive or to Shatter: The Impact of Cosmic Rays
                  on the Fate of Stripped Cold Clouds}},
  year =	 2025
}

@article{Ruszkowski2017ApJ,
  adsurl =
                  {https://ui.adsabs.harvard.edu/abs/2017ApJ...844...13R},
  archiveprefix ={arXiv},
  author =	 {{Ruszkowski}, Mateusz and {Yang}, H.-Y. Karen and
                  {Reynolds}, Christopher S.},
  doi =		 {10.3847/1538-4357/aa79f8},
  eid =		 13,
  eprint =	 {1701.07441},
  journal =	 {\apj},
  month =	 jul,
  number =	 1,
  pages =	 13,
  primaryclass = {astro-ph.HE},
  title =	 {{Cosmic-Ray Feedback Heating of the Intracluster
                  Medium}},
  volume =	 844,
  year =	 2017
}

@article{Ruszkowski2023A&ARv,
  adsurl =
                  {https://ui.adsabs.harvard.edu/abs/2023A&ARv..31....4R},
  archiveprefix ={arXiv},
  author =	 {{Ruszkowski}, Mateusz and {Pfrommer}, Christoph},
  doi =		 {10.1007/s00159-023-00149-2},
  eid =		 4,
  eprint =	 {2306.03141},
  journal =	 {\aapr},
  month =	 dec,
  number =	 1,
  pages =	 4,
  primaryclass = {astro-ph.HE},
  title =	 {{Cosmic ray feedback in galaxies and galaxy
                  clusters}},
  volume =	 31,
  year =	 2023
}

@article{Ryu2003ApJ,
  adsurl =
                  {https://ui.adsabs.harvard.edu/abs/2003ApJ...593..599R},
  archiveprefix ={arXiv},
  author =	 {{Ryu}, Dongsu and {Kang}, Hyesung and {Hallman},
                  Eric and {Jones}, T.~W.},
  doi =		 {10.1086/376723},
  eprint =	 {astro-ph/0305164},
  journal =	 {\apj},
  month =	 aug,
  number =	 2,
  pages =	 {599-610},
  primaryclass = {astro-ph},
  title =	 {{Cosmological Shock Waves and Their Role in the
                  Large-Scale Structure of the Universe}},
  volume =	 593,
  year =	 2003
}

@article{Salem2016MNRAS,
  adsurl =
                  {https://ui.adsabs.harvard.edu/abs/2016MNRAS.456..582S},
  archiveprefix ={arXiv},
  author =	 {{Salem}, Munier and {Bryan}, Greg L. and {Corlies},
                  Lauren},
  doi =		 {10.1093/mnras/stv2641},
  eprint =	 {1511.05144},
  journal =	 {\mnras},
  month =	 feb,
  number =	 1,
  pages =	 {582-601},
  primaryclass = {astro-ph.GA},
  title =	 {{Role of cosmic rays in the circumgalactic medium}},
  volume =	 456,
  year =	 2016
}

@article{Sampson2023MNRAS,
  adsurl =
                  {https://ui.adsabs.harvard.edu/abs/2023MNRAS.519.1503S},
  archiveprefix ={arXiv},
  author =	 {{Sampson}, Matt L. and {Beattie}, James R. and
                  {Krumholz}, Mark R. and {Crocker}, Roland M. and
                  {Federrath}, Christoph and {Seta}, Amit},
  doi =		 {10.1093/mnras/stac3207},
  eprint =	 {2205.08174},
  journal =	 {\mnras},
  month =	 feb,
  number =	 1,
  pages =	 {1503-1525},
  primaryclass = {astro-ph.GA},
  title =	 {{Turbulent diffusion of streaming cosmic rays in
                  compressible, partially ionized plasma}},
  volume =	 519,
  year =	 2023
}

@article{Sazonov2015MNRAS,
  adsurl =
                  {https://ui.adsabs.harvard.edu/abs/2015MNRAS.454.3464S},
  archiveprefix ={arXiv},
  author =	 {{Sazonov}, S. and {Sunyaev}, R.},
  doi =		 {10.1093/mnras/stv2255},
  eprint =	 {1509.08408},
  journal =	 {\mnras},
  month =	 dec,
  number =	 4,
  pages =	 {3464-3471},
  primaryclass = {astro-ph.CO},
  title =	 {{Preheating of the Universe by cosmic rays from
                  primordial supernovae at the beginning of cosmic
                  reionization}},
  volume =	 454,
  year =	 2015
}

@article{Schroer2025PhRvL,
  adsurl =
                  {https://ui.adsabs.harvard.edu/abs/2025PhRvL.134d5201S},
  archiveprefix ={arXiv},
  author =	 {{Schroer}, Benedikt and {Caprioli}, Damiano and
                  {Blasi}, Pasquale},
  doi =		 {10.1103/PhysRevLett.134.045201},
  eid =		 045201,
  eprint =	 {2409.02230},
  journal =	 {\prl},
  month =	 jan,
  number =	 4,
  pages =	 045201,
  primaryclass = {astro-ph.HE},
  title =	 {{Role of Nonlinear Landau Damping for Cosmic-Ray
                  Transport}},
  volume =	 134,
  year =	 2025
}

@article{Shalchi2020SSRv,
  adsurl =
                  {https://ui.adsabs.harvard.edu/abs/2020SSRv..216...23S},
  author =	 {{Shalchi}, Andreas},
  doi =		 {10.1007/s11214-020-0644-4},
  eid =		 23,
  journal =	 {\ssr},
  month =	 feb,
  number =	 2,
  pages =	 23,
  title =	 {{Perpendicular Transport of Energetic Particles in
                  Magnetic Turbulence}},
  volume =	 216,
  year =	 2020
}

@article{Shibata2011LRSP,
  adsurl =
                  {https://ui.adsabs.harvard.edu/abs/2011LRSP....8....6S},
  author =	 {{Shibata}, Kazunari and {Magara}, Tetsuya},
  doi =		 {10.12942/lrsp-2011-6},
  eid =		 6,
  journal =	 {Living Reviews in Solar Physics},
  month =	 dec,
  number =	 1,
  pages =	 6,
  title =	 {{Solar Flares: Magnetohydrodynamic Processes}},
  volume =	 8,
  year =	 2011
}

@article{Strong2007ARNPS,
  adsurl =
                  {https://ui.adsabs.harvard.edu/abs/2007ARNPS..57..285S},
  archiveprefix ={arXiv},
  author =	 {{Strong}, Andrew W. and {Moskalenko}, Igor V. and
                  {Ptuskin}, Vladimir S.},
  doi =		 {10.1146/annurev.nucl.57.090506.123011},
  eprint =	 {astro-ph/0701517},
  journal =	 {Annual Review of Nuclear and Particle Science},
  month =	 nov,
  number =	 1,
  pages =	 {285-327},
  primaryclass = {astro-ph},
  title =	 {{Cosmic-Ray Propagation and Interactions in the
                  Galaxy}},
  volume =	 57,
  year =	 2007
}

@article{Tanimura2020A&A,
  adsurl =
                  {https://ui.adsabs.harvard.edu/abs/2020A&A...637A..41T},
  archiveprefix ={arXiv},
  author =	 {{Tanimura}, H. and {Aghanim}, N. and {Bonjean},
                  V. and {Malavasi}, N. and {Douspis}, M.},
  doi =		 {10.1051/0004-6361/201937158},
  eid =		 {A41},
  eprint =	 {1911.09706},
  journal =	 {\aap},
  month =	 may,
  pages =	 {A41},
  primaryclass = {astro-ph.CO},
  title =	 {{Density and temperature of cosmic-web filaments on
                  scales of tens of megaparsecs}},
  volume =	 637,
  year =	 2020
}

@article{Tsung2023MNRAS,
  adsurl =
                  {https://ui.adsabs.harvard.edu/abs/2023MNRAS.526.3301T},
  archiveprefix ={arXiv},
  author =	 {{Tsung}, Tsun Hin Navin and {Oh}, S. Peng and
                  {Bustard}, Chad},
  doi =		 {10.1093/mnras/stad2720},
  eprint =	 {2305.14432},
  journal =	 {\mnras},
  month =	 dec,
  number =	 3,
  pages =	 {3301-3334},
  primaryclass = {astro-ph.GA},
  title =	 {{The impact of cosmic rays on thermal and
                  hydrostatic stability in galactic haloes}},
  volume =	 526,
  year =	 2023
}

@article{Vacca2018Galax,
  adsurl =
                  {https://ui.adsabs.harvard.edu/abs/2018Galax...6..142V},
  author =	 {{Vacca}, Valentina and {Murgia}, Matteo and
                  {Govoni}, Federica and {En{\ss}lin}, Torsten and
                  {Oppermann}, Niels and {Feretti}, Luigina and
                  {Giovannini}, Gabriele and {Loi}, Francesca},
  doi =		 {10.3390/galaxies6040142},
  eid =		 142,
  journal =	 {Galaxies},
  month =	 dec,
  number =	 4,
  pages =	 142,
  title =	 {{Magnetic Fields in Galaxy Clusters and in the
                  Large-Scale Structure of the Universe}},
  volume =	 6,
  year =	 2018
}

@article{Vazza2014MNRAS,
  adsurl =
                  {https://ui.adsabs.harvard.edu/abs/2014MNRAS.439.2662V},
  archiveprefix ={arXiv},
  author =	 {{Vazza}, F. and {Gheller}, C. and {Br{\"u}ggen}, M.},
  doi =		 {10.1093/mnras/stu126},
  eprint =	 {1401.4454},
  journal =	 {\mnras},
  month =	 apr,
  number =	 3,
  pages =	 {2662-2677},
  primaryclass = {astro-ph.CO},
  title =	 {{Simulations of cosmic rays in large-scale
                  structures: numerical and physical effects}},
  volume =	 439,
  year =	 2014
}

@article{Vazza2015MNRAS,
  adsurl =
                  {https://ui.adsabs.harvard.edu/abs/2015MNRAS.451.2198V},
  archiveprefix ={arXiv},
  author =	 {{Vazza}, F. and {Eckert}, D. and {Br{\"u}ggen},
                  M. and {Huber}, B.},
  doi =		 {10.1093/mnras/stv1072},
  eprint =	 {1505.02782},
  journal =	 {\mnras},
  month =	 aug,
  number =	 2,
  pages =	 {2198-2211},
  primaryclass = {astro-ph.HE},
  title =	 {{Electron and proton acceleration efficiency by
                  merger shocks in galaxy clusters}},
  volume =	 451,
  year =	 2015
}

@article{Vazza2025A&A,
  adsurl =
                  {https://ui.adsabs.harvard.edu/abs/2025A&A...696A..58V},
  archiveprefix ={arXiv},
  author =	 {{Vazza}, F. and {Gheller}, C. and {Zanetti}, F. and
                  {Tsizh}, M. and {Carretti}, E. and {Mtchedlidze},
                  S. and {Br{\"u}ggen}, M.},
  doi =		 {10.1051/0004-6361/202451709},
  eid =		 {A58},
  eprint =	 {2501.19041},
  journal =	 {\aap},
  month =	 apr,
  pages =	 {A58},
  primaryclass = {astro-ph.HE},
  title =	 {{The evolution of cosmic ray electrons in the cosmic
                  web: Seeding by active galactic nuclei, star
                  formation, and shocks}},
  volume =	 696,
  year =	 2025
}

@article{Vernstrom2021MNRAS,
  adsurl =
                  {https://ui.adsabs.harvard.edu/abs/2021MNRAS.505.4178V},
  archiveprefix ={arXiv},
  author =	 {{Vernstrom}, T. and {Heald}, G. and {Vazza}, F. and
                  {Galvin}, T.~J. and {West}, J.~L. and {Locatelli},
                  N. and {Fornengo}, N. and {Pinetti}, E.},
  doi =		 {10.1093/mnras/stab1301},
  eprint =	 {2101.09331},
  journal =	 {\mnras},
  month =	 aug,
  number =	 3,
  pages =	 {4178-4196},
  primaryclass = {astro-ph.CO},
  title =	 {{Discovery of magnetic fields along stacked cosmic
                  filaments as revealed by radio and X-ray emission}},
  volume =	 505,
  year =	 2021
}

@article{Vernstrom2023SciA,
  adsurl =
                  {https://ui.adsabs.harvard.edu/abs/2023SciA....9E7233V},
  archiveprefix ={arXiv},
  author =	 {{Vernstrom}, Tessa and {West}, Jennifer and {Vazza},
                  Franco and {Wittor}, Denis and {Riseley},
                  Christopher John and {Heald}, George},
  doi =		 {10.1126/sciadv.ade7233},
  eid =		 {eade7233},
  eprint =	 {2302.08072},
  journal =	 {Science Advances},
  month =	 feb,
  number =	 7,
  pages =	 {eade7233},
  primaryclass = {astro-ph.CO},
  title =	 {{Polarized accretion shocks from the cosmic web}},
  volume =	 9,
  year =	 2023
}

@article{Vurm2023A&A,
  adsurl =
                  {https://ui.adsabs.harvard.edu/abs/2023A&A...673A..62V},
  archiveprefix ={arXiv},
  author =	 {{Vurm}, I. and {Nevalainen}, J. and {Hong},
                  S.~E. and {Bah{\'e}}, Y.~M. and {Dalla Vecchia},
                  C. and {Hein{\"a}m{\"a}ki}, P.},
  doi =		 {10.1051/0004-6361/202243904},
  eid =		 {A62},
  eprint =	 {2303.03244},
  journal =	 {\aap},
  month =	 may,
  pages =	 {A62},
  primaryclass = {astro-ph.CO},
  title =	 {{Cosmic gas highways in C-EAGLE simulations}},
  volume =	 673,
  year =	 2023
}

@article{Waterval2025,
  adsurl =
                  {https://ui.adsabs.harvard.edu/abs/2025MNRAS.537.2726W},
  archiveprefix ={arXiv},
  author =	 {{Waterval}, Stefan and {Cannarozzo}, Carlo and
                  {Macci{\`o}}, Andrea V.},
  doi =		 {10.1093/mnras/staf198},
  eprint =	 {2501.19009},
  journal =	 {MNRAS},
  month =	 mar,
  number =	 3,
  pages =	 {2726-2751},
  primaryclass = {astro-ph.GA},
  title =	 {{Gas accretion at high redshift: cold flows all the
                  way}},
  volume =	 537,
  year =	 2025
}

@article{Weber2025A&A,
  adsurl =
                  {https://ui.adsabs.harvard.edu/abs/2025A&A...698A.125W},
  archiveprefix ={arXiv},
  author =	 {{Weber}, M. and {Thomas}, T. and {Pfrommer}, C. and
                  {Pakmor}, R.},
  doi =		 {10.1051/0004-6361/202553954},
  eid =		 {A125},
  eprint =	 {2501.18678},
  journal =	 {\aap},
  month =	 jun,
  pages =	 {A125},
  primaryclass = {astro-ph.GA},
  title =	 {{CRexit: How different cosmic ray transport modes
                  affect thermal instability in the circumgalactic
                  medium}},
  volume =	 698,
  year =	 2025
}

@article{Werhahn2021MNRAS,
  adsurl =
                  {https://ui.adsabs.harvard.edu/abs/2021MNRAS.505.3273W},
  archiveprefix ={arXiv},
  author =	 {{Werhahn}, Maria and {Pfrommer}, Christoph and
                  {Girichidis}, Philipp and {Puchwein}, Ewald and
                  {Pakmor}, R{\"u}diger},
  doi =		 {10.1093/mnras/stab1324},
  eprint =	 {2105.10509},
  journal =	 {\mnras},
  month =	 aug,
  number =	 3,
  pages =	 {3273-3294},
  primaryclass = {astro-ph.HE},
  title =	 {{Cosmic rays and non-thermal emission in simulated
                  galaxies - I. Electron and proton spectra compared
                  to Voyager-1 data}},
  volume =	 505,
  year =	 2021
}

@article{Wu2024Univ,
  adsurl =
                  {https://ui.adsabs.harvard.edu/abs/2024Univ...10..287W},
  archiveprefix ={arXiv},
  author =	 {{Wu}, Kinwah and {Owen}, Ellis R. and {Han}, Qin and
                  {Inoue}, Yoshiyuki and {Luo}, Lilian},
  doi =		 {10.3390/universe10070287},
  eid =		 287,
  eprint =	 {2407.01208},
  journal =	 {Universe},
  month =	 jul,
  number =	 7,
  pages =	 287,
  primaryclass = {astro-ph.HE},
  title =	 {{Energetic Particles and High-Energy Processes in
                  Cosmological Filaments and Their Astronomical
                  Implications}},
  volume =	 10,
  year =	 2024
}

@article{Xu2013ApJ,
  adsurl =
                  {https://ui.adsabs.harvard.edu/abs/2013ApJ...779..140X},
  archiveprefix ={arXiv},
  author =	 {{Xu}, Siyao and {Yan}, Huirong},
  doi =		 {10.1088/0004-637X/779/2/140},
  eid =		 140,
  eprint =	 {1307.1346},
  journal =	 {\apj},
  month =	 dec,
  number =	 2,
  pages =	 140,
  primaryclass = {astro-ph.HE},
  title =	 {{Cosmic-Ray Parallel and Perpendicular Transport in
                  Turbulent Magnetic Fields}},
  volume =	 779,
  year =	 2013
}

@article{Yamasaki2024MNRAS,
  adsurl =
                  {https://ui.adsabs.harvard.edu/abs/2024MNRAS.528.3854Y},
  archiveprefix ={arXiv},
  author =	 {{Yamasaki}, Shotaro and {Sarkar}, Kartick C. and
                  {Li}, Zhaozhou},
  doi =		 {10.1093/mnras/stae281},
  eprint =	 {2309.17451},
  journal =	 {\mnras},
  month =	 feb,
  number =	 2,
  pages =	 {3854-3863},
  primaryclass = {astro-ph.HE},
  title =	 {{Are odd radio circles virial shocks around massive
                  galaxies? Implications for cosmic-ray diffusion in
                  the circumgalactic medium}},
  volume =	 528,
  year =	 2024
}

@article{Yan2008ApJ,
  adsurl =
                  {https://ui.adsabs.harvard.edu/abs/2008ApJ...673..942Y},
  archiveprefix ={arXiv},
  author =	 {{Yan}, Huirong and {Lazarian}, A.},
  doi =		 {10.1086/524771},
  eprint =	 {0710.2617},
  journal =	 {\apj},
  month =	 feb,
  number =	 2,
  pages =	 {942-953},
  primaryclass = {astro-ph},
  title =	 {{Cosmic-Ray Propagation: Nonlinear Diffusion
                  Parallel and Perpendicular to Mean Magnetic Field}},
  volume =	 673,
  year =	 2008
}

@article{Yang2023NatAs,
  adsurl =
                  {https://ui.adsabs.harvard.edu/abs/2023NatAs...7..351Y},
  archiveprefix ={arXiv},
  author =	 {{Yang}, Rui-zhi and {Li}, Guang-Xing and {Wilhelmi},
                  Emma de O{\~n}a and {Cui}, Yu-Dong and {Liu}, Bing
                  and {Aharonian}, Felix},
  doi =		 {10.1038/s41550-022-01868-9},
  eprint =	 {2301.06716},
  journal =	 {Nature Astronomy},
  month =	 mar,
  pages =	 {351-358},
  primaryclass = {astro-ph.HE},
  title =	 {{Effective shielding of {\ensuremath{\lesssim}}10
                  GeV cosmic rays from dense molecular clumps}},
  volume =	 7,
  year =	 2023
}

@article{Yao2025,
  adsurl =
                  {https://ui.adsabs.harvard.edu/abs/2025MNRAS.536.3053Y},
  archiveprefix ={arXiv},
  author =	 {{Yao}, Zhiyuan and {Mandelker}, Nir and {Oh},
                  S. Peng and {Aung}, Han and {Dekel}, Avishai},
  doi =		 {10.1093/mnras/stae2771},
  eprint =	 {2410.12914},
  journal =	 {\mnras},
  month =	 jan,
  number =	 3,
  pages =	 {3053-3089},
  primaryclass = {astro-ph.GA},
  title =	 {{Effects of cloud geometry and metallicity on
                  shattering and coagulation of cold gas, and
                  implications for cold streams penetrating virial
                  shocks}},
  volume =	 536,
  year =	 2025
}

@article{Zhang2023Sci,
  adsurl =
                  {https://ui.adsabs.harvard.edu/abs/2023Sci...380..494Z},
  archiveprefix ={arXiv},
  author =	 {{Zhang}, Shiwu and {Cai}, Zheng and {Xu}, Dandan and
                  {Shimakawa}, Rhythm and {Arrigoni Battaia}, Fabrizio
                  and {Prochaska}, Jason Xavier and {Cen}, Renyue and
                  {Zheng}, Zheng and {Wu}, Yunjing and {Li}, Qiong and
                  {Dou}, Liming and {Wu}, Jianfeng and {Zabludoff},
                  Ann and {Fan}, Xiaohui and {Ai}, Yanli and
                  {Golden-Marx}, Emmet Gabriel and {Li}, Miao and
                  {Lu}, Youjun and {Ma}, Xiangcheng and {Wang}, Sen
                  and {Wang}, Ran and {Yuan}, Feng},
  doi =		 {10.1126/science.abj9192},
  eprint =	 {2305.02344},
  journal =	 {Science},
  month =	 may,
  number =	 6644,
  pages =	 {494-498},
  primaryclass = {astro-ph.GA},
  title =	 {{Inspiraling streams of enriched gas observed around
                  a massive galaxy 11 billion years ago}},
  volume =	 380,
  year =	 2023
}

@article{Zhu2015ApJ,
  adsurl =
                  {https://ui.adsabs.harvard.edu/abs/2015ApJ...811...94Z},
  archiveprefix ={arXiv},
  author =	 {{Zhu}, Weishan and {Feng}, Long-long},
  doi =		 {10.1088/0004-637X/811/2/94},
  eid =		 94,
  eprint =	 {1508.06875},
  journal =	 {\apj},
  month =	 oct,
  number =	 2,
  pages =	 94,
  primaryclass = {astro-ph.CO},
  title =	 {{Vortical Motions of Baryonic Gas in the Cosmic Web:
                  Growth History and Scaling Relation}},
  volume =	 811,
  year =	 2015
}

@article{Zweibel2017PhPl,
  adsurl =
                  {https://ui.adsabs.harvard.edu/abs/2017PhPl...24e5402Z},
  author =	 {{Zweibel}, Ellen G.},
  doi =		 {10.1063/1.4984017},
  eid =		 055402,
  journal =	 {Physics of Plasmas},
  month =	 may,
  number =	 5,
  pages =	 055402,
  title =	 {{The basis for cosmic ray feedback: Written on the
                  wind}},
  volume =	 24,
  year =	 2017
}

@article{bib_MHD_Berlok_2019b,
  author =	 {Berlok, Thomas and Pfrommer, Christoph},
  doi =		 {10.1093/mnras/stz2347},
  eprint =	 {1904.02167},
  issn =	 13652966,
  journal =	 {MNRAS},
  number =	 3,
  pages =	 {3368--3384},
  publisher =	 {Oxford University Press},
  title =	 {{The impact of magnetic fields on cold streams
                  feeding galaxies}},
  volume =	 489,
  year =	 2019
}

@article{deRoo2025MNRAS,
  adsurl =
                  {https://ui.adsabs.harvard.edu/abs/2025MNRAS.540L..78D},
  archiveprefix ={arXiv},
  author =	 {{de Roo}, W. and {Vegetti}, S. and {Powell},
                  D.~M. and {Ndiritu}, S.~W. and {Pakmor}, R. and
                  {McKean}, J.~P.},
  doi =		 {10.1093/mnrasl/slaf030},
  eprint =	 {2412.08705},
  journal =	 {\mnras},
  month =	 jun,
  number =	 1,
  pages =	 {L78-L83},
  primaryclass = {astro-ph.GA},
  title =	 {{A grand-design spiral galaxy with an ordered
                  magnetic field at redshift 2.6 as resolved with ALMA
                  and gravitational lensing}},
  volume =	 540,
  year =	 2025
}

@article{vanWeeren2019SSRv,
  adsurl =
                  {https://ui.adsabs.harvard.edu/abs/2019SSRv..215...16V},
  archiveprefix ={arXiv},
  author =	 {{van Weeren}, R.~J. and {de Gasperin}, F. and
                  {Akamatsu}, H. and {Br{\"u}ggen}, M. and {Feretti},
                  L. and {Kang}, H. and {Stroe}, A. and {Zandanel},
                  F.},
  doi =		 {10.1007/s11214-019-0584-z},
  eid =		 16,
  eprint =	 {1901.04496},
  journal =	 {\ssr},
  month =	 feb,
  number =	 1,
  pages =	 16,
  primaryclass = {astro-ph.HE},
  title =	 {{Diffuse Radio Emission from Galaxy Clusters}},
  volume =	 215,
  year =	 2019
}

@article{2022PTEP_databook,
  adsurl =
                  {https://ui.adsabs.harvard.edu/abs/2022PTEP.2022h3C01W},
  author =	 {{Workman}, R.~L. and {Burkert}, V.~D. and {Crede},
                  V. and {Klempt}, E. and {Thoma}, U. and {Tiator},
                  L. and {Agashe}, K. and {Aielli}, G. and {Allanach},
                  B.~C. and {Amsler}, C. and {Antonelli}, M. and
                  {Aschenauer}, E.~C. and {Asner}, D.~M. and {Baer},
                  H. and {Banerjee}, Sw and {Barnett}, R.~M. and
                  {Baudis}, L. and {Bauer}, C.~W. and {Beatty},
                  J.~J. and {Belousov}, V.~I. and {Beringer}, J. and
                  {Bettini}, A. and {Biebel}, O. and {Black},
                  K.~M. and {Blucher}, E. and {Bonventre}, R. and
                  {Bryzgalov}, V.~V. and {Buchmuller}, O. and
                  {Bychkov}, M.~A. and {Cahn}, R.~N. and {Carena},
                  M. and {Ceccucci}, A. and {Cerri}, A. and
                  {Chivukula}, R. Sekhar and {Cowan}, G. and
                  {Cranmer}, K. and {Cremonesi}, O. and {D'Ambrosio},
                  G. and {Damour}, T. and {de Florian}, D. and {de
                  Gouv{\^e}a}, A. and {DeGrand}, T. and {de Jong},
                  P. and {Demers}, S. and {Dobrescu}, B.~A. and
                  {D'Onofrio}, M. and {Doser}, M. and {Dreiner},
                  H.~K. and {Eerola}, P. and {Egede}, U. and
                  {Eidelman}, S. and {El-Khadra}, A.~X. and {Ellis},
                  J. and {Eno}, S.~C. and {Erler}, J. and {Ezhela},
                  V.~V. and {Fetscher}, W. and {Fields}, B.~D. and
                  {Freitas}, A. and {Gallagher}, H. and {Gershtein},
                  Y. and {Gherghetta}, T. and {Gonzalez-Garcia},
                  M.~C. and {Goodman}, M. and {Grab}, C. and
                  {Gritsan}, A.~V. and {Grojean}, C. and {Groom},
                  D.~E. and {Gr{\"u}newald}, M. and {Gurtu}, A. and
                  {Gutsche}, T. and {Haber}, H.~E. and {Hamel},
                  Matthieu and {Hanhart}, C. and {Hashimoto}, S. and
                  {Hayato}, Y. and {Hebecker}, A. and {Heinemeyer},
                  S. and {Hern{\'a}ndez-Rey}, J.~J. and {Hikasa},
                  K. and {Hisano}, J. and {H{\"o}cker}, A. and
                  {Holder}, J. and {Hsu}, L. and {Huston}, J. and
                  {Hyodo}, T. and {Ianni}, Al and {Kado}, M. and
                  {Karliner}, M. and {Katz}, U.~F. and {Kenzie},
                  M. and {Khoze}, V.~A. and {Klein}, S.~R. and
                  {Krauss}, F. and {Kreps}, M. and {Kri{\v{z}}an},
                  P. and {Krusche}, B. and {Kwon}, Y. and {Lahav},
                  O. and {Laiho}, J. and {Lellouch}, L.~P. and
                  {Lesgourgues}, J. and {Liddle}, A.~R. and {Ligeti},
                  Z. and {Lin}, C.-J. and {Lippmann}, C. and {Liss},
                  T.~M. and {Littenberg}, L. and {Louren{\c{c}}o},
                  C. and {Lugovsky}, K.~S. and {Lugovsky}, S.~B. and
                  {Lusiani}, A. and {Makida}, Y. and {Maltoni}, F. and
                  {Mannel}, T. and {Manohar}, A.~V. and {Marciano},
                  W.~J. and {Masoni}, A. and {Matthews}, J. and
                  {Mei{\ss}ner}, U.-G. and {Melzer-Pellmann},
                  I.-A. and {Mikhasenko}, M. and {Miller}, D.~J. and
                  {Milstead}, D. and {Mitchell}, R.~E. and
                  {M{\"o}nig}, K. and {Molaro}, P. and {Moortgat},
                  F. and {Moskovic}, M. and {Nakamura}, K. and
                  {Narain}, M. and {Nason}, P. and {Navas}, S. and
                  {Nelles}, A. and {Neubert}, M. and {Nevski}, P. and
                  {Nir}, Y. and {Olive}, K.~A. and {Patrignani},
                  C. and {Peacock}, J.~A. and {Petrov}, V.~A. and
                  {Pianori}, E. and {Pich}, A. and {Piepke}, A. and
                  {Pietropaolo}, F. and {Pomarol}, A. and {Pordes},
                  S. and {Profumo}, S. and {Quadt}, A. and {Rabbertz},
                  K. and {Rademacker}, J. and {Raffelt}, G. and
                  {Ramsey-Musolf}, M. and {Ratcliff}, B.~N. and
                  {Richardson}, P. and {Ringwald}, A. and {Robinson},
                  D.~J. and {Roesler}, S. and {Rolli}, S. and
                  {Romaniouk}, A. and {Rosenberg}, L.~J. and {Rosner},
                  J.~L. and {Rybka}, G. and {Ryskin}, M.~G. and
                  {Ryutin}, R.~A. and {Sakai}, Y. and {Sarkar}, S. and
                  {Sauli}, F. and {Schneider}, O. and {Sch{\"o}nert},
                  S. and {Scholberg}, K. and {Schwartz}, A.~J. and
                  {Schwiening}, J. and {Scott}, D. and {Sefkow},
                  F. and {Seljak}, U. and {Sharma}, V. and {Sharpe},
                  S.~R. and {Shiltsev}, V. and {Signorelli}, G. and
                  {Silari}, M. and {Simon}, F. and {Sj{\"o}strand},
                  T. and {Skands}, P. and {Skwarnicki}, T. and
                  {Smoot}, G.~F. and {Soffer}, A. and {Sozzi},
                  M.~S. and {Spanier}, S. and {Spiering}, C. and
                  {Stahl}, A. and {Stone}, S.~L. and {Sumino}, Y. and
                  {Syphers}, M.~J. and {Takahashi}, F. and
                  {Tanabashi}, M. and {Tanaka}, J. and
                  {Ta{\v{s}}evsk{\'y}}, M. and {Terao}, K. and
                  {Terashi}, K.},
  doi =		 {10.1093/ptep/ptac097},
  eid =		 {083C01},
  journal =	 {Progress of Theoretical and Experimental Physics},
  month =	 aug,
  number =	 8,
  pages =	 {083C01},
  title =	 {{Review of Particle Physics}},
  volume =	 2022,
  year =	 2022
}

\appendix

\section{Cosmic rays in cold streams}

\subsection{Cosmic ray cooling and interaction processes}
\label{sec:CR_processes}

\subsubsection{Cosmic ray protons}
\label{sec:transport_CR_protons}

For CR protons, the general transport Eq.~(\ref{eq:transport}) can be simplified because we do not include volumetric proton injection within the stream. 
The proton population is supplied only through the outer boundary, while continuous energy changes arise from adiabatic compression or expansion and streaming-related losses. 
Attenuation is dominated by inelastic pp collisions, which we treat as a catastrophic loss process. 
In the quasi-one-dimensional stream geometry, with cross-sectional area $A(r)=\pi r_{\rm s}^2(r)$, the proton distribution $n_p(E,r)$ is described by: 
\begin{align}
-\frac{1}{A(r)}\frac{\partial}{\partial r}
\left[ 
A(r)\,D(E,r)\frac{\partial n_p}{\partial r}
\right]
+&\frac{1}{A(r)}\frac{\partial}{\partial r}
\left[
A(r)\,u_{\rm eff}(r)\,n_p
\right] \\ \nonumber
&-\frac{\partial}{\partial E}\!\left[b_p(E,r)\,n_p\right]
=
-\frac{n_p(E,r)}{t_{pp}(E,r)} \ , 
\end{align} 
where $u_{\rm eff}$ denotes the effective radial transport velocity. 

The dominant catastrophic loss process is inelastic pp scattering. 
The corresponding interaction timescale in the stream is 
\begin{equation}
t_{pp}(E,r)=\left[n_{\rm H, s}(r)\,\sigma_{pp}(E)\,c\right]^{-1},
\label{eq:had_time}
\end{equation} 
where $\sigma_{pp}(E)$ is introduced as the inelastic pp cross-section, for which we adopt the parametrization of~\cite{Kafexhiu2014PhRvD}. 

The continuous energy-change term is written as an effective proton cooling coefficient 
\begin{equation}
b_p(E,r)=E\left[t_{\rm ad}^{-1}(r)+t_{\rm st}^{-1}(r)\right] \, , 
\label{eq:adiabatic}
\end{equation} 
where $b_p=-{\rm d}E/{\rm d}t$ may include either losses or gains depending on whether the adiabatic term is positive or negative, which is set by the divergence of the effective transport velocity: 
\begin{equation}
t_{\rm ad}^{-1}(r)=\frac{1}{3}\,\nabla\cdot \mathbf{u}_{\rm eff}
\approx
\frac{1}{3A(r)}\frac{{\rm d}}{{\rm d}r}\left[A(r)u_{\rm eff}(r)\right] ,
\end{equation} 
while the streaming term is estimated from the local CR pressure gradient: 
\begin{equation}
t_{\rm st}^{-1}(r) = (\gamma_{\rm cr} - 1) v_{\rm A}(r)\,
\left|\frac{\partial \ln P_p}{\partial r}\right| ,
\end{equation}
where $P_p = (\gamma_{\rm cr} - 1) U_p(r)$, $\gamma_{\rm cr}=4/3$ is the CR adiabatic index, and $U_p(r)$ is the CR proton energy density. 

\subsubsection{Cosmic ray electrons}
\label{sec:transport_CR_electrons}

The treatment of CR electrons differs from that of the protons. 
This is because (i) the electron population
is subject to different energy loss processes, in particular more rapid radiative and collisional cooling losses, 
and (ii) a secondary component develops, which is generated locally as a product of the inelastic pp interactions of 
the CR proton population. 
The energy losses are reflected by the use of different cooling terms (and the absence of any catastrophic loss term). In the quasi-steady, quasi-one-dimensional limit, the electron distribution $n_e(E,r)$ is obtained from Eq.~\ref{eq:transport} as: 
\begin{align} 
\label{eq:electron_transport_eq}
-\frac{1}{A(r)}\frac{\partial}{\partial r}
\left[
A(r)\,D(E,r)  \frac{\partial n_e}{\partial r}
\right] 
+&\frac{1}{A(r)}\frac{\partial}{\partial r}
\left[
A(r)\,u_{\rm eff}(r)\,n_e
\right] \\ \nonumber
& -\frac{\partial}{\partial E}\!\left[b_e(E,r)\,n_e\right]
=
S_e(E,r) \, . 
\end{align}
Here we use the same effective diffusion and advection prescriptions as for CR protons, while the boundary condition imposed at the virial radius is described in Appendix~\ref{app:cr_profile}. 
The term $S_e(E,r)$ denotes the local volumetric source term for electrons. 
We compute the secondary production rate from the CR proton distribution using the {\tt AAFrag} tool~\citep{Kachelriess2019CoPhC}. 
This provides interpolated production spectra based on the pp event generator {\tt QGSJET-II-04m} above interaction energies of 4 GeV~\citep{Ostapchenko2011PhRvD, Ostapchenko2013EPJWC, Kachelriess2015ApJ}, with the analytical cross-sections of \citet{Kamae2006ApJ} used at lower energies down to the interaction threshold. 

The explicit energy-advection term includes contributions from adiabatic losses, $t_{\rm ad}$ (which takes the same form as introduced for protons), and microphysical loss processes, $t_{\rm micro}$ which can be written in terms of the corresponding timescales: 
\begin{equation}
b_e(E,r)=E\, \left[ t_{\rm ad}^{-1}(r)  + t_{\rm micro}^{-1}(r) \right]\, ,
\end{equation}
and where the microphysical loss timescale is given by 
\begin{equation}
t_{\rm micro}^{-1}(E,r)=
t_{\rm syn}^{-1}(E,r)
+t_{\rm IC}^{-1}(E,r)
+t_{{\rm Coul},e}^{-1}(E,r)
+t_{\rm ff}^{-1}(E,r) \, , 
\end{equation}
which accounts for all radiative and collisional losses. These include synchrotron cooling, given by: 
\begin{equation}
t_{\rm syn}(E,r)=
\frac{E}
{\frac{4}{3}\sigma_{\rm T} c \gamma_e^2 U_B(r)} \, ,
\end{equation}
where $\sigma_{\rm T}$ is introduced as the Thomson cross-section, and $U_B(r)$ is the magnetic energy density at a position $r$ in the stream. 
Inverse Compton cooling in ambient radiation fields operates over a timescale of  
\begin{equation}
t_{\rm IC}(E,r)=
\frac{E}
{\frac{4}{3}\sigma_{\rm T} c \gamma_e^2 U_{\rm CMB}(z)} \, ,
\end{equation}
where $U_{\rm CMB}= U_{\rm CMB,0} \;\! (1+z)^4$ is the redshift-dependent CMB energy density, with a present value of $U_{\rm CMB,0} = 0.26\;\! {\rm eV}\;\! {\rm cm}^{-3}$~\citep{2022PTEP_databook}. 
Inverse Compton losses can also arise with photons emitted from stellar populations in the host galaxy, however 
we found that for all model parameter choices we considered, 
their contribution is 
sub-dominant 
compared to the cooling brought about by the CMB in regions of the cold streams where CRs have an important effect. We therefore do not include this component in our calculations. 
The Coulomb loss time is approximated by
\begin{equation}
t_{{\rm Coul},e}(E,r)\approx
\frac{E}
{m_e c^3 \sigma_{\rm T} n_{\rm H, s}(r)\ln\Lambda} \, ,
\end{equation}
where $\ln\Lambda=30$ is the Coulomb logarithm, and the free-free cooling time is given by 
\begin{equation}
t_{\rm ff}(E,r)\approx
\left[\alpha \sigma_{\rm T} c\,n_{\rm H, s}(r)\right]^{-1} \ ,  
\end{equation}
where $\alpha \approx 1/137$ is the fine structure constant.  

\subsection{Cosmic ray heating rates}
\label{sec:cr_heating_rates}

The CR population deposits energy through its interactions, which heats the gas in the cold stream. We write the total volumetric CR
heating rate as: 
\begin{equation}
Q_{\rm tot}(r) = Q_{\rm Coul,p}(r) + Q_{\rm had,p}(r) + Q_{\rm stream,p}(r) + Q_{e}(r)  \ , 
\label{eq:Qheat_total} 
\end{equation}
where the terms on the right-hand side respectively describe
heating by proton Coulomb losses, direct thermalisation in
hadronic pp interactions, proton streaming losses (where the proton energy is lost through self-confinement
and Alfv\'en-wave excitation is assumed to be dissipated
locally in the gas), and 
collisional heating by CR electrons.  
In our calculations, each of these terms is evaluated
separately. For a continuous loss process \(\ell\), the corresponding
local heating rate for a CR species \(j\) is written as
\begin{equation}
Q_{\ell,j}(r)
=
\int_{E_{\min, j}}^{E_{\max, j}}
n_j(E,r)\,b_{\ell,j}(E,r)\,{\rm d}E 
\label{eq:Qi_general}
\end{equation}
where $b_{\ell,j}(E,r)$ 
retains its earlier definition as the energy-loss rate 
associated with that process, assuming 
the energy lost from the CR is thermalised locally. The 
integration limits are chosen according to the energy range of
the relevant CR species, and the expression is applied
separately to the proton and electron distributions obtained in
Section~\ref{sec:cr_transport}.

For protons, we include both collisional Coulomb/ionisation
heating and the gas heating associated with CR streaming losses. The electron contribution is evaluated from 
the full electron distribution, including both the externally
supplied component and the secondaries produced locally by
pp interactions. For electrons, we retain only the collisional
channels that can heat the gas efficiently, namely Coulomb
losses and the additional low-energy collisional term
parameterized through \(t_{\rm ff}\). Other losses (e.g. synchrotron and
inverse-Compton) are treated as radiative emission that does 
not contribute directly to the heating term. 

A separate treatment is required for direct hadronic (pp-mediated) heating by
CR protons. In an inelastic pp interaction, only part of the
energy removed from the CR proton population is thermalised
locally. A substantial fraction instead escapes in the form of
\(\gamma\)-rays and neutrinos, while some is transferred into
secondary electrons and positrons, which are already treated
explicitly through the electron transport equation and their
subsequent collisional heating. We therefore define an
energy-dependent direct-heating efficiency as 
\begin{equation}
f_{\rm heat}(E_p)
=
\max\!\left[0,\,
\kappa(E_p)-f_{\rm sec}(E_p)
\right] ,
\label{eq:fheat_def}
\end{equation} 
where \(\kappa(E_p)\) is the inelasticity of the pp interaction and
\(f_{\rm sec}(E_p)\) is the summed fraction of the proton energy
per collision lost to photons, secondary \(e^\pm\), and neutrinos.
To ensure consistency with our treatment of electron injection,
we evaluate \(f_{\rm sec}\) using the same inclusive secondary
production spectra as adopted to calculate \(S_e\) in
Eq.~\ref{eq:electron_transport_eq}. 
To avoid overestimating the thermalisation at high energy, we
adopt a conservative estimate for the deposited energy per
collision by capping it at the recoil energy of the struck target
nucleon,
\begin{equation}
E_{\rm dep}^{\rm pp}(E_p)
=
f_{\rm kin}\,
K_{\rm recoil}(E_p)
\label{eq:Edep_had}
\end{equation} 
where \(K_{\rm recoil}(E_p)\) is estimated from pp collision
kinematics and \(f_{\rm kin}\) is the fraction of this recoil energy
that ultimately thermalises rather than being lost to
escaping line or recombination radiation. For simplicity, we set
\(f_{\rm kin}=1\), which is appropriate for ionised gas.\footnote{In partially ionized gas near $10^4$ K, $f_{\rm kin}$ may fall below unity. Since this choice does not affect our conclusions, a detailed recoil-thermalisation model is deferred to future work.} 
The resulting direct hadronic heating rate is then: 
\begin{equation}
Q_{\rm had,p}(r)
= \int n_p(E,r)\,
\frac{E_{\rm dep}^{\rm pp}(E)}{t_{\rm pp}(E,r)}
\,{\rm d}E \, .
\label{eq:Q_had_p}
\end{equation} 
This differs from simplified system-averaged treatments that are often adopted, where a fixed fraction (typically 1/6; e.g.~\citealt{Mannheim1994A&A, Guo2008MNRAS, Lin2023MNRAS}) of pp losses is assumed to heat the gas. Our more detailed approach is necessary because the fraction of CR proton energy that thermalises locally is energy-dependent, and because the proton spectrum and collision rate vary along the flow. In a more complete treatment, the final thermalisation efficiency may also depend on local gas conditions through the ionisation state and radiative escape. 

\subsection{CR spectral evolution along the stream}
\label{app:cr_profile}

\begin{figure}[t]
    \centering
    \includegraphics[width=1.\linewidth]{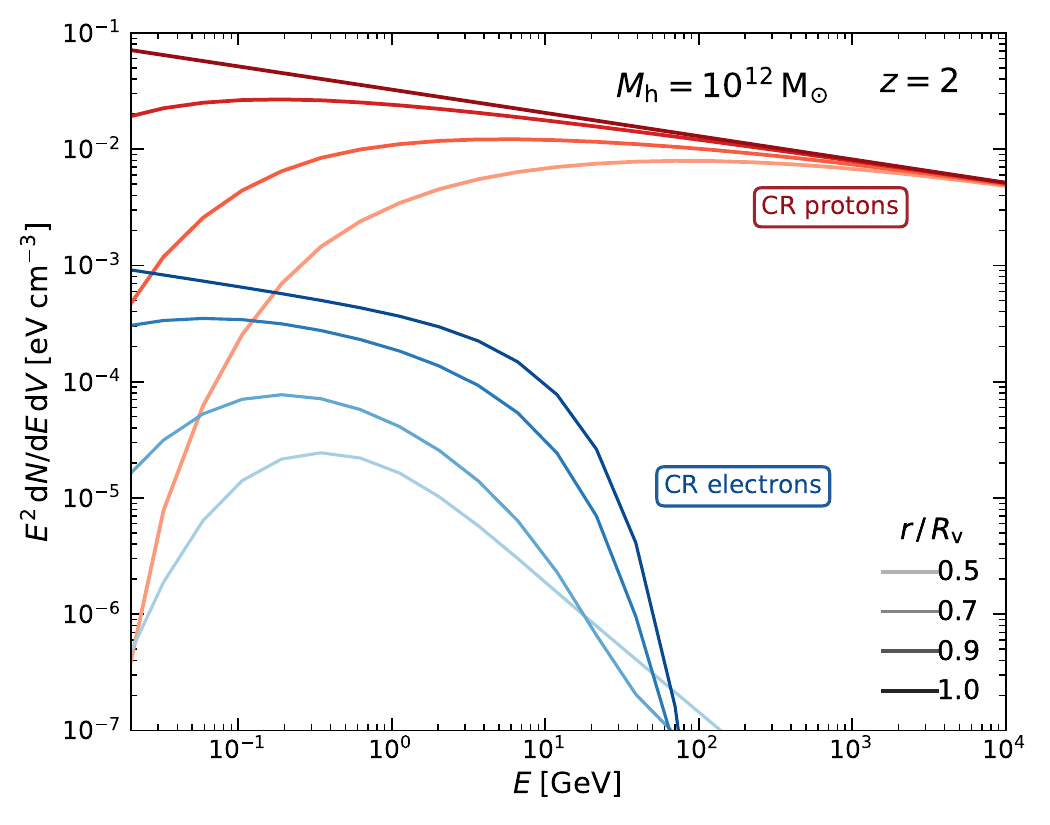}
    \caption{CR proton (red) and electron (blue) energy spectra at four representative positions along the cold stream for the fiducial model ($M_{\rm h}=10^{12}\,\Msun$; $z=2$). Line shading indicates radial position, with lighter tones corresponding to regions closer to the host galaxy.}
    \label{fig:cr_spectrum}
\end{figure}

In our model, the CR proton and electron spectra along a cold stream are set by the fixed boundary spectrum imposed at the virial radius, and are then modified by transport, energy losses, attenuation, and secondary production as the particles propagate inward. 
Because these processes differ for protons and electrons, the two populations evolve differently through the flow. 
This spectral evolution underlies the radial structure of the heating channels discussed in Sec.~\ref{sec:CR_heating_profile}. 
Figure~\ref{fig:cr_spectrum} shows the resulting CR proton and electron distributions at four galactocentric positions in the fiducial model, with $M_{\rm h}=10^{12}\,\Msun$ and $z=2$.

At the outer boundary, we impose the externally supplied CR population as a fixed power-law spectrum,
\begin{equation}
n_j(E,R_v)
=
N_{0,j}
\left(\frac{E}{E_{\min,j}}\right)^{-q} \exp\left(-\frac{E}{E_{{\rm cut},j}}\right) \ ,
\end{equation}
for species $j\in\{\rm p,\rm e\}$, 
where the normalisation $N_{0,j}$ is chosen separately for protons and electrons to reproduce the adopted boundary energy density of each component. 
The total CR energy density at the virial radius is written as
$U_{\mathrm{CR,ext}} = U_p + U_e$, and its partition between protons and electrons is parametrized by
$K_{p/e}\equiv U_p/U_e$. 
At the inner boundary, we impose a reflecting condition. In practice, this choice does not affect our conclusions, since the main CR heating and attenuation effects arise before the CR population reaches the inner edge of the computational domain.

We adopt a source spectral index of $q=2.2$ for both CR protons and electrons, consistent with values commonly inferred for Galactic CRs, although the effects explored in this work are not strongly sensitive to this choice. For electrons, we set the exponential cut-off as $E_{{\rm cut},e} = 10$ GeV to account for their faster cooling prior to entrainment at the virial radius. For protons, we consider there to be no clear spectral cut-off, and set $E_{{\rm cut},p}$ to be sufficiently large that the exponential term is practically absent. 
The proton-to-electron energy-density ratio is uncertain in environments beyond galaxies and is unlikely to be universal. 
For a charge-balanced CR population with equal proton and electron power-law indices at injection, the expected energy-density ratio is 
$U_p/U_e\sim (m_p/m_e)^{(3-q)/2}$~\citep{Persic2014A&A, Persic2015mgm}, which gives $K_{p/e}\sim 20$ for $q\simeq 2.2$. 
Virial-shock models can imply a more electron-rich partition, for example $K_{p/e}\sim 10$~\citep{Yamasaki2024MNRAS}. 
By contrast, standard diffusive-shock-acceleration arguments for galaxy-cluster shocks tend to favour more proton-rich ratios, $K_{p/e}\gtrsim 100$~\citep{Vazza2015MNRAS}. 
In the absence of a firmer constraint, we adopt the more proton-rich value $K_{p/e}=100$ as a conservative fiducial choice, while noting that lower values remain plausible.  

CR protons, shown by the red-shaded curves in Fig.~\ref{fig:cr_spectrum}, are injected at $r/R_{\mathrm{v}}=1$. 
As they are transported inwards, the spectral shape is progressively modified by Coulomb and ionisation losses, adiabatic and streaming-related energy changes, and attenuation by pp interactions. 
At low energies ($E\lesssim 1$ GeV), Coulomb and ionisation losses deplete the spectrum, producing a flattening that becomes more pronounced at smaller radii where the gas density (and, hence, the accumulated column density) is larger. 
At intermediate energies, up to $E\sim 100$ GeV, pp interactions further attenuate the spectrum. 
This suppression accumulates inward and is stronger for lower-energy protons, because their smaller diffusion coefficients lead to longer residence times and larger effective traversed columns. 
These hadronic interactions also inject secondary CR electrons, which become increasingly visible in the inner electron spectra. 
By contrast, at higher energies ($E\gtrsim 100$ GeV), the proton spectrum is modified less strongly, and a substantial CR proton population can survive deeper into the flow. 

The CR electron spectrum, shown by the blue-shaded curves in Fig.~\ref{fig:cr_spectrum}, is also injected at $r/R_{\mathrm{v}}=1$. 
A secondary electron component is additionally produced throughout the flow by pp interactions of the CR protons. 
At high energies, the electron spectrum is strongly shaped by synchrotron and inverse-Compton losses, which steepen the distribution above $E\gtrsim 1$ GeV and generate an effective cooling break that shifts to lower energies at smaller radii, where the magnetic field is stronger. 
At lower energies ($E\lesssim 1$ GeV), the electron spectrum is instead shaped by the competition between Coulomb losses and ongoing secondary injection. 
The secondary component becomes increasingly important toward smaller radii, where the primary electron population has already been depleted by cooling, while the local pp production rate rises with gas density.

\section{Cold stream profiles}
\label{app:stream_prof}

\begin{figure}
    \centering
    \includegraphics[width=.85\linewidth]{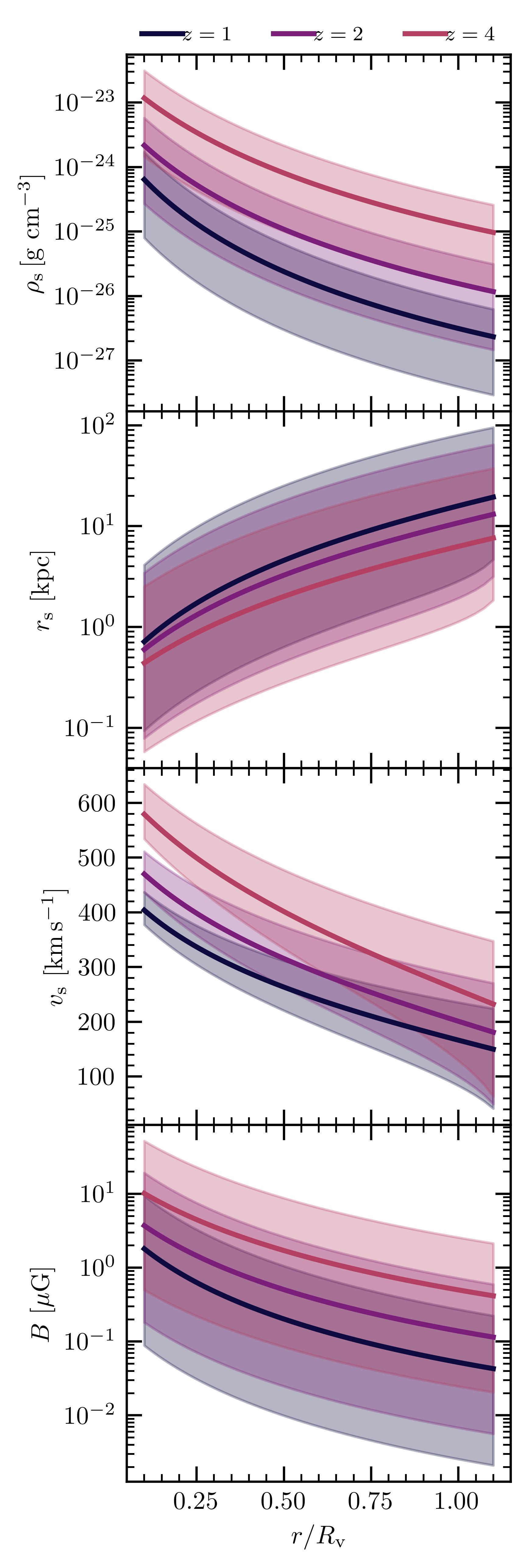}
    \caption{Radial profiles of the cold-stream density, radius, velocity, and magnetic-field strength for a representative $M_{\rm h}=10^{12}\,\rm \Msun$ halo. Results are shown at redshifts $z=1.5$, $2$, and $5$, including the lower and upper stream-density bounds.}
    \label{fig:stream_prof}
\end{figure}

Figure~\ref{fig:stream_prof} shows the radial profiles of the stream properties used in the model described in Sec.~\ref{sec:inflow_halo_model}. 
Results are shown for a representative $M_{\rm h}=10^{12}\,\rm \Msun$ halo at redshifts $z=1.5$, $2$, and $5$, including the lower and upper stream-density bounds adopted in the main text.

\section{Cooling and Heating rates}
\label{app:cooling_net}

\begin{figure}
    \centering
    \includegraphics[width=.92\linewidth]{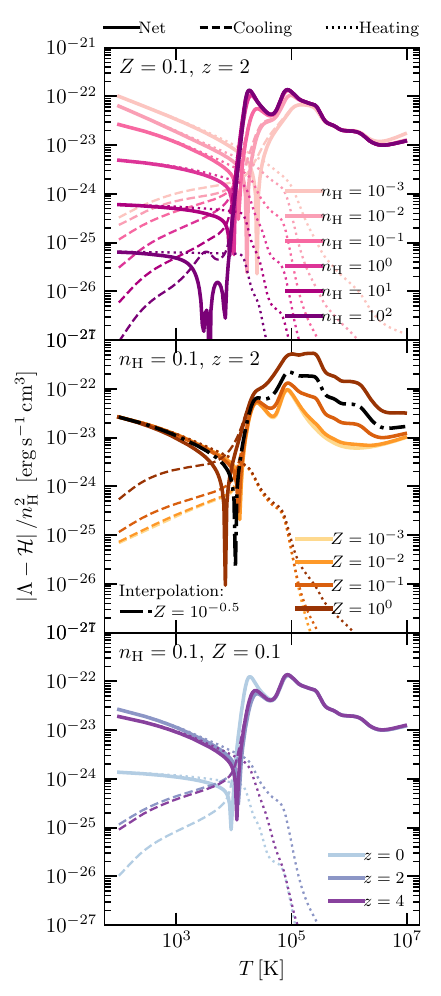}
    \caption{Net cooling (plain line), cooling (dashed line) and heating (dotted line) rates normalised by hydrogen number density squared from the {\tt CLOUDY} model, as a function of temperature. {Top panel:} Curves shown for varying hydrogen number density at fixed metallicity and redshift. {Middle panel:} Curves shown for varying metallicity at fixed hydrogen number density and redshift. The dashed-dotted black line shows an example of the interpolation function we use, here for $Z=10^{-0.5}\, Z_\odot$. {Bottom panel:} Curves shown for varying redshift at fixed hydrogen number density and metallicity.}
    \label{fig:cooling_curves}
\end{figure}

Figure~\ref{fig:cooling_curves} shows the gas heating (photoheating from the UV background, without CRs), cooling and their combined net rates ($\Lambda_\mathrm{net}$) as a function of temperature used in our calculations, obtained from the {\tt CLOUDY} models described in Sec.~\ref{sec:timescales}. Curves at intermediate parameter values, for which no dedicated {\tt CLOUDY} model was run, are obtained by interpolation. An example of such an interpolated curve is shown by the black dash-dotted line in the middle panel.  

\section{Cosmic ray energies and timescales} 
\label{app:cr_timescale_energy}

The diagnostic timescale analysis in Sec.~\ref{sec:timescales} is evaluated at a representative CR kinetic energy of $E = 10$ MeV, chosen to highlight the low-energy proton population that most efficiently deposits energy through Coulomb and ionisation losses. However, the CR heating efficiency depends on particle energy, and the full transport calculations presented in the main text are spectrally resolved.
Here, we repeat the fixed-energy timescale analysis at higher representative CR energies to illustrate how the balance between CR heating, radiative cooling, and stream advection changes across the low-energy CR spectrum.   

\begin{figure}
    \centering  
    \includegraphics[width=1\linewidth]{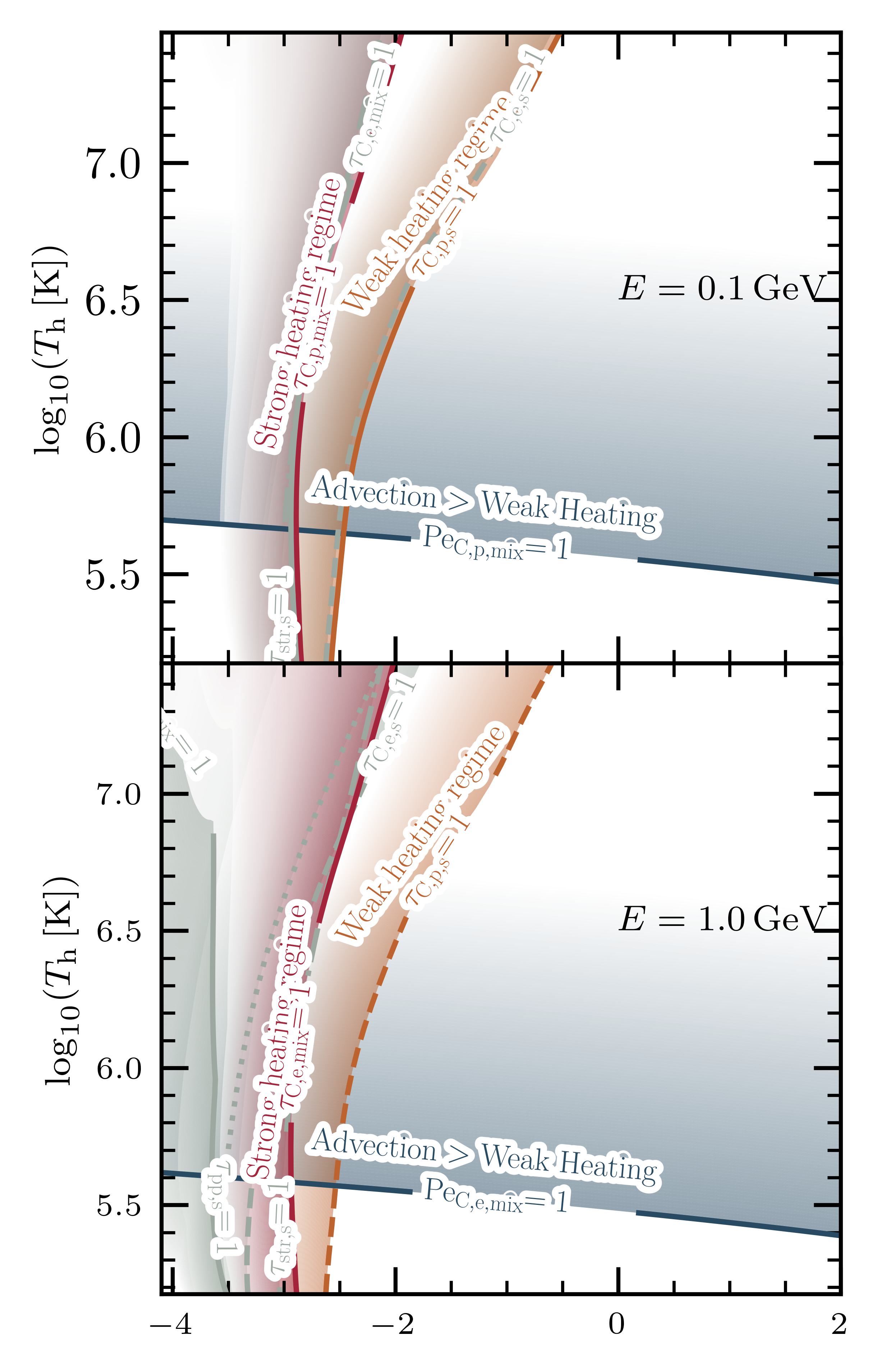}
    \caption{Same as Fig.~\ref{fig:timescale_map} for $E=0.1$ and $1.0$ GeV in the top and bottom panels, respectively. Red and orange contours show the dominant CR-heating timescale to gas cooling timescale ratios in the mixing layer and in the stream, respectively. As for $E=10$ MeV in Fig.~\ref{fig:timescale_map}, the dominant heating mechanism for $E=0.1$ GeV is proton-Coulomb losses, while for $E=1$ GeV, electron-Coulomb losses become dominant. Light grey contours show sub-dominant heating channels, including direct pp heating in the stream and streaming. 
Coloured contours indicate where CR heating is faster than radiative cooling, extending to $\tau_{\ell}=0.1$. 
The dark blue line shows the ratio between the dominant CR heating time and the local stream advection time, while dark blue contours indicate the regime where advection is faster than heating, extending to $t_{{\rm heat},C,{\rm p/e}}/t_{\rm dyn}=10$. Here, advection dominates across almost the entire parameter space. 
}
    \label{fig:timescale_map_100MeV}
\end{figure}

The top panel of Fig.~\ref{fig:timescale_map_100MeV}  shows the corresponding timescale balance for $E = 100$ MeV. Compared to the 
$E = 10$ MeV case shown in Fig.~\ref{fig:timescale_map}, proton Coulomb heating is less efficient over most of the parameter space, and the region where CR heating can act before local stream advection is reduced. At this energy, the stream is generally advected or compressed faster than it is heated in the parameter range shown. The bottom panel of Fig.~\ref{fig:timescale_map_100MeV} shows the same diagnostic at $E = 1$ GeV, which additionally begins to show the impact of hadronic pp heating, which is able to operate at these higher energies. These comparisons show that the most efficient CR heating of cold-stream gas is driven by the low-energy part of the proton spectrum.
They also motivate the spectrally resolved treatment adopted in the main calculations, since the total CR heating rate depends sensitively on how much of the externally supplied CR population survives at sub-GeV energies.  

\section{Cosmic ray diffusion coefficient} 
\label{sec:diffusion_study}

\begin{figure}
    \centering
    \includegraphics[width=1.\linewidth]{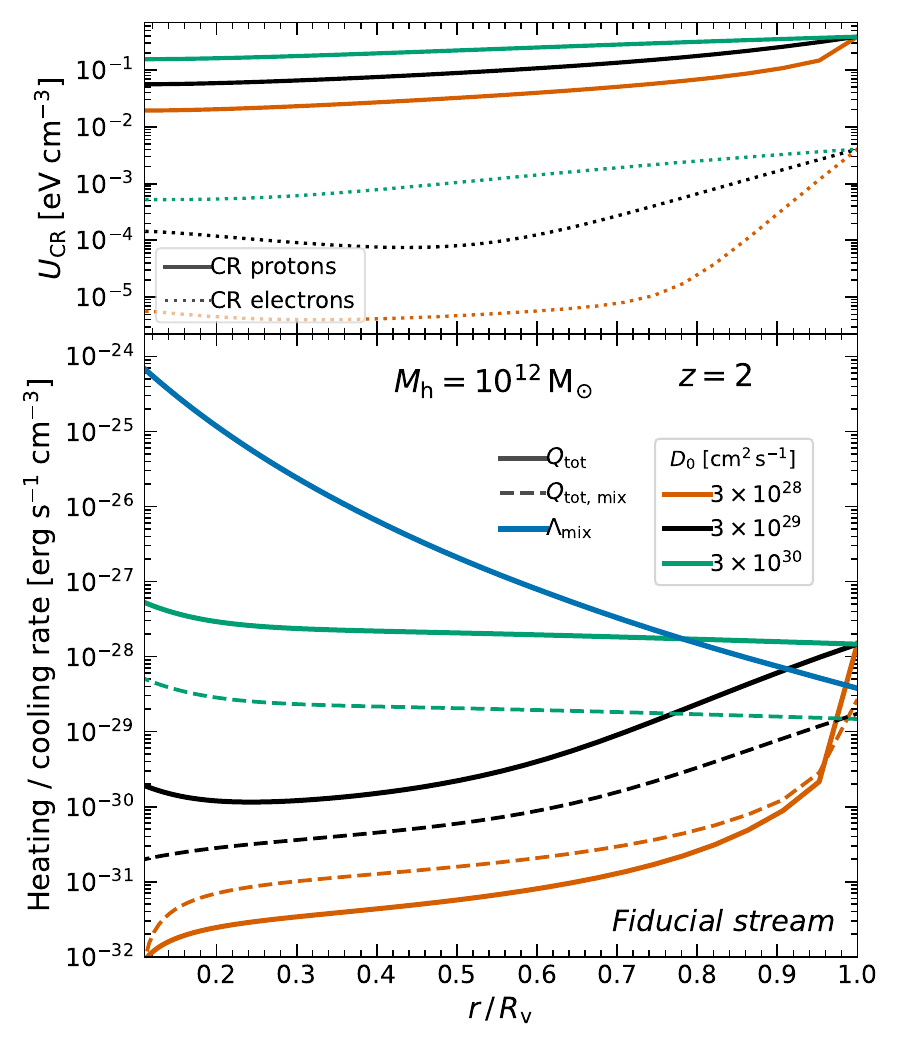}
    \caption{Top: CR distribution along the cold stream in 
    the fiducial ($M_{\rm h}=10^{12}\,\Msun$; $z=2$) model, for different choices of the normalization of the CR diffusion coefficient, $D_{0}$ (colours corresponding to values indicated by the legend in the bottom panel). 
    Bottom: Corresponding radial heating and cooling profiles with radius normalized to the halo virial radius ($R_{\rm v}$), where 
    the blue solid line shows the (volumetric) radiative cooling rate
$\Lambda_{\rm mix}(r)$ in the mixing layer, while the coloured lines show the total CR heating
rate in the stream's spine ($Q_{\rm tot}$, solid lines) and mixing layer ($Q_{\rm tot, mix}$, dashed lines) for each choice of CR diffusion coefficient, $D_{0}$. The results for a diffuse stream are shown in Fig.~\ref{fig:heating_profile_diffuse2}.} 
    \label{fig:heating_profile2}
\end{figure}

\begin{figure}
    \centering
    \includegraphics[width=1.\linewidth]{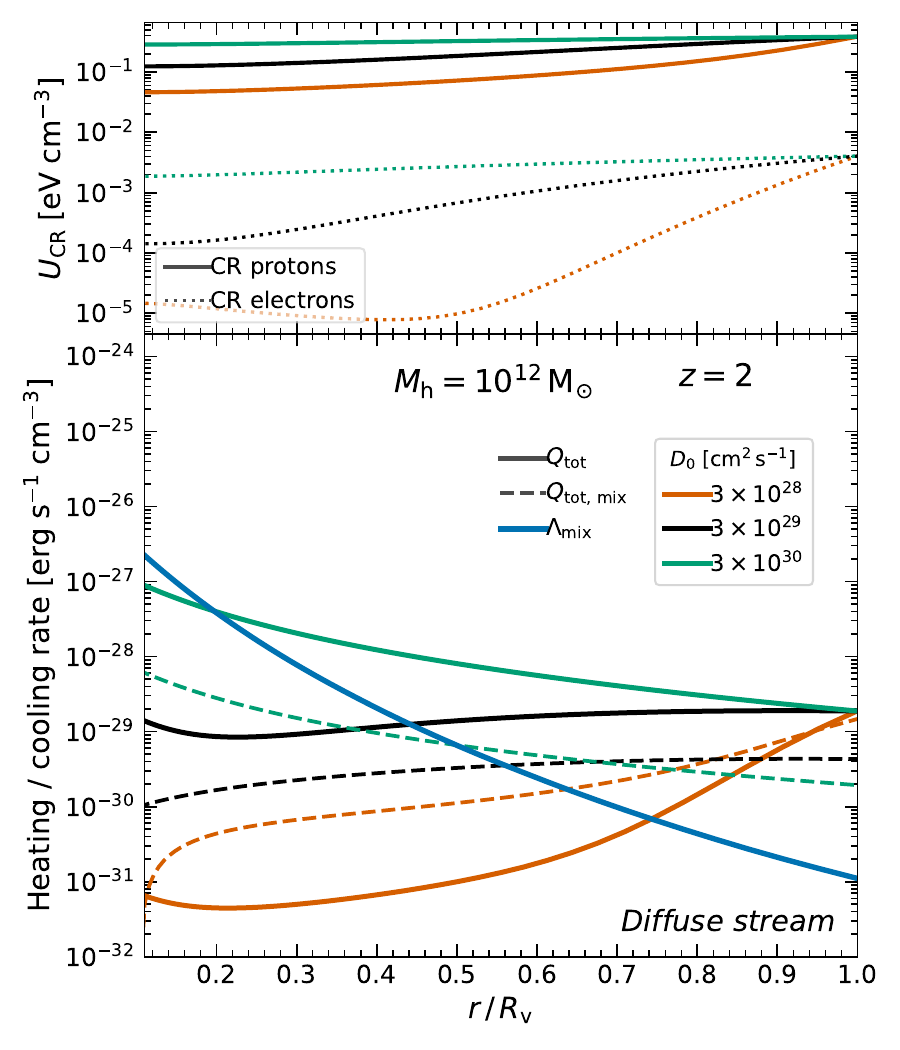}
    \caption{Same as Fig.~\ref{fig:heating_profile2}, but for a diffuse stream.} 
    \label{fig:heating_profile_diffuse2}
\end{figure}

In this work, several model parameters are set to representative values. Among these, the normalisation of the CR diffusion coefficient, $D_{0}$, is particularly important because it regulates how efficiently externally supplied CRs are redistributed along the stream before they deposit their energy. Its value is also not tightly constrained in halo environments. As a fiducial choice, we adopt $D_{0}=3\times10^{29} \;{\rm cm^{2} \; s^{-1}}$ for both CR protons and electrons. This is representative of values used in calibrated galaxy halo-scale CR transport calculations~\citep[e.g.][]{Chan2019MNRAS,Ji2020MNRAS}. However, the appropriate effective value in cold streams and their surrounding mixing layers remains uncertain. It is expected to exceed the typical Galactic ISM value of $3\times10^{28} \; {\rm cm^{2} \; s^{-1}}$ often adopted for few-GeV CR protons~\citep{Strong2007ARNPS}, but should remain below the tentative upper limit of $5\times10^{30} \; {\rm cm^{2} \; s^{-1}}$ inferred from $\gamma$-ray halo interpretations of extended high-energy emission around M31~\citep{Recchia2021ApJ}. This broad possible range motivates an explicit check of how our results depend on $D_{0}$. In Figs.~\ref{fig:heating_profile2} and~\ref{fig:heating_profile_diffuse2}, we therefore repeat the fiducial and diffuse-stream calculations with $D_{0}=3\times10^{28}$, $3\times10^{29}$, and $3\times10^{30} \; {\rm cm^{2} \; s^{-1}}$. 

These results show that changing $D_{0}$ modifies the radial distribution of CRs and therefore the detailed heating profile. Lower values of $D_{0}$ confine the externally supplied CR population closer to the outer boundary, producing a steeper radial CR profile and reducing the heating rate deeper inside the stream. Higher values of $D_{0}$ allow CRs to diffuse more efficiently inward, smoothing the CR distribution and increasing the heating rate at smaller galactocentric radii. This behaviour is visible in both the fiducial and diffuse-stream cases. The effect is strongest where the heating and cooling rates are already comparable, because modest changes in the CR residence time or penetration depth can shift the radius where CR heating becomes thermally important. The mixing-layer heating profile is affected in the same qualitative way, although the absolute impact depends on the lower gas density and cooling rate of the mixed phase.  

The qualitative conclusions of this work are unchanged across the range of $D_{0}$ explored here. 
Dense or fiducial stream material remains comparatively resilient to CR heating through most of the halo, with the strongest heating confined to the outer stream where the CR population is least attenuated and radiative cooling is weaker. Diffuse, lower-column stream material remains much more susceptible, because CRs penetrate more effectively and the cooling rate is lower. Increasing $D_{0}$ can extend the region where CR heating competes with cooling, while decreasing $D_{0}$ makes the heating more localized near the virial boundary. This demonstrates that $D_{0}$ should be regarded as an uncertainty in the spatial pattern and quantitative level of CR heating, but it is not a parameter that changes the basic physical interpretation of our results. 

\end{document}